\documentclass[twocolumn]{aastex62}

\usepackage{xcolor}

\graphicspath{{./}{figures/}}

\shorttitle{TESS RR Lyrae first light results}
\shortauthors{Moln\'ar et al.}

\begin{document}

\title{First results on RR Lyrae stars with the TESS space telescope: untangling the connections between mode content, colors and distances }

\correspondingauthor{L. Moln\'ar}
\email{lmolnar@konkoly.hu, lacalaca85@gmail.com}

\author[0000-0002-8159-1599]{L. Moln\'ar}
\affiliation{Konkoly Observatory, Research Centre for Astronomy and Earth Sciences, E\"otv\"os Lor\'and Research Network (ELKH), Konkoly Thege Mikl\'os \'ut 15-17, H-1121 Budapest, Hungary}
\affiliation{MTA CSFK Lend\"ulet Near-Field Cosmology Research Group, 1121, Budapest, Konkoly Thege Mikl\'os \'ut 15-17, Hungary}
\affiliation{ELTE E\"otv\"os Lor\'and University, Institute of Physics, 1117, P\'azm\'any P\'eter s\'et\'any 1/A, Budapest, Hungary}

\author[0000-0002-8585-4544]{A. B\'odi}
\affiliation{Konkoly Observatory, Research Centre for Astronomy and Earth Sciences, E\"otv\"os Lor\'and Research Network (ELKH), Konkoly Thege Mikl\'os \'ut 15-17, H-1121 Budapest, Hungary}
\affiliation{MTA CSFK Lend\"ulet Near-Field Cosmology Research Group, 1121, Budapest, Konkoly Thege Mikl\'os \'ut 15-17, Hungary}
\affiliation{ELTE E\"otv\"os Lor\'and University, Institute of Physics, 1117, P\'azm\'any P\'eter s\'et\'any 1/A, Budapest, Hungary}

\author[0000-0001-5449-2467]{A. P\'al}
\affiliation{Konkoly Observatory, Research Centre for Astronomy and Earth Sciences, E\"otv\"os Lor\'and Research Network (ELKH), Konkoly Thege Mikl\'os \'ut 15-17, H-1121 Budapest, Hungary}
\affiliation{ELTE E\"otv\"os Lor\'and University, Institute of Physics, 1117, P\'azm\'any P\'eter s\'et\'any 1/A, Budapest, Hungary}
\affiliation{MIT Kavli Institute for Astrophysics and Space Research, 70 Vassar St, Cambridge, MA 02109, USA}

\author[0000-0001-6147-3360]{A. Bhardwaj}
\affiliation{Korea Astronomy and Space Science Institute, Daedeokdae-ro 776, Yuseong-gu, Daejeon 34055, Republic of Korea}

\author{F--J. Hambsch}
\affiliation{Vereniging Voor Sterrenkunde (VVS), Oostmeers 122 C, 8000 Brugge, Belgium}
\affiliation{Bundesdeutsche Arbeitsgemeinschaft f\"ur Ver\"anderliche Sterne e.V. (BAV), Munsterdamm 90, D-12169 Berlin, Germany}

\author[0000-0003-3851-6603]{J.~M. Benk\H{o}}
\affiliation{Konkoly Observatory, Research Centre for Astronomy and Earth Sciences, E\"otv\"os Lor\'and Research Network (ELKH), Konkoly Thege Mikl\'os \'ut 15-17, H-1121 Budapest, Hungary}
\affiliation{MTA CSFK Lend\"ulet Near-Field Cosmology Research Group, 1121, Budapest, Konkoly Thege Mikl\'os \'ut 15-17, Hungary}

\author[0000-0002-6526-9444]{A. Derekas}
\affiliation{ELTE E\"otv\"os Lor\'and University, Gothard Astrophysical Observatory, Szent Imre h. u. 112, 9700, Szombathely, Hungary}
\affiliation{MTA-ELTE Exoplanet Research Group, Szent Imre h. u. 112, 9700 Szombathely, Hungary}
\affiliation{Konkoly Observatory, Research Centre for Astronomy and Earth Sciences, E\"otv\"os Lor\'and Research Network (ELKH), Konkoly Thege Mikl\'os \'ut 15-17, H-1121 Budapest, Hungary}

\author{M. Ebadi}
\affiliation{Institute of Geophysics, University of Tehran, Tehran, Iran}

\author[0000-0002-8717-127X]{M. Joyce}
\affiliation{Space Telescope Science Institute, 3700 San Martin Drive, Baltimore, MD 21218, USA}
\affiliation{Research School of Astronomy and Astrophysics, Australian National University, Canberra, ACT 2611, Australia}
\affiliation{ARC Centre of Excellence for All Sky Astrophysics in 3 Dimensions (ASTRO 3D), Australia}

\author[0000-0002-7286-1438]{A. Hasanzadeh} 
\affiliation{Institute of Geophysics, University of Tehran, Tehran, Iran}

\author[0000-0001-5181-5072]{K. Kolenberg}
\affiliation{Institute of Astronomy, KU Leuven, Celestijnenlaan 200D, B-3001 Heverlee, Belgium}
\affiliation{Physics Department, University of Antwerp, Groenenborgerlaan 171, B-2020
Antwerpen, Belgium}
\affiliation{Physics and Astronomy Department, Vrije Universiteit Brussel (VUB), Pleinlaan 2, 1050 Brussel}

\author[0000-0003-2527-1598]{M.~B. Lund}
\affiliation{Caltech/IPAC, 1200 E. California Blvd. Pasadena, CA 91125, USA}

\author[0000-0003-1281-5525]{J. M. Nemec}
\affiliation{Department of Physics \& Astronomy, Camosun College, Victoria, BC, Canada}

\author[0000-0001-5608-0028]{H. Netzel}
\affiliation{Konkoly Observatory, Research Centre for Astronomy and Earth Sciences, E\"otv\"os Lor\'and Research Network (ELKH), Konkoly Thege Mikl\'os \'ut 15-17, H-1121 Budapest, Hungary}
\affiliation{MTA CSFK Lend\"ulet Near-Field Cosmology Research Group, 1121, Budapest, Konkoly Thege Mikl\'os \'ut 15-17, Hungary}
\affiliation{Nicolaus Copernicus Astronomical Center, Polish Academy of Sciences, Bartycka 18, PL-00-716 Warsaw, Poland}

\author[0000-0001-8771-7554]{C.--C. Ngeow}
\affiliation{Graduate Institute of Astronomy, National Central University, Jhongli 32001, Taiwan}

\author[0000-0002-3827-8417]{J. Pepper}
\affiliation{Department of Physics, Lehigh University, 16 Memorial Drive East, Bethlehem, PA 18015, USA}

\author[0000-0002-5481-3352]{E. Plachy}
\affiliation{Konkoly Observatory, Research Centre for Astronomy and Earth Sciences, E\"otv\"os Lor\'and Research Network (ELKH), Konkoly Thege Mikl\'os \'ut 15-17, H-1121 Budapest, Hungary}
\affiliation{MTA CSFK Lend\"ulet Near-Field Cosmology Research Group, 1121, Budapest, Konkoly Thege Mikl\'os \'ut 15-17, Hungary}

\author[0000-0001-5497-5805]{Z. Prudil}
\affiliation{Astronomisches Rechen-Institut, Zentrum f\"ur Astronomie der Univ\"ersitat Heidelberg, M\"onchhofstr. 12-14, D-69120 Heidelberg, Germany}

\author[0000-0001-5016-3359]{R.~J. Siverd}
\affiliation{Gemini Observatory/NSF's NOIRLab, 670 N. A`ohoku Place, Hilo, HI, 96720, USA}

\author[0000-0002-7602-0046]{M. Skarka}
\affiliation{Department of Theoretical Physics and Astrophysics, Masaryk University, Kotl\'{a}\v{r}sk\'{a} 2, 61137 Brno, Czech Republic}
\affiliation{Astronomical Institute, Czech Academy of Sciences, Fri\v{c}ova 298, 25165, Ond\v{r}ejov, Czech Republic}

\author[0000-0001-7217-4884]{R. Smolec}
\affiliation{Nicolaus Copernicus Astronomical Center, Polish Academy of Sciences, Bartycka 18, PL-00-716 Warsaw, Poland}

\author[0000-0001-7806-2883]{\'A. S\'odor}
\affiliation{Konkoly Observatory, Research Centre for Astronomy and Earth Sciences, E\"otv\"os Lor\'and Research Network (ELKH), Konkoly Thege Mikl\'os \'ut 15-17, H-1121 Budapest, Hungary}

\author{S. Sylla}
\affiliation{Universit\`e Cheikh Anta Diop Dakar-Fann, S\`en\`egal}
\affiliation{Physics Department, University of Antwerp, Groenenborgerlaan 171, B-2020
Antwerpen, Belgium}

\author[0000-0002-5781-1926]{P. Szab\'o}
\affiliation{Konkoly Observatory, Research Centre for Astronomy and Earth Sciences, E\"otv\"os Lor\'and Research Network (ELKH), Konkoly Thege Mikl\'os \'ut 15-17, H-1121 Budapest, Hungary}
\affiliation{MTA CSFK Lend\"ulet Near-Field Cosmology Research Group, 1121, Budapest, Konkoly Thege Mikl\'os \'ut 15-17, Hungary}
\affiliation{Magdalene College, University of Cambridge, CB3 0AG, Cambridge, UK}

\author[0000-0002-3258-1909]{R. Szab\'o}
\affiliation{Konkoly Observatory, Research Centre for Astronomy and Earth Sciences, E\"otv\"os Lor\'and Research Network (ELKH), Konkoly Thege Mikl\'os \'ut 15-17, H-1121 Budapest, Hungary}
\affiliation{MTA CSFK Lend\"ulet Near-Field Cosmology Research Group, 1121, Budapest, Konkoly Thege Mikl\'os \'ut 15-17, Hungary}
\affiliation{ELTE E\"otv\"os Lor\'and University, Institute of Physics, 1117, P\'azm\'any P\'eter s\'et\'any 1/A, Budapest, Hungary}

\author[0000-0002-9037-0018]{H. Kjeldsen}
\affiliation{Stellar Astrophysics Centre, Department of Physics and Astronomy, Aarhus University, Ny Munkegade 120, DK-8000 Aarhus C, Denmark}
\affiliation{Astronomical Observatory, Institute of Theoretical Physics and Astronomy, Vilnius University, Saulėtekio av. 3, 10257, Vilnius, Lithuania}

\author[0000-0001-5137-0966]{J. Christensen-Dalsgaard}
\affiliation{Stellar Astrophysics Centre, Department of Physics and Astronomy, Aarhus University, Ny Munkegade 120, DK-8000 Aarhus C, Denmark}

\author[0000-0003-2058-6662]{G.~R. Ricker}
\affiliation{MIT Kavli Institute for Astrophysics and Space Research, 70 Vassar St, Cambridge, MA 02109, USA}

\date{Submitted: 23/06/2021; Accepted: 14/09/2021}



\begin{abstract}
The TESS space telescope is collecting continuous, high-precision optical photometry of stars throughout the sky, including thousands of RR Lyrae stars. In this paper, we present results for an initial sample of 118 nearby RR Lyrae stars observed in TESS Sectors 1 and 2. We use differential-image photometry to generate light curves and analyse their mode content and modulation properties. We combine accurate light curve parameters from TESS with parallax and color information from the \textit{Gaia} mission to create a comprehensive classification scheme. We build a clean sample, preserving RR Lyrae stars with unusual light curve shapes, while separating other types of pulsating stars. We find that a large fraction of RR Lyrae stars exhibit various low-amplitude modes, but the distribution of those modes is markedly different from those of the bulge stars. This suggests that differences in physical parameters have an observable effect on the excitation of extra modes, potentially offering a way to uncover the origins of these signals. However, mode identification is hindered by uncertainties when identifying the true pulsation frequencies of the extra modes. We compare mode amplitude ratios in classical double-mode stars to stars with extra modes at low amplitudes and find that they separate into two distinct groups. Finally, we find a high percentage of modulated stars among the fundamental-mode pulsators, but also find that at least 28\% of them do not exhibit modulation, confirming that a significant fraction of stars lack the Blazhko effect. 
\end{abstract}



\section{Introduction} \label{sec:intro}
RR Lyrae variable stars are old, low-mass stars in the core He-burning phase and thus they occupy the horizontal branch of the Hertzsprung--Russell diagram. They pulsate radially with large amplitudes and short periods (typically between 0.25-1.0 days). They can be found in large numbers throughout the Milky Way Galaxy, in the bulge, thick disk, halo, globular clusters, and in various dwarf galaxies and stellar streams in and around our Galaxy. 

The study of RR Lyrae stars began over a century ago \citep{kapteyn-1890,bailey1902,Pickering-1901}. Initially, they were considered a homogeneous group of bright variable stars that are abundant in the Milky Way and found to be useful as distance and population indicators. However, as \citet{Preston-1964} remarked in his review of the group, intriguing differences soon emerged. These included the Oosterhoff dichotomy, which separates the stars (or, at first, the globular clusters they populate) into two groups in the period--amplitude plane \citep{oosterhoff-1939}. The dichotomy has 
since been
explained through detailed evolutionary calculations and spectroscopic measurements of metal-rich and metal-poor RR Lyrae stars that 
reside in
various clusters and other structures in the Milky Way  \citep[see, e.g.,][]{Sollima-2014,Fabrizio-2019,Prudil-2019}. 

Short-period pulsators, below 0.2~d, were recognized to be different types of stars, today known as $\delta$~Sct and SX~Phe variables \citep{Smith-1955,woltjer-1956}. The long variations observed in some stars, first described by \citet{blazko-1907}, were proving difficult to explain, especially the 41--d variation in the star RR~Lyr itself. This slow modulation of the pulsation is known today as the Blazhko effect, named after its discoverer. Later, the first double-mode stars were identified and used to constrain the masses of RR Lyrae stars \citep{Cox-1980}.

With increasing computing power, pulsation models improved \citep[see, e.g.,][]{Bono-1994,kollath-2002}, and large-scale an/or dedicated surveys were initiated \citep{macho-1999,ogle-2003,jurcsik2009}. The newest generation of large photometric, astrometric and spectroscopic surveys massively expanded the number of known, observed and classified RR~Lyrae stars, which now measures in the hundreds of thousands \citep{sesar-2017,holl2018,clementini2019,soszynski-2019,Liu-2020}. Mining these large data sets led to new discoveries in numerous areas, from the structure and collisional history of the Miky Way \citep{Moretti-2014,hernitschek-2017,Iorio-2019,prudil2021} to chemical evolution \citep{Crestani-2021} to the detection of shockwaves and associated emission lines in overtone and double-mode stars \citep{Duan-1-2021,Duan-2-2021}. 

While great advances were made, some aspects of RR Lyrae stars have still remained poorly understood. Despite numerous hypotheses, the Blazhko effect has not been explained satisfactorily. Mass determinations remained controversial thanks to uncertainties in the pulsation models and the nearly complete lack of confirmed RR Lyrae binaries \citep{hajdu-2015,Liska-2016,Skarka-2018}. 
Long-term photometric and spectroscopic monitoring and high-precision astrometry have recently begun to also reveal RR Lyrae stars with companions \citep{sodor-2017,kervella-2019-2,Barnes-2021,Hajdu-2021}, but the first candidate RR Lyrae-type variable in an eclipsing binary system for which mass could be determined was ultimately identified as a low-mass binary pulsator rather than a \textit{bona fide} RR Lyrae star \citep{bep}.

Upon entering the era of photometric space missions, the very first observation of the prototype double-mode star, AQ Leo, with the MOST space telescope delivered the first discovery of low-amplitude additional modes in RR Lyrae stars \citep{gruberbauer2007}. The CoRoT and \textit{Kepler} light curves showed how diverse the appearances of the Blazhko effect can be, with a multitude of varying and multi-periodic modulations detected \citep{guggenberger-2011,guggenberger2012,benko2014}. These observations revealed further dynamical effects in the pulsation including period doubling of the fundamental mode and various additional modes in many stars \citep{chadid-2010,szabo-2010,szabo2014}. 

One drawback of the \textit{Kepler} and K2 observations, however, is that nearly all of the observed stars 1) were pre-selected based on earlier observations and surveys, 2) were limited to small areas on the sky, and 3) were mostly faint and distant targets \citep{Howell2014,Borucki-2016}. While this made it possible to reach the edge of the Galactic halo and nearby dwarf galaxies, it also meant that spectroscopic follow-up and the calibration of the photometric [Fe/H] relation for the \textit{Kepler} passband required the largest ground-based telescopes \citep{molnar-leoiv-2015,nemec-2013}. 

In contrast, the TESS (Transiting Exoplanet Survey Satellite) space telescope follows a different strategy \citep{ricker2015}. It uses four small cameras that provide an enormous, 24\degr{}$\times\,$96\degr{} field-of-view, at the cost of lower resolution and depth than those of \textit{Kepler}. Benefits and limitations of TESS concerning RR Lyrae and Cepheid stars were summarized by \citet{Plachy-TESS-2020}. Briefly, the observing strategy of the mission will cover nearly the entire sky, therefore including the brightest RR Lyrae stars. This will make it possible to combine the detailed photometric analysis with extensive spectroscopic observations and with precise geometric parallax and proper motion data from the European \textit{Gaia} mission \citep{gaia2016}. We expect the faint limit of TESS to be around 16--17~mag, depending on the crowding and the required level of precision. The number density of RR~Lyrae stars within the Galaxy peaks around this brightness, although most of them are fainter \citep{Plachy-TESS-2020}. Nevertheless, we expect that TESS will be able to characterize a few tens of thousands of stars with varying degrees of accuracy. 

TESS confers the ability to characterize the short-term variability of a large sample of nearby RR~Lyrae stars, including precise light curve shape, (sub)mmag-level additional modes and short-period modulations. Furthermore, the densely sampled light curves allow us to test classifications based on sparse photometry and to create a cleaner sample of nearby RR~Lyrae stars, which can in turn be used to construct a more precise period-luminosity (PL) relation when combined with \textit{Gaia} parallaxes. Precise PL (and period-luminosity-metallicity) relations can then be used as a separate distance scale anchor that relies on Population II stars \citep{Beaton-2016,Neeley-2019}.

This paper presents the first results obtained for RR Lyrae stars with TESS. A companion paper details the first results on various Cepheid-type stars \citep{Plachy-Cep-2020}. The paper is structured as follows: we describe our photometry method in Sect.~\ref{sect:data}. Results, including classification, detailed light curve analysis and mode contents are described in Sect.~\ref{sect:results}. We compare our first results to ground-based photometry and pulsation models in Sect.~\ref{sect:models}. Finally, we discuss the future prospects of RR Lyrae research with TESS and draw our conclusions in Sections \ref{sect:future} and \ref{sect:conclusions}.

\section{Data and methods}
\label{sect:data}
Data used in this paper come from the TESS and \textit{Gaia} space missions. We introduced TESS in the preceding section and present the data reduction step below. The other source, the \textit{Gaia} mission, is collecting high-precision astrometric and sparse photometric observations throughout the entire sky \citep{gaia2016}. Processed \textit{Gaia} data are released in batches: the most recent one, Early Data Release 3, contains astrometry and three-band average photometry for over 1.8 billion stars \citep{gaia-edr3-2020}. The previous release, DR2, also included, among other products, radial velocity (RV) measurements and variable candidate classifications \citep{gaia2018}.

\subsection{TESS observations}

TESS observations are separated into so-called \emph{sectors}. During each sector, the spacecraft pointing is kept constant with respect to the celestial reference frame. Each sector is made up 
from two consecutive \emph{orbits} when the spacecraft orbits Earth with a period of half of one sidereal month ($P_{\rm TESS}\approx 13.7\,{\rm d}$). Observations are gathered continuously for one orbit, and the collected data are downloaded at perigee. Sectors 1 and 2 lasted for 27.9 and 27.4 d, with 1.13 and 1.44 d long mid-sector gaps, respectively. During each sector, the entire field-of-view is recorded as Full-Frame Images (FFI) at a 30-minute cadence, while selected targets are measured with 2-minute cadence. 

As TESS is orbiting Earth, it is affected by scattered or direct light entering the cameras both from Earth and the Moon. Most of Sector 1 was affected by diffuse light from Earth that was modulated by the planet's rotation. Glinting from the Moon affected the last few cadences in Camera 1. Mars was outside but close to the field-of-view of Camera 1, and so it also affected some observations in Sector 1. Moreover, a 2-day long segment was affected by excess jitter in the spacecraft pointing\footnote{TESS Data Release Notes are available at \url{https://archive.stsci.edu/tess/tess_drn.html}}. 

Sector 2 was much less eventful, with scattered light from Earth only entering the cameras towards the ends of each orbit. Scattered light manifests itself as a large, diffuse, slow-moving patch of excess light over the background in the images, raising the local background level.

\subsection{Targets and data reduction}
In Sectors 1 and 2, three targets were observed as 2 min cadence targets, part of the TASC (TESS Asteroseismic Science Consortium) target list (ST~Pic in both sectors, BV~Aqr in S1, and RU~Scl in S2). The rest of the RR~Lyrae stars were FFI targets. We searched the SIMBAD database and the \textit{Gaia} DR2 variable star candidate catalogs of \citet{clementini2019} and \citet{holl2018} for RR Lyrae stars. We limited the sample to stars brighter than approximately 12.5 mag in the TESS passband for Cameras 1--3 and brighter than 14.0 mag in Camera 4. We discarded a few targets that showed severe instrumental noise and/or contamination in their light curves.  

\subsubsection{2-minute targets}
TESS 2 min cadence targets are observed in dedicated postage stamp images, and both the corrected target pixel files and reduced light curves are released by the SPOC (Science Processing Operations Center). The same target pixel files have been reduced by TASOC (TESS Asteroseismic Science Operations Center\footnote{\url{https://tasoc.dk}}) as well. However, their initial release utilized an older pipeline that was developed for solar-like oscillators observed by the \textit{Kepler} mission and hence does not handle large-amplitude variables well. We investigated both sets of light curves and decided to create our own photometry based on custom pixel apertures with the \textsc{lightkurve} tool, and, in particular, its interactive pixel selection option \citep{lightkurve}. This approach provided full control over selecting the pixel apertures, and yielded good-quality light curves for use in the frequency analysis.

\begin{figure*}
\begin{center}
\noindent
\resizebox{53mm}{!}{\includegraphics{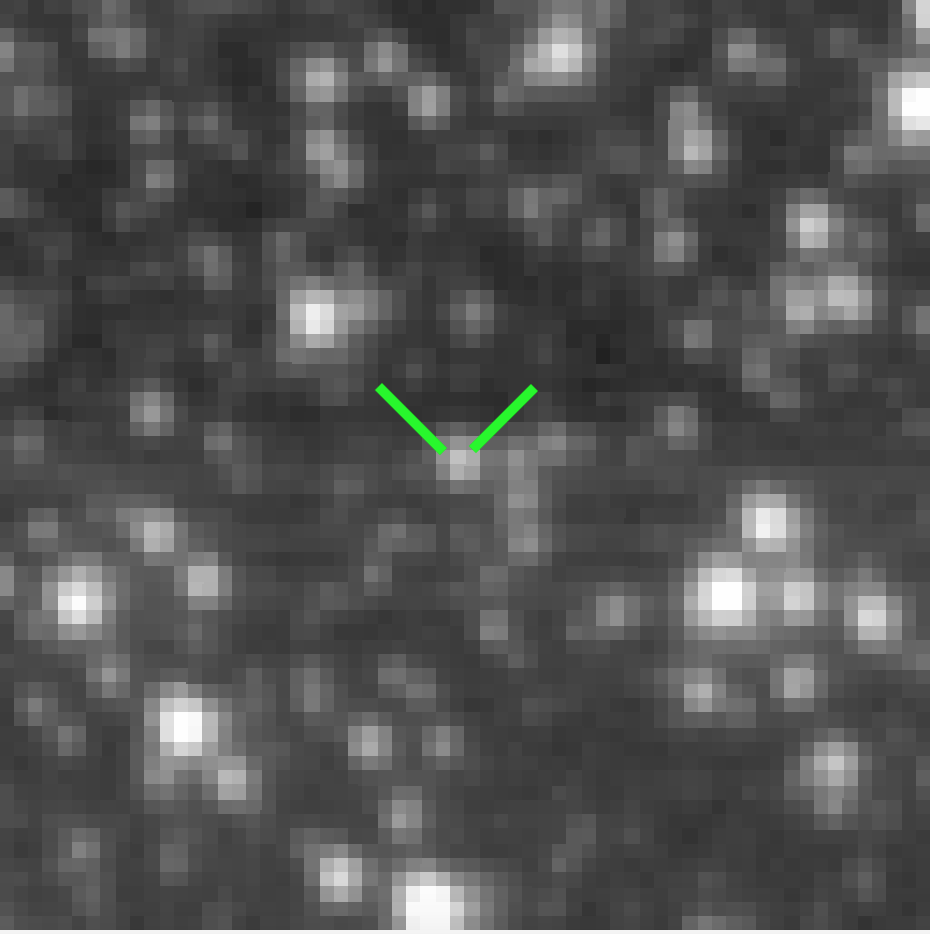}}\hspace*{8mm}%
\resizebox{53mm}{!}{\includegraphics{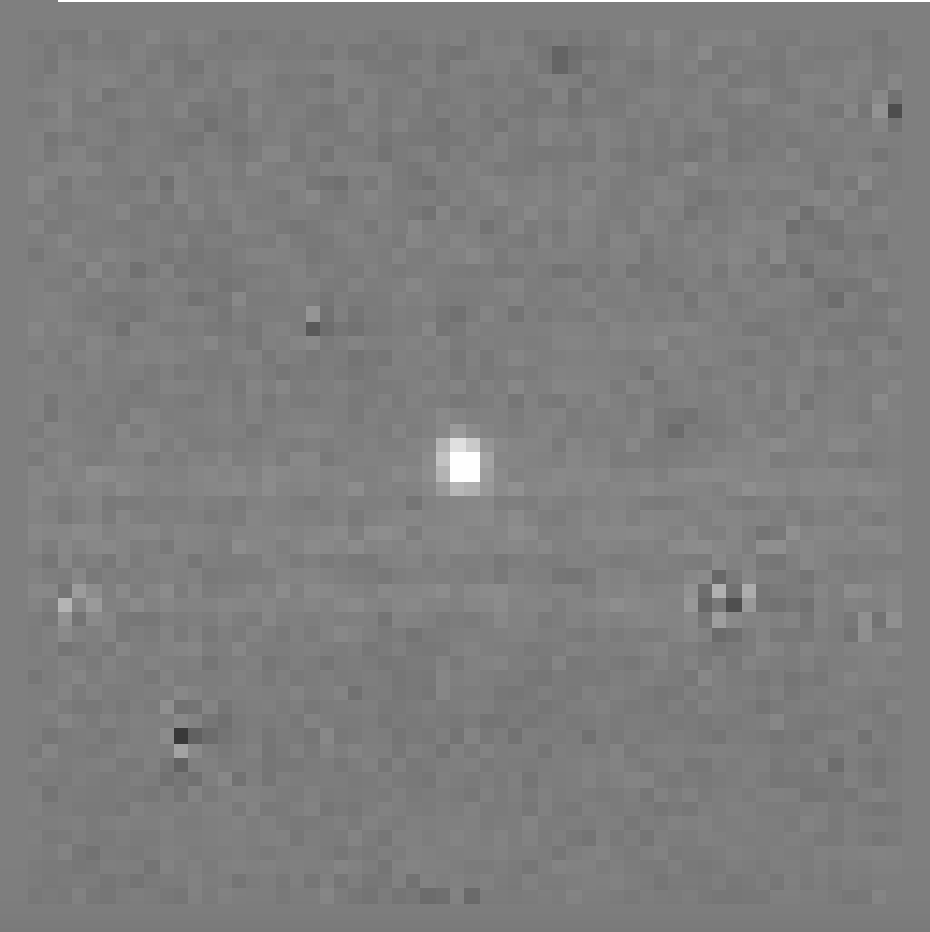}}\hspace*{8mm}%
\resizebox{53mm}{!}{\includegraphics{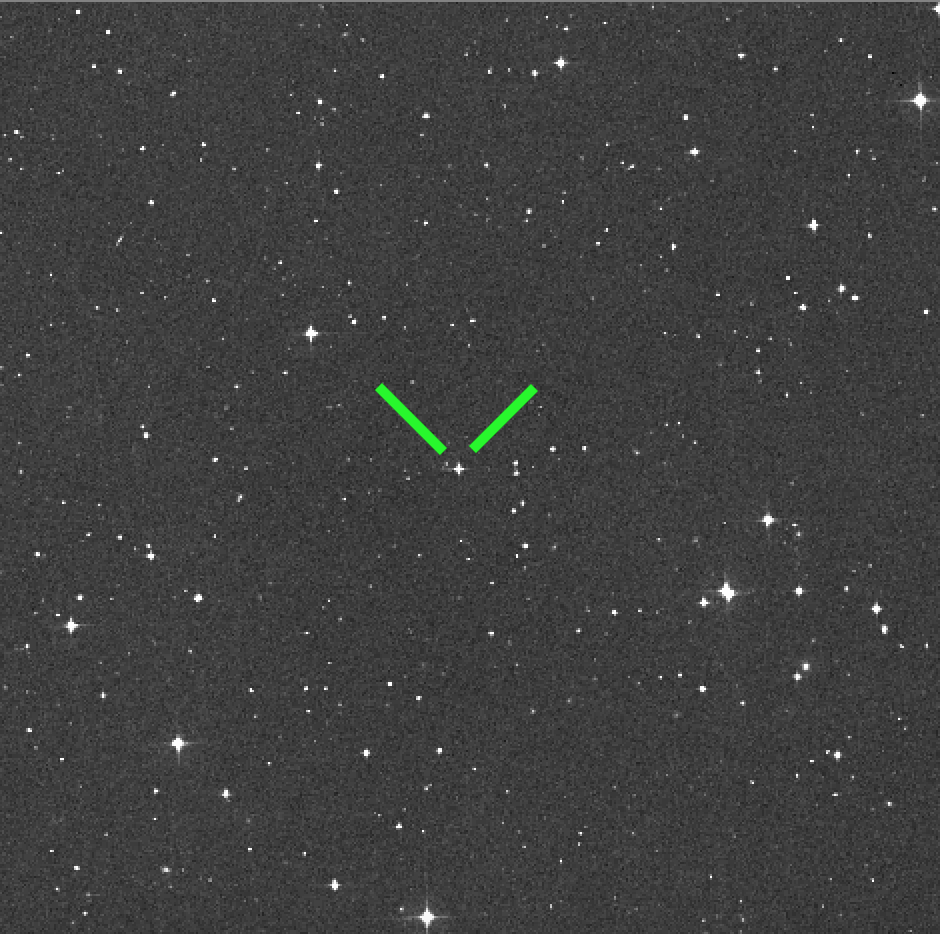}}
\vspace*{4mm}

\noindent
\resizebox{53mm}{!}{\includegraphics{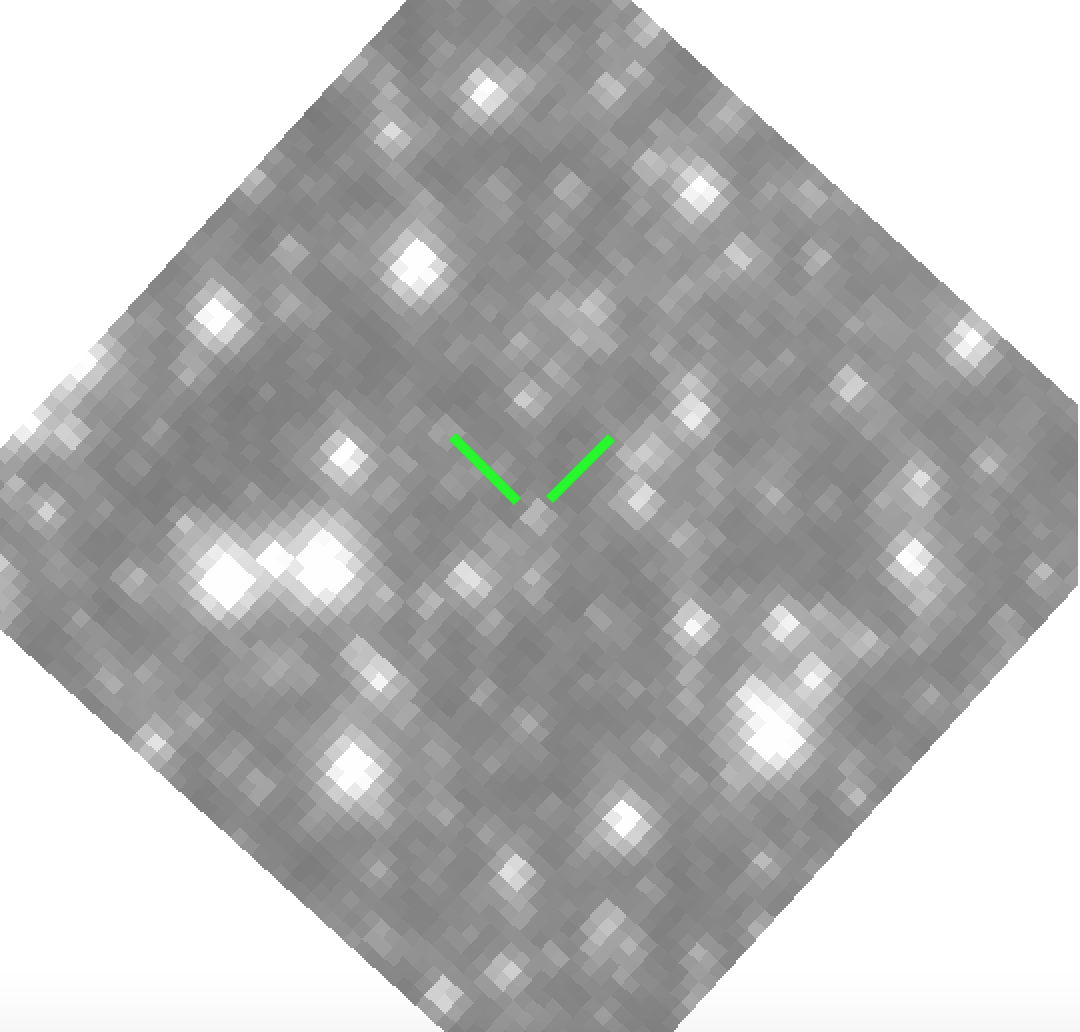}}\hspace*{8mm}%
\resizebox{53mm}{!}{\includegraphics{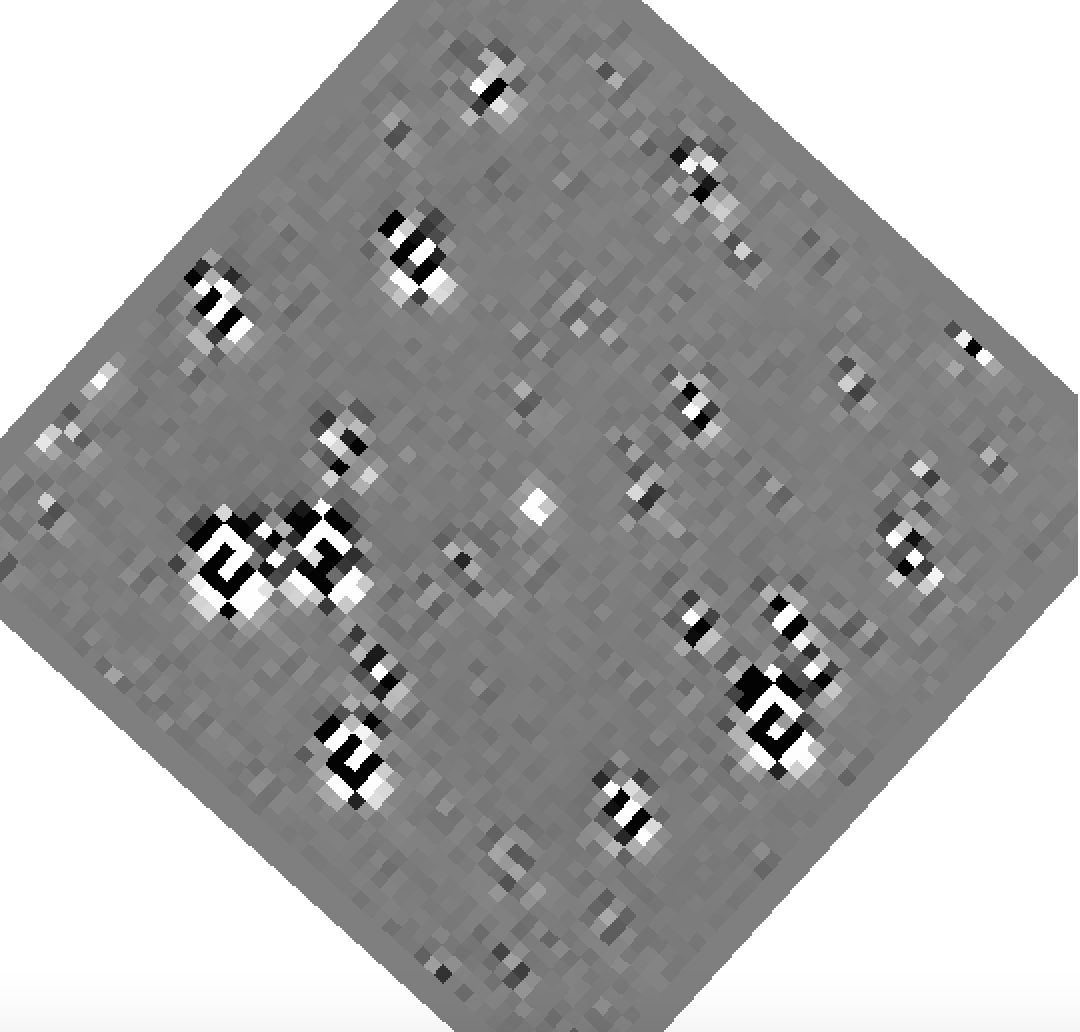}}\hspace*{8mm}%
\resizebox{53mm}{!}{\includegraphics{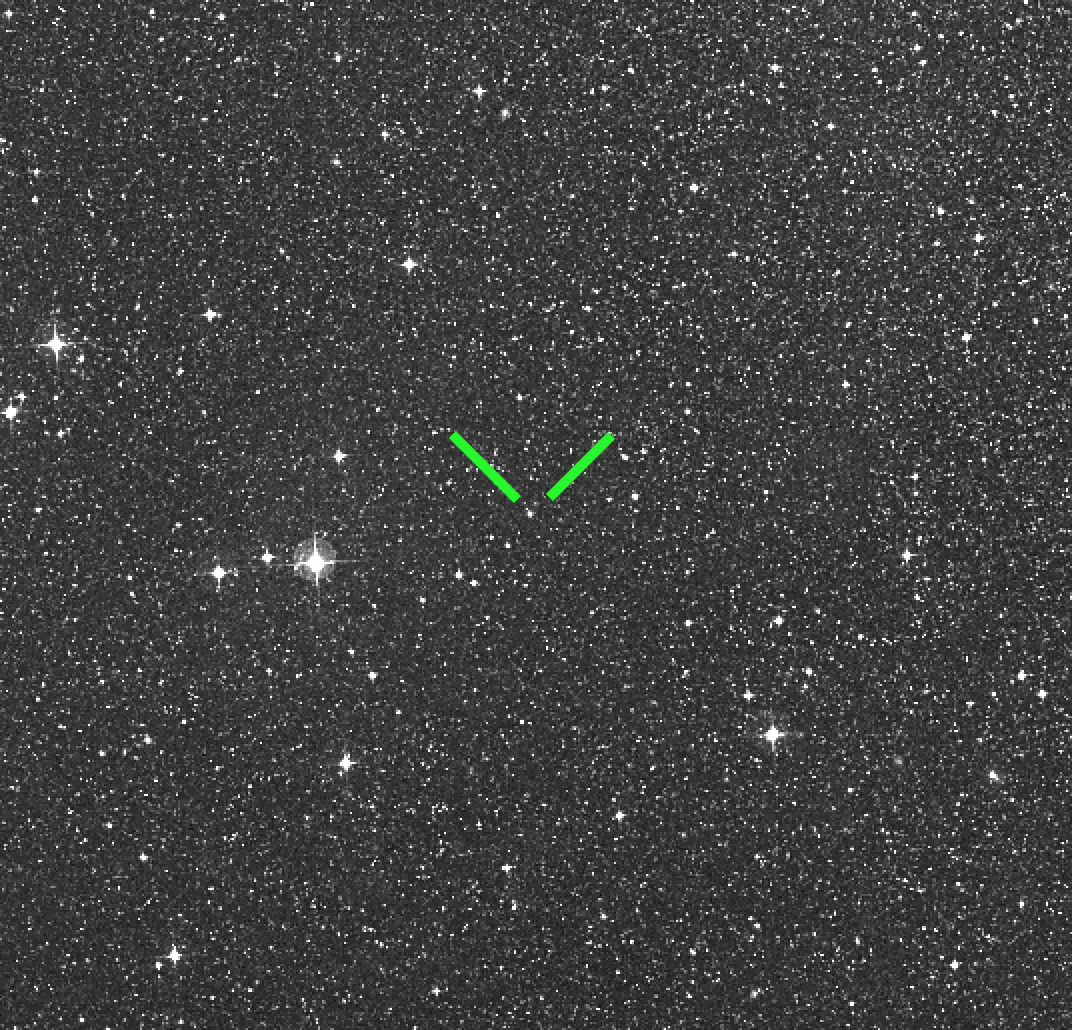}}
\end{center}
\caption{Images of two RR Lyrae stars, marked with green lines in the images. Upper panels: RV Hor, a star located in a relatively sparse stellar field; bottom panels: OGLE--LMC--RRLYR--23457, at the outskirts of the LMC. Left panels: a single TESS cadence from Sector 1. Middle panels: a single differential frame. The images of RV Hor are affected by reflections from the stripes of CCD electronics but otherwise almost entirely clean. The LMC image shows many residuals after image subtraction but the variable is clearly visible in the center. Right panels: corresponding DSS images. All images are oriented to celestial directions (north up, east to the left). The TESS cutouts are 64$\times$64 px or 22.4'$\times$22.4' large. }
\label{fig:tessdiff}
\end{figure*}

\subsubsection{Full-frame image targets}
We processed the FFI targets with the \textsc{fitsh} software package \citep{pal2012}. The description here is largely identical to that in the Cepheid first light paper \citep{Plachy-Cep-2020}, and the same method has been employed in other works as well \citep[e.g.,][]{borkovits-2020,szegedi-elek-2020,merc-2021,nagy-2021,Ripepi-2021,szabozs-2021}. 

The data reduction process has been split into two distinct procedures in our pipeline. First, we compute the plate solution for the calibrated FFIs, and we perform the astrometric cross-matching, using the \textit{Gaia} DR2 catalogue as well as various tasks (\texttt{fistar}, \texttt{grmatch}, \texttt{grtrans}) of the \textsc{fitsh} package. In this step, we also derive the flux zero-point. This is calculated with respect to the \textit{Gaia} $G_\mathrm{RP}$ magnitudes as the throughput of TESS  
\citep[see fig.~1 in][]{ricker2015} is very similar to the $G_\mathrm{RP}$ passband of the \textit{Gaia} photometric system \citep[see fig.~3 in][]{jordi2010}. This flux zero-point level is obtained for various TESS fields from Sector 1 and Sector 2. We find an RMS residual of $0.015\,{\rm mag}$, which indicates high accuracy. 

We note that the plate solution model applied in this work is more sophisticated than what the WCS-related FITS keywords would enable. Namely, besides the application of the gnomonic (tangential) projection, we apply a third-order Brown-Conrady model \citep{Brown71close-rangecamera} with respect to the optical axis of the images (which do not fall on silicon due to the focal plane design of the TESS cameras). A third-order polynomial correction is then applied afterwards in order to account for all other optical aberrations and the differential velocity aberration throughout a sector.

The steps described above are performed completely independently for all calibrated full-frame images for each of 4 CCD units in each of 4 cameras. The total number of available cadences was 1282 and 1245 for TESS Sectors 1 and 2, respectively, while 15 and 17 frames were flagged as low-quality ones due to the presence of TESS reaction wheel momentum dumps, respectively. 

In the second step, we cut out small images centered on the target stars and perform the differential image analysis only on those sub-frames, as shown in Fig.~\ref{fig:tessdiff}. With the astrometric solution at hand, we can simply query for the presence of any given object at any given instance. Any type of differential image analysis requires a reference frame, from which the deviations of the individual images are (expected to be) minimal and which itself has a good signal-to-noise ratio. In our analysis, we use the median average of $9$ or $11$ individual subsequent sub-frames with a size of $64\times64$ pixels to create this reference frame. The individual frames were selected around the mid-times of the observation series (i.e. at the very end of the first orbit or at the very beginning of the second orbit) in order to minimize the differential velocity aberration, and also at a  point when Earth was below the horizon of TESS.

We can then use the reference frame to obtain both the image convolution coefficients and a set of reference fluxes (using the \textsc{fitsh} task \texttt{fiphot}, and see \citealt{pal2009,pal2012}). Image convolution can correct many of the instrumental and/or intrinsic differences between the target frames and the reference frame, including the slight drift caused by the differential velocity aberration, spacecraft jitter, and background and stray light variations. In addition, convolution can help to eliminate or significantly decrease the effects caused by the momentum wheel dumps. We then determine the differential flux values on the convolved and subtracted images by simple aperture photometry. Two examples for the effectiveness of image subtraction are shown in Fig.~\ref{fig:tessdiff}, one for a sparse stellar field and one at the outskirts of the LMC.

Since reference fluxes are hard to accurately obtain even for moderately confused stellar fields with TESS (due to the large pixel size of 21\arcsec/px), we instead elected to use the \textit{Gaia} GDR2 $G_\mathrm{RP}$ (\texttt{phot\_rp\_mean\_mag}) magnitudes of the targets. We note that the \textit{Gaia} EDR3 definitions of the \textit{Gaia} passbands are slightly different from the DR2 ones, but differences are minimal, in most cases below 0.05 mag \citep{gaia-edr3-2020,gaia-riello-2020}. The final fluxes (i.e., the sums of the reference fluxes and the respective differential fluxes) still needed to be adjusted, as the reference frame we subtract is not averaged out over the variation of the star but rather contains a prior unknown segment of the light curve only. At the final step, we rescaled the average flux and associated errors to the $G_\mathrm{RP}$ value. For targets observed in both sectors, we also applied further corrections ($10^{-3}$ to $10^{-2}$ relative flux level shifts and percent-level scalings) when necessary in order to stitch the light curve segments together. A sample of the data file containing measurements of all FFI targets is shown in the Appendix.

\begin{figure*}[]
\includegraphics[width=1.0\textwidth]{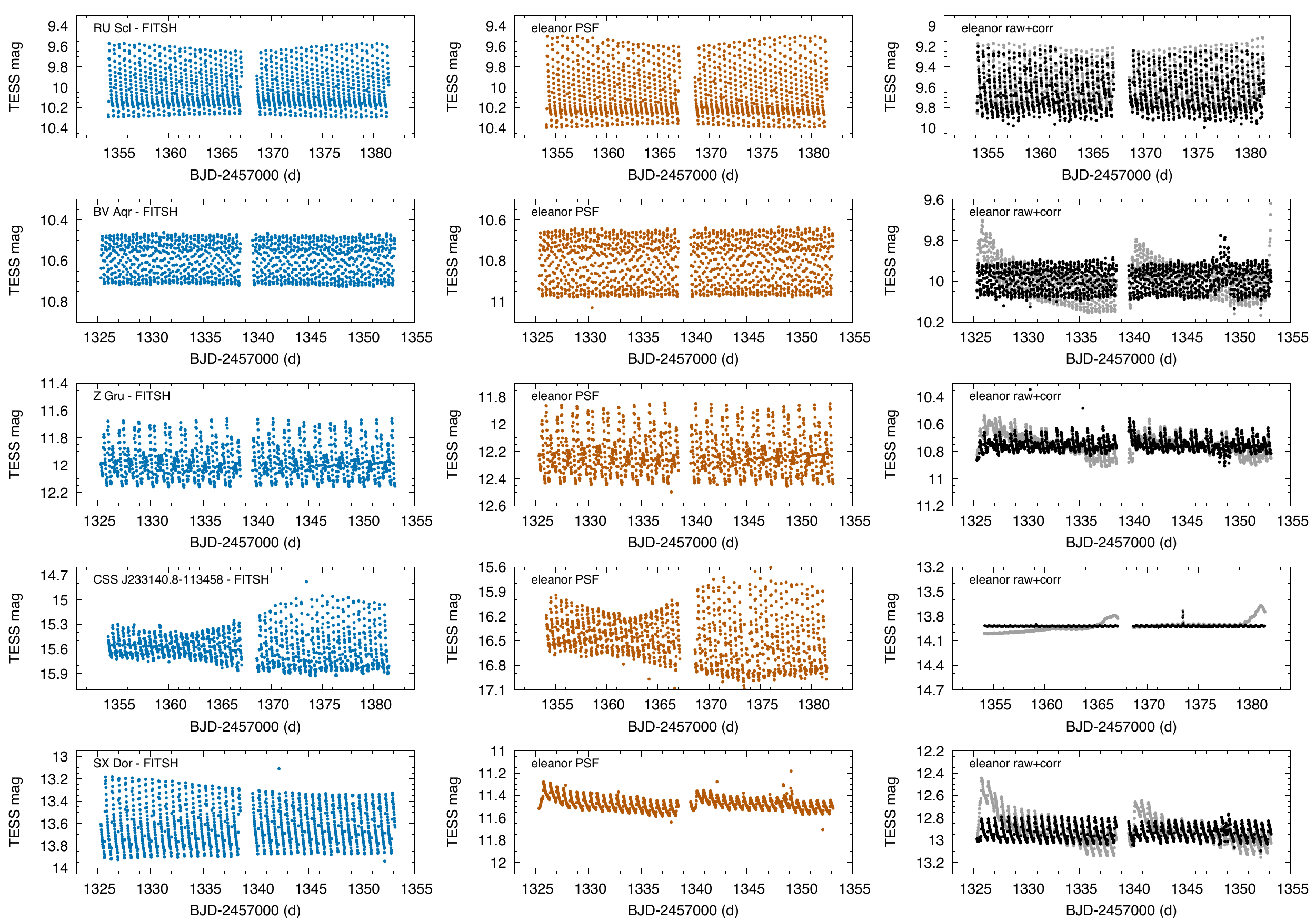}
\caption{Comparison of the \textsc{fitsh} differential-imaging aperture photometry (left row) with the \textsc{eleanor} photometries. Center: PSF fitting; right: raw (grey) and corrected (black) pixel aperture photometry. From top to bottom, three relatively bright stars from each subclass (RRab, RRc, RRd); a faint RR Lyrae star; SX Dor, from a crowded region near the Large Magellanic Cloud (LMC). Plots are on the same scale in each row. }\label{fig:eleanor}
\end{figure*}

The light curves were then analyzed by several co-authors separately, using predominantly the \textsc{Period04} software \citep{period04} to identify pulsation periods, harmonics and various other frequency components in the data. Temporal variations in the Fourier amplitudes and phases were also investigated.
 
\subsection{Comparison with \textsc{eleanor} photometry}
The \textsc{eleanor} open-source software tool was developed by \citet{feinstein2019} to produce light curves from the TESS FFI data. The software is capable to do pixel aperture and PSF photometry, and decorrelating systematics and co-trending signals from the light curves. We tested the performance of \textsc{eleanor} on a small selection of RR Lyrae stars. Classical correction methods, such as regressing the photometry against position changes, are difficult to apply to RR Lyrae stars since the high-amplitude and rapidly-changing signal often overwhelms the comparatively smaller systematics. We found that the simple pixel-photometry light curves, corrected for pixel position-related changes (the CORR\_FLUX photometry), are of inferior quality compared to our results, as shown in Fig.~\ref{fig:eleanor}. This result was not unexpected as multiple correction methods developed for general use in the K2 mission fared even worse when applied to RR Lyrae stars \citep{plachy2019}.

In contrast, the PSF photometry produced by \textsc{eleanor}, which fits a two-dimensional Gaussian function as PSF model to each frame, usually resulted in light curves that matched the quality of our data. This was true even for faint targets---down to 16 mag stars---as long as the star was well separated. One example is CSS J233140.8--113458, shown in the fourth row of Fig~\ref{fig:eleanor}. Blending and nearby stars, however, can confuse the algorithm and may result in a poor light curve: this happened to SX Dor, which is in front of the outskirts of the LMC (last row of Fig~\ref{fig:eleanor}). We concluded that limiting the size of the image cutout to exclude other sources can be beneficial but the code requires a minimum size of 9$\times$9 px image for a reliable fit. We also observed differences between the average brightness and pulsation amplitude values produced by \textsc{eleanor} and our code: the \textsc{eleanor} light curves are usually, but not always, fainter by up to 0.3 mag, or 33\% in mean flux. Lower mean flux suggests that the background level may be overestimated in the PSF photometry. This again highlights the difficulties caused by crowding and blending in the TESS images. 

\section{Results}
\label{sect:results}
In this section we present various results obtained from the TESS light curves and \textit{Gaia} measurements. First we study the kinematics of the targets, and classify the stars based on their absolute brightness, color and light curve shape information to create a clean RR Lyrae sample. We then study how accurately can we identify the Blazhko effect in the relatively short TESS light curves. We also evaluate the mode content of each light curve, and compare those results to earlier findings. 

\subsection{Kinematics}
\label{sect:kinematics}
We searched for available radial velocities (RVs) for the selected RR~Lyrae stars and found data for 57 targets in \textit{Gaia} EDR3, in the Radial Velocity Experiment \citep[RAVE,][]{Steinmetz2020Rave}, in the Galactic Archaeology with HERMES Survey \citep[GALAH DR3,][]{Buder2020GalahDR3}, and in the observations by \citet{Layden1994}. The \textit{Gaia}, GALAH and RAVE surveys do not take into account stellar pulsation in their radial velocity measurements, hence their values can be off by up $\approx$ 75\,km\,s$^{-1}$ based on Eq.~1 in \cite{Liu1991} and a projection factor of $1.37$ (the ratio between intrinsic pulsation velocities and disk-averaged RVs: value from \citealt{Sesar2012}). On the other hand, radial velocities derived by \cite{Layden1994} take into account stellar pulsation and they report errors of order 15\,km\,s$^{-1}$. Multiple stars in our sample have radial velocities both in the \textit{Gaia} EDR3 and in the \citet{Layden1994} catalogues, and the difference between the two ranges from 10\,km\,s$^{-1}$ to 30\,km\,s$^{-1}$. Therefore we assumed a conservative error of 75\,km\,s$^{-1}$ on the \textit{Gaia}, GALAH and RAVE RVs. 

We did not attempt to phase RV values for stars that are present in multiple databases: while this could, in principle, help us to determine the systemic velocity of the star more accurately, the phasing of multiple 
RV measurements may not be a trivial task. The alignment of these data would be affected not only by unknown phase shifts and possible modulation present in the pulsation, but also by the fact that different surveys use different spectral lines that sample regions with different kinematics in the atmosphere \citep{Braga-2021}. Further, some databases do not provide precise times of measurement.

We used the \textsc{gaiadr3\_zeropoint} software to correct for the zero point offset present in the \textit{Gaia} EDR3 parallaxes \citep{Lindegren2020Astrometry,Lindegren2020}. We note that almost all RVs in \textit{Gaia} DR2 are propagated into EDR3 unchanged, with the exception of the most discrepant values in DR2, which were removed from the new catalog \citep{Soubiran2018Gaia}.

\begin{figure}[]
\centering
\includegraphics[width=1.0\columnwidth]{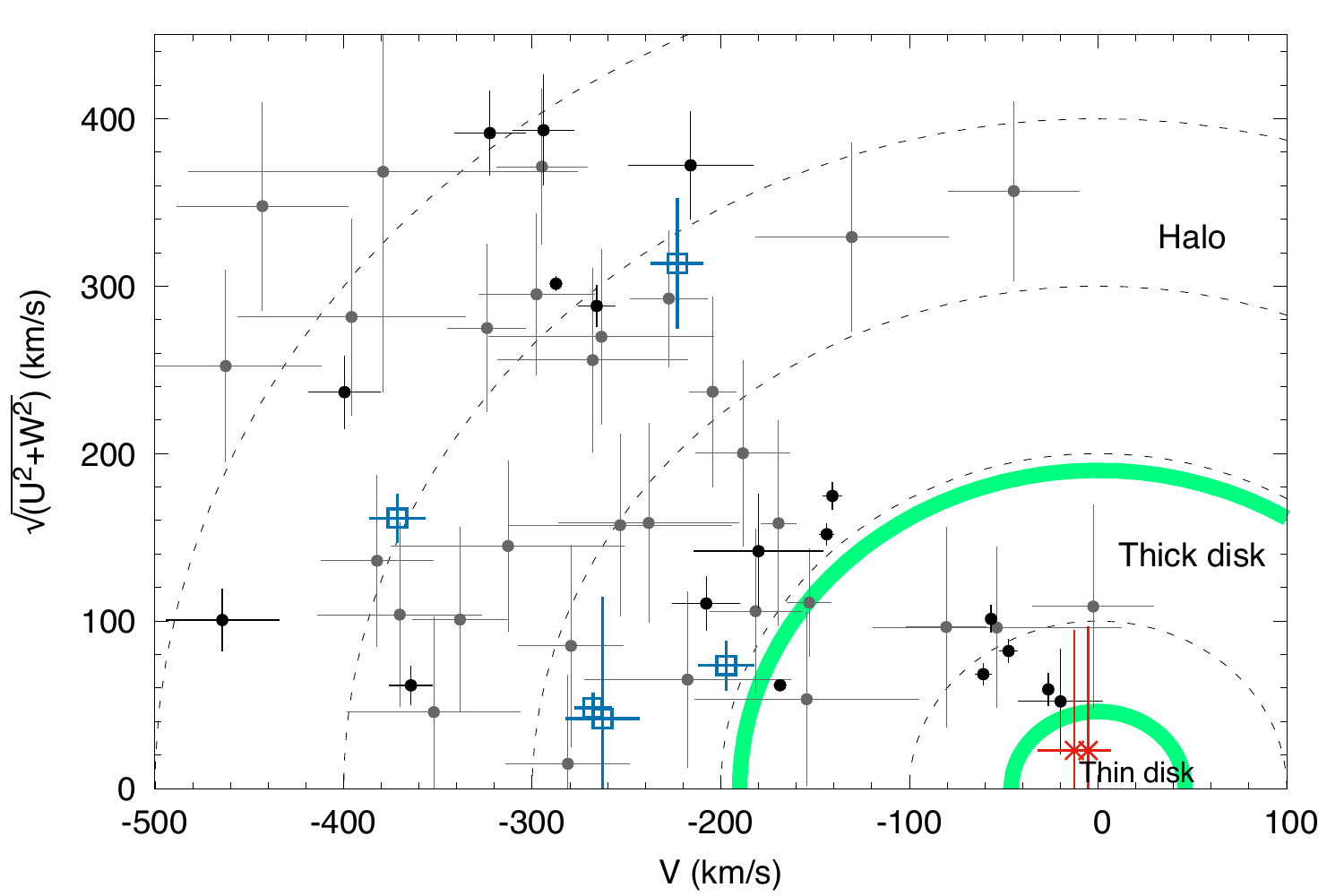}
\caption{Toomre diagram of 42 stars from our sample. Black dots are halo RR Lyrae stars, blue squares are variables that may be members of \textit{Gaia}-Enceladus structure according to \citet{prudil2020}. Dashed lines mark constant total space velocities while green lines show the approximate thin--thick disk and disk--halo boundaries. The red crosses are two non-pulsating stars we identify in Section \ref{sec:classification}. }\label{fig:toomre}
\end{figure}

We then computed the full 6D orbital solution of the stars using the \textit{Gaia} EDR3 proper motions and parallaxes in concert with collected radial velocities using the \textsc{galpy} library \citep{galpy2015}. The whole calculation was performed through a Monte Carlo error analysis with varying the proper motions, parallaxes, and radial velocities within their errors. Uncertainties and covariances among
all parameters were taken into account in calculating the velocity uncertainties. The final values and their errors were taken as the medians and absolute median deviations from the generated distributions. We corrected for the solar motion with respect to the local standard of rest with the velocities $\left(U_{\odot},V_{\odot},W_{\odot} \right) = \left(11.1, 12.24, 7.25\right)$\,km\,s$^{-1}$, where the U component is defined pointing towards the anticenter direction \citep{Schonrich2010,Schonrich2012}. We used $z_{\odot} = 20.8\pm0.3$~pc and $r_{\odot} = 8.122\pm0.031$~kpc for the solar position above the plane and the distance to the Galactic center, respectively \citep{Bennett2019,GravityColab2018}. The resulting velocity components $U$, $V$ and $W$ are the rectangular Heliocentric velocity components.

Stars spread out in Fig.~\ref{fig:toomre} that effectively shows orbital energy and angular momentum compared to the Local Standard of Rest. Stars with high total velocities, above 180--200 km\,s$^{-1}$ are halo members. Stars with velocities below that are part of the disk, with the thin--thick boundary being at around 46 km\,s$^{-1}$ \citep{bensby-2003,buder-2019}. While most of the stars in the sample are either part of the halo or the thick disk, two objects, marked red, are close to zero, i.e., they clearly belong to the thin disk and follow galactic orbits similar to that of the Sun. We later show that these stars are in fact not pulsating stars. 

We crossmatched our list of objects with the 314 objects studied by \citet{prudil2020}. Since Sectors 1--2 are far from the Galactic plane, we found no objects that were identified as possible thin-disk RR Lyrae stars. However, four of our targets (UW~Gru, VW~Scl, YY~Tuc and XZ~Mic) are potentially members of the \textit{Gaia}-Enceladus structure, also known as the \textit{Gaia} Sausage, the remnant of a galaxy merged into the Milky Way \citep{helmi2018}. This sample is currently too small to analyze separately but highlights the fact that we will be able to use TESS to study different RR~Lyrae populations.

\subsection{Classification}
\label{sec:classification}
The stars selected for  this study have been classified as RR Lyrae at least once already by prior studies. However, these classifications are sometimes based on observations of limited quality and/or quantity. We therefore verified the classifications based on the TESS light curves, complemented by \textit{Gaia} EDR3 brightness and parallax data \citep{gaia-edr3-2020}.

\subsubsection{Absolute magnitudes and distances}

 \begin{figure*}[]
\includegraphics[width=1.0\textwidth]{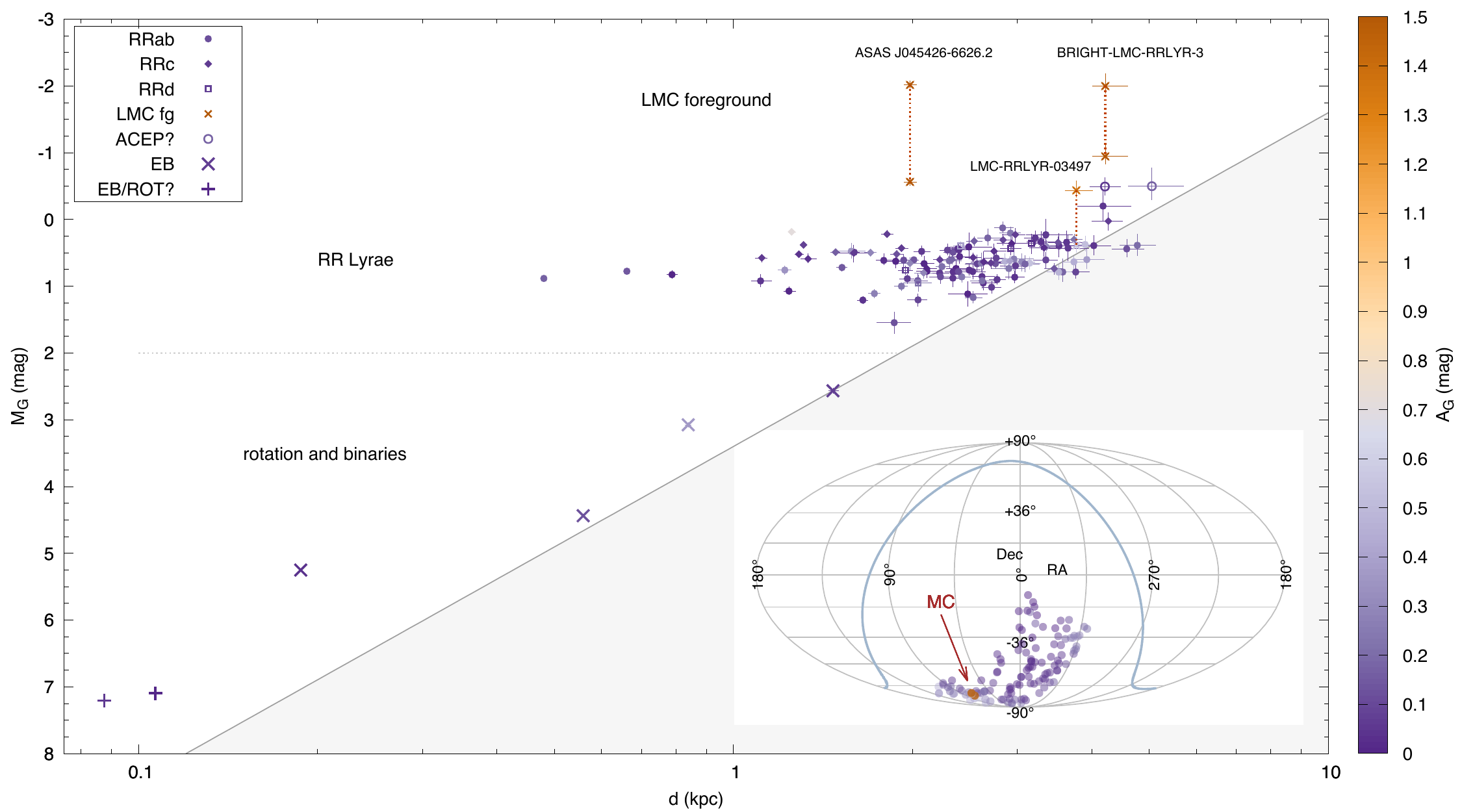}
\caption{Main: the \textit{Gaia} EDR3 $M_{G}$ absolute magnitudes and distances of the sample, color-coded with the $A_G$ absorption coefficients calculated with \textsc{mwdust}. The diagonal line and light-grey area below it mark the approximate brightness cut of the sample. Large crosses mark eclipsing stars that we were able to classify based on light curve shapes. The large pluses at the bottom are two stars where the light curve shape was not decisive but their positions clearly rule them out. Different extinction corrections for the three LMC foreground stars are connected with the dashed lines. Insert: map of the targets with and the $A_G$ absorption coefficients in the EDR3 \textit{G} band. The position of the Magellanic Clouds is marked with the arrow. }\label{fig:rrl_dist}
\end{figure*}

 \begin{figure*}[]
\includegraphics[width=1.0\textwidth]{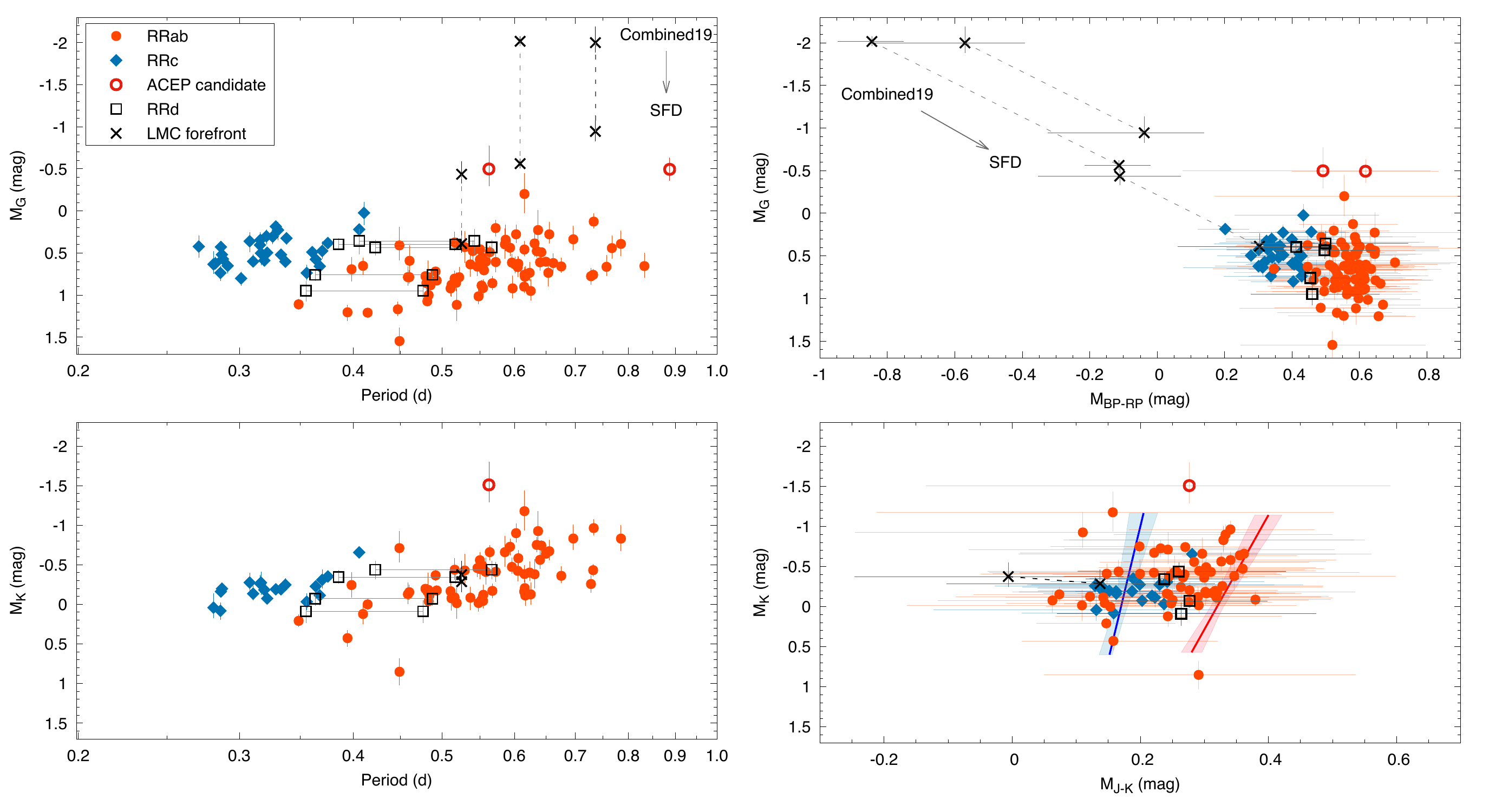}
\caption{Left: period-luminosity relation of the sample. Brightness is in $M_{G}$ (top) and $M_{K}$ (bottom) absolute magnitudes. RRab and RRc stars are represented with blue diamonds and red dots, respectively. For RRd stars (black empty rectangles) we marked both periods. The two anomalous Cepheid (ACEP) candidates are marked with empty circles. Stars with erroneous extinction correction are marked with black crosses. Right: distribution of the sample in the absolute \textit{Gaia} and \textit{J--K} (bottom) color-magnitude planes: here we marked the blue and red edges of the instability strip, as calculated by \citet{Marconi-2015}, with the blue and red solid lines. Differences between the two extinction corrections are indicated with dashed lines for the three LMC foreground stars. }\label{fig:rrl_pl}
\end{figure*}

We computed the absolute magnitudes in the \textit{Gaia} EDR3 \textit{G}, $G_{\rm BP}$ and $G_{\rm RP}$ bands, using the geometric \textit{Gaia} EDR3 distances calculated by \citet{bailer-jones2020}, and correcting for extinction with the \textsc{mwdust} code \citep{bovy2016}. We used the combined maps created by \citet{bovy2016}. This method provided good separation between the intrinsically fainter rotational variables and binaries and the brighter, evolved HB stars where we expect RR Lyrae variables to appear. The calculated absolute brightness values are plotted in Fig.~\ref{fig:rrl_dist}, along with the distribution of the absorption coefficients in the sky. We also calculated absolute magnitudes in the \textit{V, J, H} and \textit{K} bands, based on the brightness values listed in the SIMBAD database for each star \citep{simbad}. Geometric \textit{Gaia} distances for RR Lyrae stars have been verified previously for the DR2 data by \citet{hernitschek-2019}. They found that they are accurate to about 5 kpc, but 
lose accuracy
beyond about 10 kpc, at which point they become dominated by the distance prior used by \citet{bailer-jones2018}. 
We can therefore 
expect the EDR3 distances for our sample, which only extends to 5 kpc, to be accurate as well. 

The $M_G$ brightness of most of the stars is between 1.5 and 0.0 mag, where we expect horizontal-branch stars to cluster \citep{gaiadr2-hrd-2018}. Six stars fall clearly below this range. The variations of the two faintest (marked with blue plus signs), ASAS J212045--5649.2 and ASAS J225559--2709.9, look superficially like RR Lyrae-type pulsation: these are likely rotational and/or ellipsoidal variables. Four more are cataloged as RR Lyrae stars but the TESS light curves make it clear that these are eclipsing binaries (blue cross signs). These are, in decreasing absolute brightness: CRTS J215543.7-500050, AZ~Pic, UW~Dor and UY~Scl. We note that UY~Scl has been listed previously as a possible binary RR~Lyrae based on its large proper motion anomaly in DR2 \citep{kervella-2019}. We calculated velocity components for two of these six targets in Sect.~\ref{sect:kinematics}. ASAS J212045--5649.2 and UY~Scl are the two red crosses in Fig.~\ref{fig:toomre}, closest to zero, indicating that they are part of the thin disk.

Three other stars stand out due to high absorption coefficient values ($A_G$ = 1.4, 2.6 and 3.0 mag, for OGLE LMC--RRLYR--03497 (also known as SW~Dor), OGLE BRIGHT--LMC--RRLYR--3 and ASAS J045426--6626.2, respectively): all three lie in the direction of the LMC, as the sky map in Fig.~\ref{fig:rrl_dist} indicates. OGLE BRIGHT--LMC--RRLYR--3 was newly identified in the OGLE--III LMC Shallow Survey as a Galactic RR Lyrae: it is known as OGLE LMC--RRLYR--28980 in their main catalog \citep{Ulaczyk2013}. Since all three stars are clearly nearby stars that are in the foreground of the LMC, these extinction corrections appear to be excessive. To test this we then calculated the interstellar absorption using the SFD dust maps instead \citep{SFD2011}: for these three stars the SFD map gave about 0.7--1.5 mag smaller absorption but two of the three still remained overluminuos. 

\subsubsection{Period-luminosity relations and colors}

With the pulsation periods from the TESS light curves in hand, we created the period-luminosity plot of our TESS first light sample in $M_{G}$ and $M_{K}$ bands, as shown in the left panels of Fig.~\ref{fig:rrl_pl}. With this sample size the slopes of the PL relations are not entirely apparent, especially for the RRc stars, but the increase in brightness with the period can be seen. \textit{K}-band brightness is much less sensitive to interstellar extinction than shorter wavelengths, limiting the spread of the PL relation, although for RR Lyrae stars the spread of the PL relation is largely driven by metallicity \citep{Catelan-2004,Layden-2019,bhardwaj-2020,Gilligan-2021}. Furthermore,  not all stars have their $K$ brightness measured (here, 32 of the 118 targets are missing from the lower plots), and even when they do, single-epoch measurements may not represent mean brightness values. 

For RRd stars both of their periods are indicated by the black empty squares. Two of the three overcorrected stars, ASAS J045426--6626.2 and BRIGHT--LMC--RRLYR--3, stand out with either correction methods, but the third, LMC--RRLYR--03497, moves into the RRab locus with the lower, SFD-based correction. 

We also calculated the extinction-corrected EDR3 $G_{\rm BP} - G_{\rm RP}$ and $J-K$ colors for every star. In the color-magnitude diagrams (CMDs, right panels of Fig.~\ref{fig:rrl_pl}) the three stars that are in front of the LMC are clearly shifted bluewards (black crosses), confirming that they are over-corrected for interstellar extinction. Tracing them back along the reddening vector to the expected colors indicate that they would be part of the RR Lyrae group. We note that the star LMC--RRLYR--03497---which apparently falls into the RRc group with the SFD correction---is in fact an RRab pulsator, as indicated by its period and light curve shape. We therefore conclude that both the Combined19 dust map of \citet{bovy2016} and the SFD map overestimate the required extinction for stars that are directly in front of the LMC. In the near-infrared CMD (lower right panel of Fig.~\ref{fig:rrl_pl}), we were able to overlay the theoretical blue and red edges of the instability strip, as calculated by \citet{Marconi-2015}. In this CMD the RRab and RRc stars appear to be mixed in color, likely because neither magnitudes represent the mean brightness. However, multiple stars fall outside of the first overtone blue edge, but almost none are redder than the fundamental mode red edge. 

We identified two stars that have the expected colors but still appear to be more luminous than the bulk of the sample: ASAS J221052--5508.0 ($P=0.887$~d) and SX~PsA ($P=0.563$~d) are both above the RR Lyrae locus, although only marginally in  the latter case. $K$ brightness is only available for the latter, but this also appears to exceed the brightnesses of the rest of the stars. We consider the possibility that these are anomalous Cepheids in Sect.~\ref{sect:acep}. The intrinsically faintest RR Lyrae star in the sample is XZ~Mic. 

The color-magnitude diagram also shows that the RRc and RRab subclasses (as determined in the next section) clearly cluster towards the blue and red sides of the group in the \textit{Gaia} passbands, with the RRd stars falling in the middle. The apparent color values themselves are typically precise to less than 0.05 mag, which would make color-based classification possible, but extinction correction imparts further uncertainties typically in the range of 0.1--0.3 mag. This means that although the RRc and RRab data points clearly separate in the \textit{Gaia} CMD (but not in the near-infrared one), their uncertainty ranges overlap, which prevent us from using color as a primary classification metric. 

\subsubsection{Relative Fourier parameters}
\begin{figure*}[]
\includegraphics[width=1.0\textwidth]{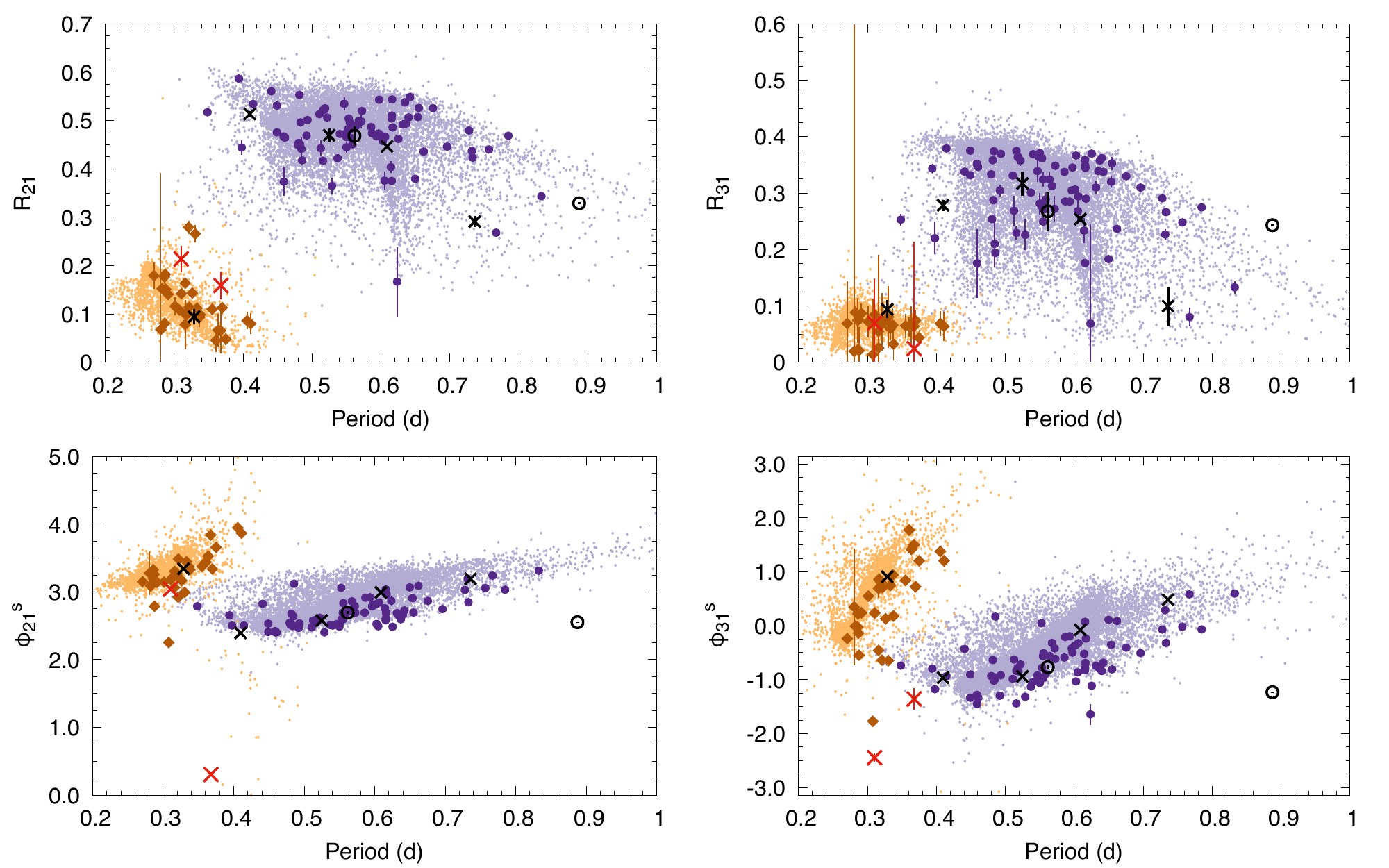}
\caption{Relative Fourier parameters $R_{i1}$ and $\phi_{i1}$ of the TESS first light sample (large symbols), overlaid on the OGLE \textit{I}-band values (dots). Orange and purple points mark RRc and RRab stars. The two empty circles are the anomalous Cepheid candidates, the three black crosses in the RRab group are the three LMC foreground stars. The two red crosses in the RRc group are two false positive, non-pulsating stars.}\label{fig:rrl_fourparam}
\end{figure*}

With the Fourier parameters of the main pulsation periods and harmonics fitted, we can compute the relative Fourier parameters $R_{i1} = A_i/A_1$ and $\phi_{i1}=\phi_i-i\,\phi_1$ to classify the stars based on their light curve shapes \citep{simonteays1982}. We use sine-based Fourier series in the form of $m = m_0 + \sum_{i}A_i\,\sin (2\pi\,if_0 t+\phi_i)$, where $m_0$ is the average brightness, $f_0$ is the dominant pulsation frequency, $A_i$ and $\phi_i$ are the amplitudes and phases of each harmonic and $i$ is the harmonic order. This method does not necessarily separate non-pulsating stars from pulsating stars if their variations look similar but it is capable of differentiating between pulsation modes among the RR~Lyrae stars. We compared the TESS results to those of the OGLE stars as a reference, as shown in Fig.~\ref{fig:rrl_fourparam}. Since the TESS passband is centered on the $I$ band, we expect only minor differences in the distribution of the Fourier parameters of the OGLE $I$ and TESS measurements. The \textit{R} parameters indeed show very good agreement for both the RRab and RRc loci, but we see differences in the $\phi$ parameters, with the TESS data clustering at slightly lower values than the OGLE data. A systematic shift between the $\phi_{31}$ parameters in the \textit{I} band and TESS pass-band light curves indicate that the we cannot adopt existing relations to calculate the photometric metallicities \citep{skowron2016}. We estimate the shifts in phase difference to be $\phi_{21,I}-\phi_{\rm 21,TESS} \approx 0.15$ and $\phi_{31,I}-\phi_{\rm 31,TESS} \approx 0.25$ rad. Calibration of photometric [Fe/H] indices for the TESS passband will be discussed in a separate paper.  

We indicated the three LMC foreground stars with black crosses. Each appears within the RRab loci of each parameter, confirming that these are RRab stars and that their intrinsic luminosities and colors must be lower than what we computed with the \texttt{mwdust} code. We also indicate the anomalous Cepheid candidates with black circles. 

Although we removed obvious eclipsing binary and rotational variable light curves from the sample prior to classification, the distance-luminosity and period-luminosity plots in Fig.~\ref{fig:rrl_dist} revealed two further dwarf stars in the sample (blue plus signs in Fig.~\ref{fig:rrl_dist}). Their Fourier parameters are at least in marginal agreement with the distribution of RRc stars: the R$_{i1}$ parameters agree, the $\phi_{21}$ is inconclusive since at longer overtone periods \textit{bona fide} RRc stars span the whole range, and only the $\phi_{31}$ value appears to be discrepant. Therefore, classification of these two stars would have remained ambiguous from the light curve shapes alone, and the accurate parallax data were necessary to exclude them from the sample. 

Based on the absolute brightness, light curve shape and period ratio information, we identified 82 RRab, 31 RRc, and 5 RRd stars among the selected stars. This amounts to 118 RR Lyrae targets in total.

\begin{figure}[]
\centering
\includegraphics[width=1.0\columnwidth]{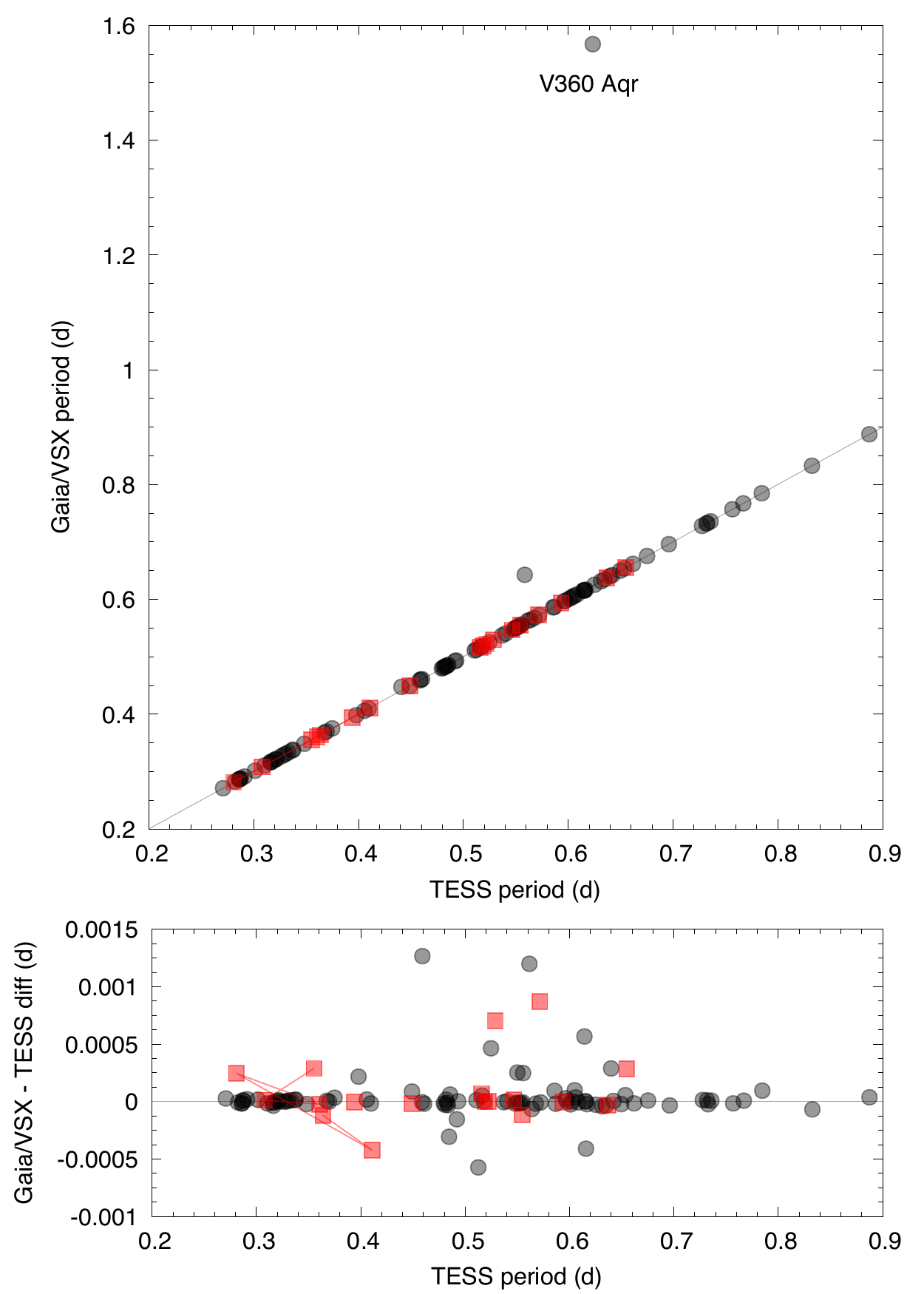}
\caption{Comparison of the pulsation periods we derived from the TESS light curves and those calculated from the \textit{Gaia} DR2 epoch photometry (grey) or present in VSX (red). Stars where two periods were identified only by either TESS or \textit{Gaia} are connected with lines. The lower panel shows the difference values. Uncertainties are smaller than the symbols.}\label{fig:per_comp}
\end{figure}

Finally, we compared the pulsation periods we obtained with those in the \textit{Gaia} DR2 catalog (specifically, the RR Lyrae and Cepheid periods calculated by \citet{clementini2019}). Of the 118+2 stars, 21 have not been identified as either type of variable: in those cases we queried the International Variable Star Index (VSX). As Fig.~\ref{fig:per_comp} shows, almost all \textit{Gaia} or VSX periods agree with those we derived from the TESS light curves. The most distant point is V360~Aqr, an RRab that was classified as a Type II Cepheid (T2CEP) by the \textit{Gaia} algorithm. A smaller but clear difference is observed for LMC--RRLYR--23457. The rest of the stars show only small differences, although in multiple cases such differences exceed the uncertainties of both the TESS and \textit{Gaia} periods. Most of the stars have been correctly classified in \textit{Gaia}: aside from V360~Aqr, only four more stars are erroneous. SU~Hyi, the shortest-period RRab star, and an double-mode star, \textit{Gaia} DR2 6529889228241771264 were classified as RRC stars, whereas two RRab stars, AE~Tuc and ASAS J215601--6129.2 were classified as RRD stars in the \textit{Gaia} notation. These statistics agree with earlier validation tests based on \textit{Kepler} data, which found that class mismatches in the \textit{Gaia} classification are rare among the bright RR Lyrae stars \citep{molnar2018}.

\subsubsection{Anomalous Cepheid candidates}
\label{sect:acep}
\begin{figure}[]
\centering
\includegraphics[width=1.0\columnwidth]{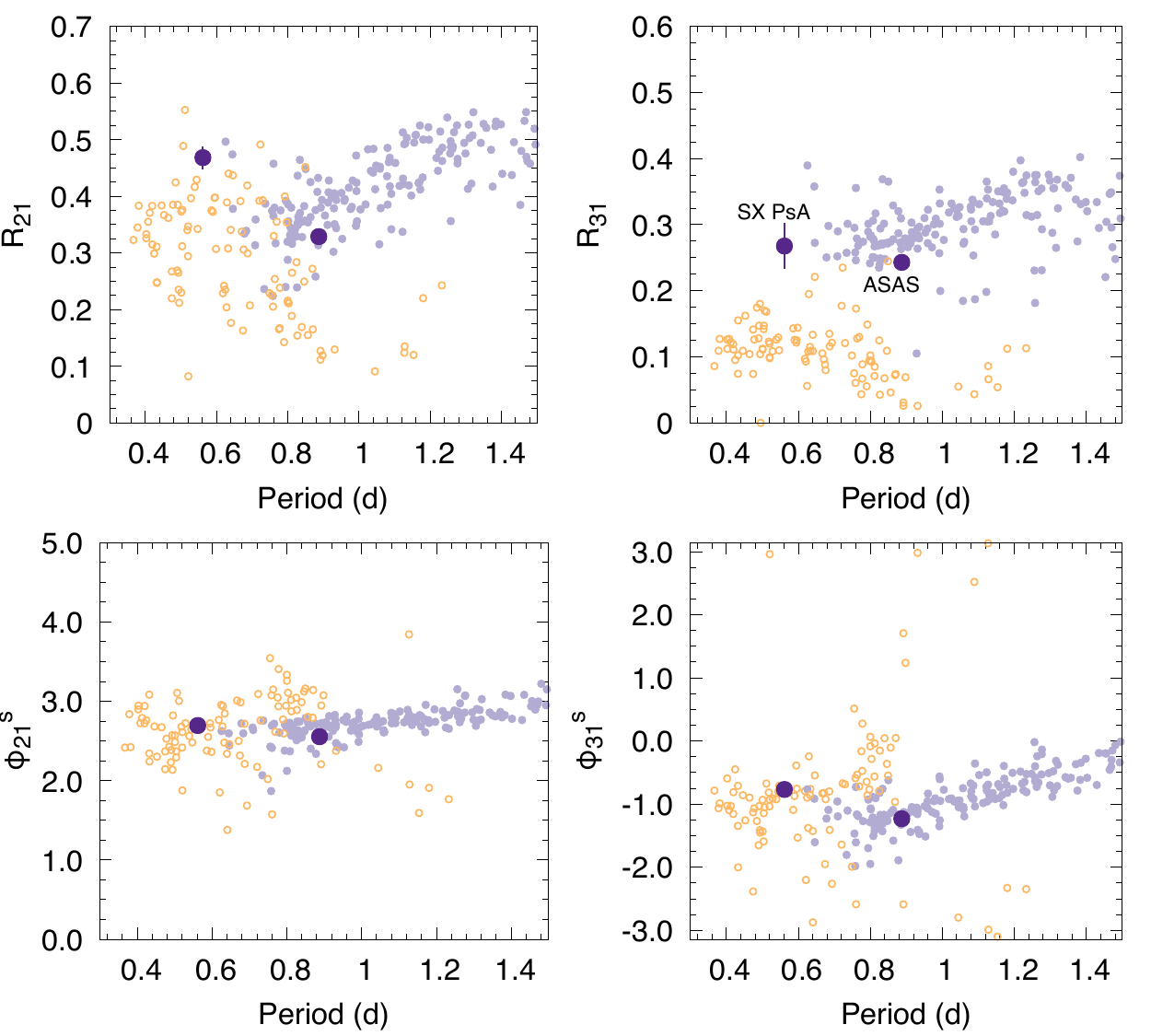}
\caption{Relative Fourier parameters of the two anomalous Cepheid candidates, compared to the OGLE anomalous Cepheid samples \citep{soszynski-2015,soszynski-2020}.}\label{fig:acep_fourparam}
\end{figure}

In addition to the 118 RR Lyrae stars, we also found two over-luminous outliers, SX~PsA and ASAS J221052-5508.0, both at $M_G = -0.5$ mag. The higher absolute brightness and long period ($P=0.887$~d) suggest that ASAS J221052-5508.0 is an anomalous Cepheid (ACEP) star rather than an RR Lyrae. The classification of SX~PsA is more uncertain, given its shorter period. We compared the two stars' Fourier parameters to those of the OGLE ACEPs in Fig.~\ref{fig:acep_fourparam}, and they both fit into the fundamental-mode ACEP locus, although the latter is at the short period end of the distribution. At the shortest periods, however, the distribution of RR Lyrae and ACEP Fourier parameters strongly overlap, and as Fig.~\ref{fig:rrl_fourparam} shows, SX~PsA fits into the RRab locus as well. 

We propose that the more luminous one, ASAS J221052--5508.0 is indeed a newly identified, Galactic anomalous Cepheid. The classification of SX~PsA is less certain, and we consider it an ACEP candidate. As SX~PsA also displays strong amplitude modulation, this would be another ACEP star showing the Blazhko effect, after the identification of the first modulated ACEP candidate in the K2 observations by \citet{plachy2019}. Although we prefer the ACEP classification, SX~PsA could possibly be an unusually luminous RR Lyrae star, too. 

Unfortunately, most anomalous Cepheids identified in \textit{Gaia} DR2 are too far away to have accurate geometric parallaxes. We therefore do not yet have a well-calibrated PL relation for Galactic ACEP stars, and too few have been studied so far with TESS \citep{clementini2019,Plachy-Cep-2020}. The importance of distance determination was also highlighted by \citet{Braga-2020}, who showed that short-period ACEP and Type II Cepheids can overlap in period with long-period RR Lyrae stars. 
Period itself is therefore not a unique classifier, and a reliable separation of Galactic short-period ACEP stars from long-period RR Lyrae stars will require accurate light curve shape information, distances and colors. Once we can separate these with confidence, we will be able to construct a PL relation for Galactic ACEP stars, too.

\subsection{The Blazhko effect}

Modulation has been detected in all subtypes of RR~Lyrae stars, even in RRd stars \citep[see, e.g.,][]{Smolec-RRd-2015,Plachy-RRd-2017,Carrell-2021}. Amplitude and phase modulation in RR Lyrae stars happen over periods extending from a few days to years, therefore one or two Sectors' worth of data are not well-suited to studying the Blazhko effect. Nevertheless, we were able to identify 
the signs of amplitude and phase changes in several stars. Although we cannot rule out the possibility of non-cyclic amplitude and phase shifts in RR Lyrae stars over short time scales, evidence for such events is scant in long-term observations. We therefore consider contamination from non-Blazhko objects to be negligible and classify all targets with smooth changes to the pulsation properties as Blazhko stars. 

Currently the best hypothesis for the Blazhko effect is the non-linear mode resonances model \citep{bk2011,kollath2018}. It agrees best with observations; however, the model is based on amplitude equation calculations, and the accurate reproduction of modulated light curves in hydrodynamic simulations is still in its infancy \citep[see, e.g.,][]{molnar-stothers-2012,smolec-2012,Goldberg-2020,Joyce-2020}.

\subsubsection{What is the Blazhko effect -- for an observer?}

Since we do not have a universally accepted model for the Blazhko effect, we have no \textit{a priori} constraints on the distributions and limits of the modulation periods and amplitudes. With the precision achieveable with \textit{Kepler} and TESS, we can now detect millimagnitude-level modulations: however, we must ask whether we should label all quasi-periodic modulations as the Blazhko effect. \citet{moskalik2015}, for example, stated that the small amplitude and phase fluctuations in the \textit{Kepler} RRc stars do not resemble the classical picture of the (semi-)coherent modulations we know as the Blazhko effect. Similarly, \citet{benko2019} found that cycle-to-cycle variations in \textit{Kepler} RRab stars can cause small side peaks, but the true incidence rate of quasi-periodic modulation is 51--55\%. In contrast, the idea that all RR Lyrae stars are modulated was recently suggested by \citet{kovacs2018}, based on his processing of K2 data. It is therefore important to test whether this hypothesis can be confirmed with the TESS sample. 

The classical Blazhko effect manifests itself as sidepeaks around the pulsation frequency and its harmonics in frequency space. In principle, searching for significant sidepeaks is a straightforward exercise. However, contamination from stellar or instrumental sources, as well as systematics introduced by the photometric and post-processing pipelines can inject variations mimicking small modulations. Slow changes in the average flux, for example, can be caused both by changes in the amount of captured flux (e.g., intra- and interpixel sensitivity changes) or by contamination from a separate source, or both, but they affect the flux amplitude differently. Choices to correct these via scaling and/or subtracting flux can be a degenerate problem. Improper handling of systematics and contamination may either introduce artificial amplitude changes that appear as sidepeaks in the frequency spectra, or remove slow variations intrinsic to the star. The same issues were found to affect some of the K2 light curves as well \citep{plachy-EAP-2017,plachy2019}. In the TESS data we found that systematics tend to generate sidepeaks surrounding only the pulsation frequency or very few harmonics, whereas Blazhko stars have a clear series of sidepeaks along the harmonics. 

Furthermore, the \textit{Kepler} data demonstrated that stars where no coherent amplitude and phase variation were detected still showed irregular amplitude and phase fluctuations, similar to those seen in Cepheids \citep{benko2019}. While over long timescales this manifests as red noise around the main harmonic series, over the much shorter TESS observations, stochastic variations can conceivably appear as a single or a few significant frequency components in some cases. 

Another problem we encountered stems from the fact that the amplitude of the main frequency peak is several orders of magnitude larger than the photometric noise level. This can cause some algorithms to be able to reach sufficiently small pre-set residual levels even with a slightly offset frequency fit. This results in an apparent phase shift between the data and the fit which can be mistaken for intrinsic modulation. We noticed this behavior when fitting RRc stars in \textsc{Period04}. We therefore double-checked all stars where an apparent linear phase shift was observed without any counterpart in the pulsation amplitude and then adjusted the pulsation frequency manually to minimize the phase shift over the length of the data. Here we also note that nearly linear phase shifts in short data sets can be caused by a very long modulation cycle as well, but we cannot separate those based on the TESS data alone, if no significant amplitude change is detectable alongside with them.

Therefore we refrain from claiming that any star that shows sidepeaks to the main pulsation frequency and its harmonics must automatically be a Blazhko star and set more strict rules. However, in lieu of a detailed model, reasons either for or against such distinction are based on phenomenological and methodological considerations alone. 

\begin{figure*}[]
\includegraphics[width=0.98\textwidth]{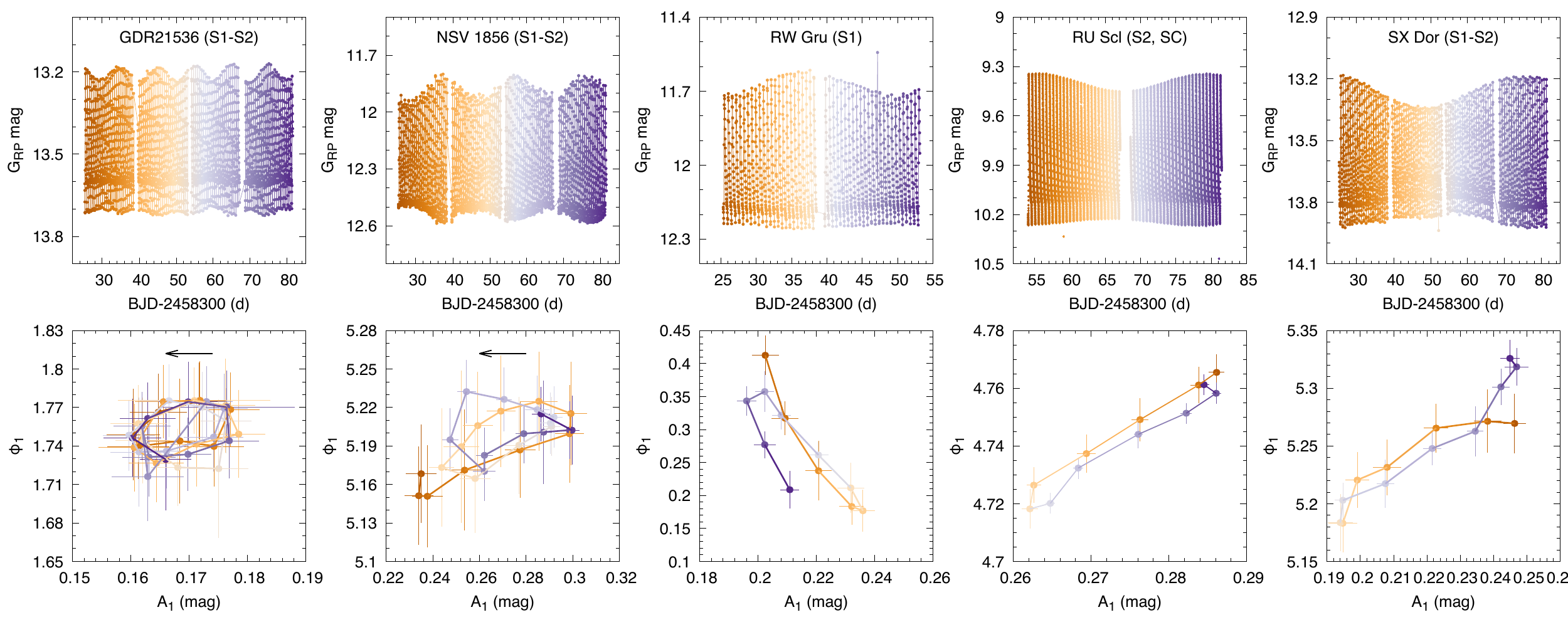}
\caption{Five examples of modulated RRab stars. Upper row: light curves. Bottom row: loop diagrams of the $A_1$ and $\phi_1$ Fourier terms. Arrows mark the direction of progression, where applicable, and color follows progression in time.}\label{fig:rrab-bl}
\end{figure*}

\begin{figure*}[]
\includegraphics[width=0.98\textwidth]{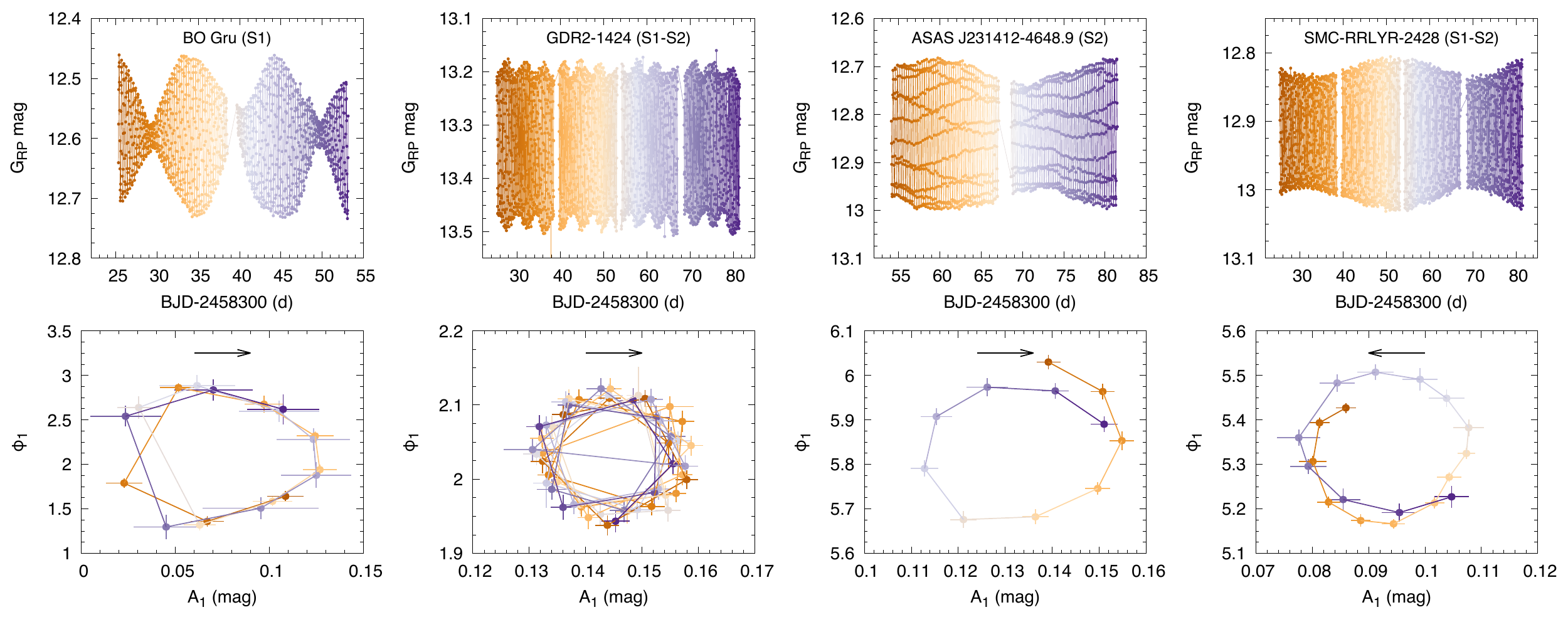}
\caption{Same as Fig.~\ref{fig:rrab-bl} but for modulated RRc stars.}\label{fig:rrc-bl}
\end{figure*}

\vfill

\subsubsection{Blazhko RRab stars}
Uncertainties in the amplitude and phase variations can be mitigated by requiring the simultaneous detection of both. This approach worked well for the original \textit{Kepler} data \citep{benkoszabo-2015}. We followed the same technique in our study, but did not limit the modulation periods, i.e., we classified clear but non-repeating amplitude and phase changes as likely long-period Blazhko stars.

Of the 82 RRab stars, we were able to identify Blazhko cycles or amplitude and phase changes compatible with partial cycles in 39 cases. In 9 more stars, we detect some temporal changes but cannot say conclusively whether those variations are generated by intrinsic modulation or contamination from blending and/or stray light. In a further 9 cases, contamination is present, or the calculated amplitude drops similarly throughout the sector with the phase not changing; therefore we cannot rule out that low-level modulation is present. This provides an occurrence rate of at least 47.5\% (39/82), and, at most, 70.7\% (58/82) which agrees with results from multiple different surveys \citep[see, e.g.,][]{jurcsik2009,benko2014}. The results are summarized in Appendix Table \ref{tab:bl}. Most of the detections are only partial modulation cycles, making it impossible to determine the modulation period(s). We were able to determine or estimate the Blazhko periods in five stars. The shortest one is SU~Hyi, with a period of $P_{\rm BL} = 5.55$~d, which is among the shortest values ever recorded \citep{skarka2020}. We investigate the light curve of SU Hyi in more detail in Sect.~\ref{sect:contam}. In two more cases we identify new, shorter modulation periods in known Blazhko stars. 

The Blazhko effect comes in many shapes and forms. Recenty, \citet{skarka2020} separated the modulated OGLE stars into six morphological groups. Our TESS light curves are unfortunately too short for similar exercises, but we can still examine how the amplitudes and phases vary over time. We calculated the temporal variations both by fitting a template light curve to the segments of the data and by just fitting the $A_1$ and $\phi_1$ parameters in each segment. The results of the two methods agreed with each other. We present five examples in Fig.~\ref{fig:rrab-bl}. The loop diagrams show a variety of $A_1-\phi_1$ relations, with the two parameters changing in a correlated, anti-correlated or circular pattern.

\subsubsection{Blazhko RRc stars}

Among the RRc stars we found four modulated stars, giving us an occurrence rate of approximately 13\%. This is more than double than the 5.6\% rate found by \citet{netzel2018} among the overtone stars in the Bulge, and closer to the results of targeted studies of globular clusters \citep{jurcsik2014,smolec2017}. It is, however, based on a very small sample size. We note that a fifth star, IY~Eri, also shows low-level amplitude variation and side peaks but we did not detect unambiguous phase variations. This suggests that the star is more likely to suffer from contamination than being modulated. 

These four stars paint a complex picture of Blazhko morphology among RRc stars. ASAS J231412--4648.9 and OGLE--SMC--RRLYR--2428 are archetypal modulated stars with moderate amplitude changes and roughly month-long cycles ($22.2\pm0.7$~d and $38\pm3$~d, respectively). \textit{Gaia} DR2 549562557969199\-1424 (hereafter GDR2--1424) has a modulation period of only $4.340\pm0.010$ d, which is among the fastest Blazhko cycles known: only a handful of stars are known with periods below 5 d, and all were recently found in the OGLE bulge collection \citep{netzel2018}. The last one, BO Gru, has a nearly sinusoidal light curve with extreme amplitude changes and with a period of $10.07\pm 0.08$~d, based on the TESS data in themselves. We initially classified this as a false positive detection, but its luminosity places this star among the RR Lyrae variables. 

We observe large differences in the amplitudes of the modulation sidepeaks in three out of four cases. In BO~Gru, the two sidepeaks have amplitudes of 55.5 and 1.0 mmag, respectively. The asymmetry parameter, $Q$, defined as the ratio of the difference and sum of the sidepeak amplitudes by \citet{Alcock2003}, is --0.97 for BO Gru, and 0.88 for SMC-RRLYR-2428. We find only one sidepeak in GDR2--1424, but a detection limit of 0.5 mmag suggests a lower limit of $Q < -0.93$. The fourth star, ASAS J231412-4648.9 has more symmetric peaks with $Q=-0.13$. These findings are in agreement with that of \citet{netzel2018}, who measured strong sidepeak asymmetries in many OGLE RRc stars. If we compare these results to the phase diagrams in Fig.~\ref{fig:rrc-bl}, we observe that the stars with negative \textit{Q} values have $A_1-\phi_1$ loop diagrams with clockwise progression, whereas positive \textit{Q} pairs with counter-clockwise progression. This correlation between the sidepeak asymmetry and the loop direction is in agreement with the analytical modulation formalism presented by \citet{Benko-2011}.

\subsubsection{Clearly non-modulated stars}
\begin{figure}[]
\centering
\includegraphics[width=1.0\columnwidth]{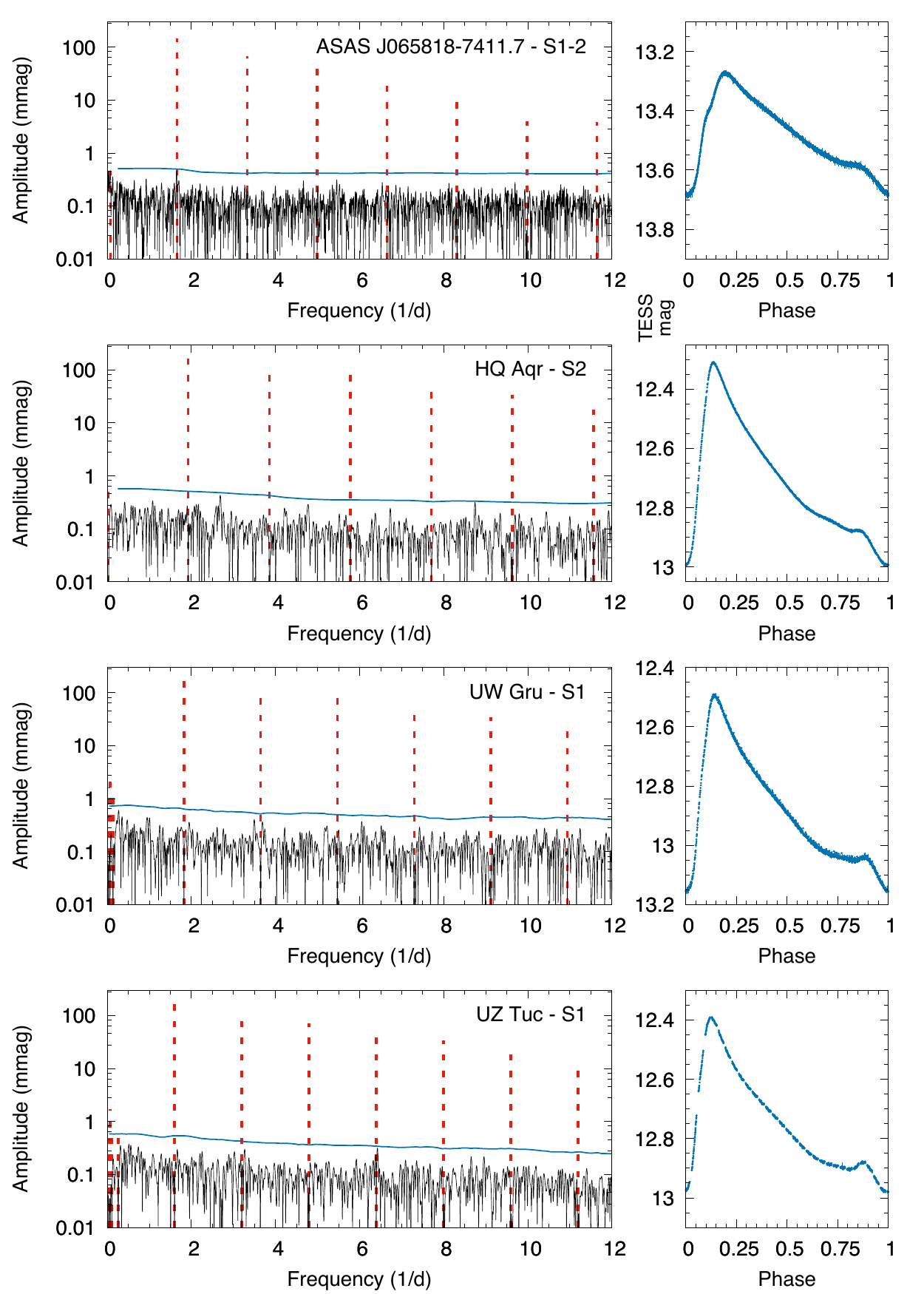}
\caption{Non-Blazhko RRab stars. Left: residual spectra with the positions of the pulsation peak and its harmonics (red dashed lines) and the 4.0 SNR level (blue line) marked. Right: folded light curves.}\label{fig:non-bl}
\end{figure}

We identified 22 RRab stars whose frequency spectra do not show any significant sidepeaks (SNR $>$ 4 against the smoothed spectra) near the main peak or its harmonics. This indicates a lower limit of 28.2\% for non-modulated stars at the photometric precision of TESS. Four examples are shown in Fig.~\ref{fig:non-bl}. Red dashed lines mark the fitted and subtracted main peaks, plotted over the residual frequency spectra. Blue lines show the SNR = 4 levels. These four stars are clearly not modulated, with an upper limit of 0.5--0.6 mmag for any sidepeaks in the spectra. While even these stars can conceivably experience modulation below the sensitivity or over considerably longer periods than the length of one or two Sectors, the result indicates that after rigorous processing, RRab stars without sidepeaks can be identified in the sample. 

\subsection{Additional modes and asteroseismology}
Extensive, high-quality data revealed an abundance of additional low-amplitude modes in RR Lyrae stars over the last decade. Interestingly, in RRab stars these almost always coincide with the presence of the Blazhko effect, whereas non-modulated stars seem to be pulsating purely in the fundamental mode. Additional modes were also detected in some RRab OGLE stars. A preliminary analysis of the K2 observations of RRab stars, however, indicated a surprising dichotomy between the OGLE Bulge sample and the field stars observed by space telescopes \citep{molnar2017}. Additional modes were found only in distinct segments of the Petersen diagram in the OGLE light curves: a larger group at the shortest periods (0.28--0.45 d) \citep{prudil2017}, from which three stars were then observed by K2 as well \citep{Nemec-2021}; a few similar, but long-period ($> 0.65$ d) stars \citep{smolec2016}; and the group of anomalous RRd (aRRd) stars, offset in period ratio from the normal RRd stars, between 0.45--0.55 d \citep{soszynski-2016}. In contrast, virtually all of the CoRoT, \textit{Kepler}, and K2 detections group in the middle (0.42--0.7 d). The only exceptions are the field stars V1127~Aql and V1125~Sgr, observed by CoRoT and K2, respectively, which fall into the short-period group \citep{chadid-2010,Nemec-2021}.

If this discrepancy is real, it could signal physical differences between the Bulge and halo/field populations that manifest themselves in the asteroseismology of RR Lyrae stars as well. We note that the analysis of the OGLE sample by \citet{prudil2017} focused primarily on strong extra modes in the short-period, non-Blazhko stars, whereas that of \citet{smolec2016} looked exclusively at long-period stars. The mid-period range was not searched thoroughly, so the modulated stars in this range could still hide Bulge stars with additional modes in them (Prudil, Smolec, priv.comm.). 

Detailed analysis of RRc and RRd stars via space-based photometry has been scant so far. About a dozen stars have been published, based on data from a variety of missions, ranging from MOST, CoRoT and \textit{Kepler} to K2, respectively \citep{gruberbauer2007,szabo2014,molnar2015,moskalik2015,kurtz2016,sodor-2017}. Many more were identified in the K2 Campaigns by \citet{molnar2018} and \citet{plachy2019} but they still await closer inspection. On the other hand, the RRc population in the Galactic bulge has been thoroughly studied in the OGLE-III and IV surveys \citep{netzel2015a,netzel2015b}, as well as the members of the globular clusters M3 and NGC~6362 \citep{jurcsik2015,smolec2017}. A recurring theme of these works has been the presence of the 0.61-type or $f_{\rm X}$ modes that seem to have a strong association with the first radial overtone in RRc and RRd stars both. In contrast, so far there is no indication of the modes associated with RRab stars appearing in double-mode stars.  

\subsubsection{Additional modes in RRab stars}

We detected additional modes in 29 RRab stars, which represents 35\% of the sample. The majority of these are also modulated or possibly modulated, with only two stars falling into the contaminated, and hence uncertain category, and another into the non-modulated group. This result agrees with findings based on \textit{Kepler} and CoRoT data, where additional modes were also identified almost exclusively in Blazhko RRab stars, suggesting a connection between modulation and excitation \citep{benko-2010,szabo2014,molnar2017}.

According to the recent results, additional modes in RRab stars appear in three broad regions, as we show in Fig.~\ref{fig:rrab-petersen} \citep{molnar2017}. Each of the three types of signals may appear alone, in pairs, or together in the stars.  One is near the expected value of the second radial overtone between $P_2/P_0 \approx 0.58-0.60$. These signals, usually labelled as $f_2$--type modes, form the most well-defined group in the Petersen-diagram. Another set of peaks, near $P/P_0\simeq 2/3$ or, $f/f_0 \simeq 1.5$, is the potential sign of period doubling (PD) \citep{szabo-2010,kollath-2011}. However, the peaks here do not form a well-defined line around the half-integer frequencies as one would expect from the presence of period doubling only. Instead they spread out to lower frequencies, up to ratios $P_2/P_0 \approx 0.7$. While some spread in the observed frequencies can be expected from the variable nature of period doubling, the peaks clearly group at $P/P_0 \ge 2/3$ ratios instead of around the 2/3 value. This suggests that we do, in fact, see non-radial modes in this range, which makes differentiating \textit{bona fide} period-doubled RRab stars from stars with nearby non-radial modes very difficult. 

The third group, the $f_1$-type modes, is loosely connected to the second, between $P_2/P_0 \approx 0.71-0.78$, encompassing the RRd ridge and the aRRd stars. Some of these may potentially be the appearance of the first overtone, and model calculations offer two ways in which the mode can become excited through mode resonances outside the normal RRd regime. When the fundamental mode goes through a period-doubling bifurcation, hydrodynamic models can become unstable against the first overtone too \citep{molnar-2012}. Alternatively, linear model calculations suggest that a parametric resonance between the fundamental, first and second overtones could potentially excite the first overtone in aRRd stars, if their metallicity is high enough \citep{soszynski-2016}. However, some of our stars are too far away from the RRd and aRRd groups to accommodate a radial mode, and must be exciting non-radial modes near the first overtone instead. Linear calculations of non-radial mode RR Lyrae models done by \citet{dziem-1999} indicate that $\ell=1$ modes have the highest linear growth rates near the frequencies of the radial modes ($\ell=0$), such as the fundamental mode or the first overtone. Therefore, it is plausible that the peaks we detect near the first overtone position are various $\ell=1$ modes.

Our findings are generally in good agreement with the preliminary results obtained from the K2 sample (and the few CoRoT stars). While our first collection of TESS stars with additional signals covers a wider period range than that of K2 (0.39--0.7 day instead of 0.42--0.65 days) the discrepancies with the OGLE Bulge sample are still present. Extending the TESS sample with further Sectors and fainter stars will be crucial to understanding the difference between the Bulge population and those in the vicinity of the Sun. TESS will also enable us to compare the distribution of modes between other populations, such as the halo and disk stars. 

We also detect three signals in the anomalous Cepheid candidate SX~PsA. These fit right into the distribution of additional modes of the RRab stars. This could mean either that SX~PsA is also an RRab star, or that ACEP stars have very similar additional modes excited. An example of the latter has been observed in the ACEP prototype XZ~Cet, a first overtone star, in which the $f_{\rm X}$-type modes were found in the TESS data. The signals in XZ Cet line up perfectly with the modes detected in RRc and first-overtone classical Cepheids \citep{Plachy-Cep-2020}.

\subsubsection{Additional modes in RRc/RRd stars}
\label{sect:rrc}

As mentioned previously, a common phenomenon among RRc stars is the presence of peculiar additional modes. One class of modes appears between period ratios $P/P_{\rm O1} \in (0.6,0.64)$, forming two distinct loci at ratios $\approx 0.615$ and $\approx 0.63$, as shown in Fig.~\ref{fig:rrc-rrd-fx}. These modes, called either $f_{\rm X}$ or 0.61-type modes in previous works, appear to have a very strong connection to the first radial overtone. They appear in RRd stars and in classical Cepheids \citep{moskalik2009,molnar2017}, and have been recently identified in an overtone anomalous Cepheid star as well, as described in the companion paper \citep{Plachy-Cep-2020}. We now know that the mode detected in AQ Leo, the first additional mode observed in an RR Lyrae star, also belongs to this group \citep{gruberbauer2007}. The other distinct group is less frequent and has a period longer than the first overtone, at $P_{\rm O1}/P \approx 0.686$ (or, $P/P_{\rm O1} \approx 1.458$). The latter mode is, notably, longer than the expected fundamental period of a normal RR Lyrae star that would have a period ratio of $P_{\rm O1}/P_{\rm FM} > 0.725$ \citep{netzel2015b}. 

\begin{figure}[]
\centering
\includegraphics[width=1.0\columnwidth]{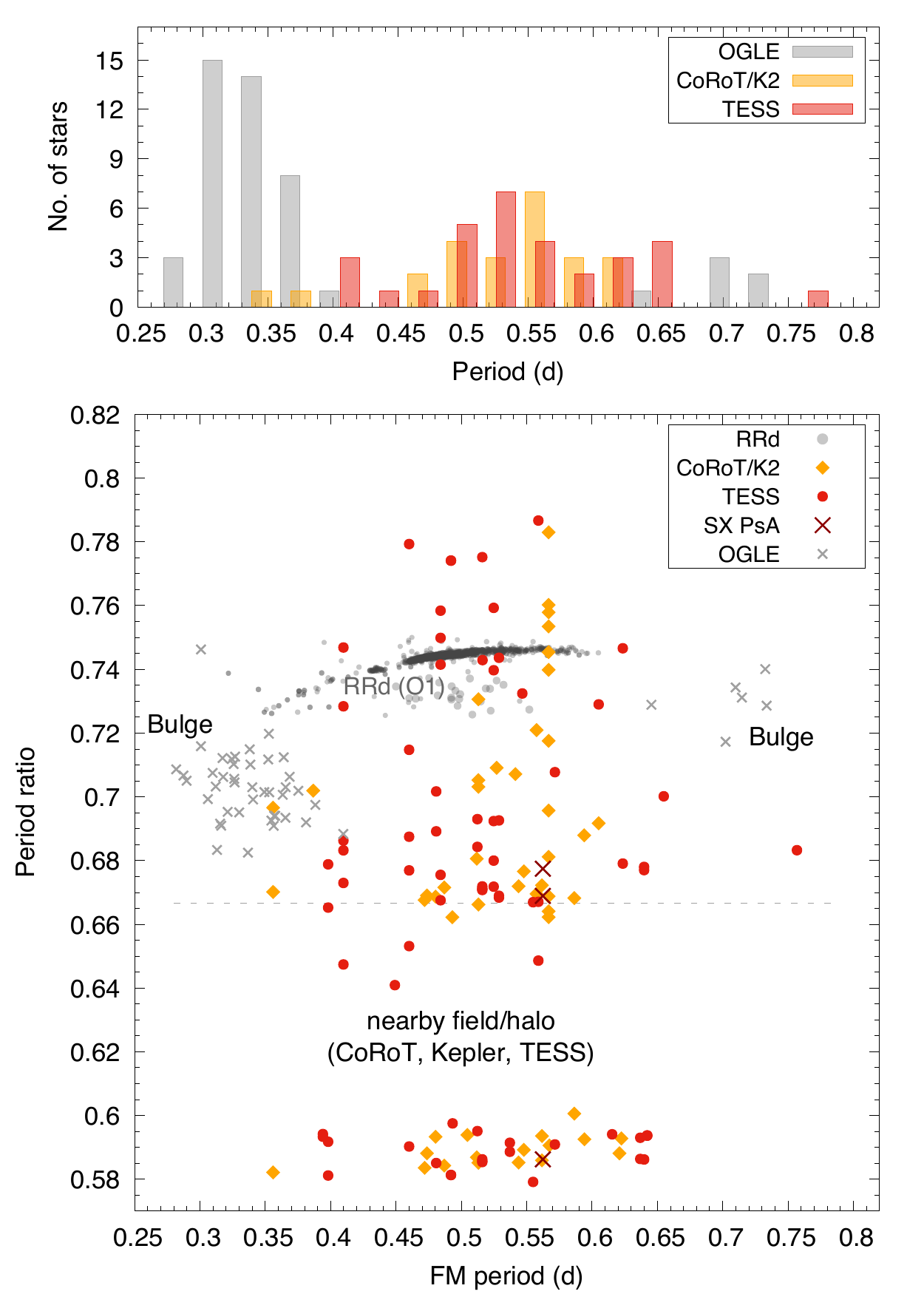}
\caption{Top: period distribution of individual RRab stars in which additional modes were detected: OGLE (grey), CoRoT/\textit{Kepler}/K2 (orange) and TESS (red). Bottom: Petersen diagram for the RRab stars. Grey crosses are the OGLE Bulge results, orange diamonds are from CoRoT, \textit{Kepler}, and the K2-E2 engineering test run, red points are the TESS observations. Small grey dots mark the distribution of the RRd stars and thus the position of the first overtone. The dashed line indicates the position of the half-integer frequency indicating period doubling. Large red crosses mark the ACEP candidate SX~PsA. Note that stars may have multiple additional signals detected in them. }\label{fig:rrab-petersen}
\end{figure}

\begin{figure}[]
\centering
\includegraphics[width=1.0\columnwidth]{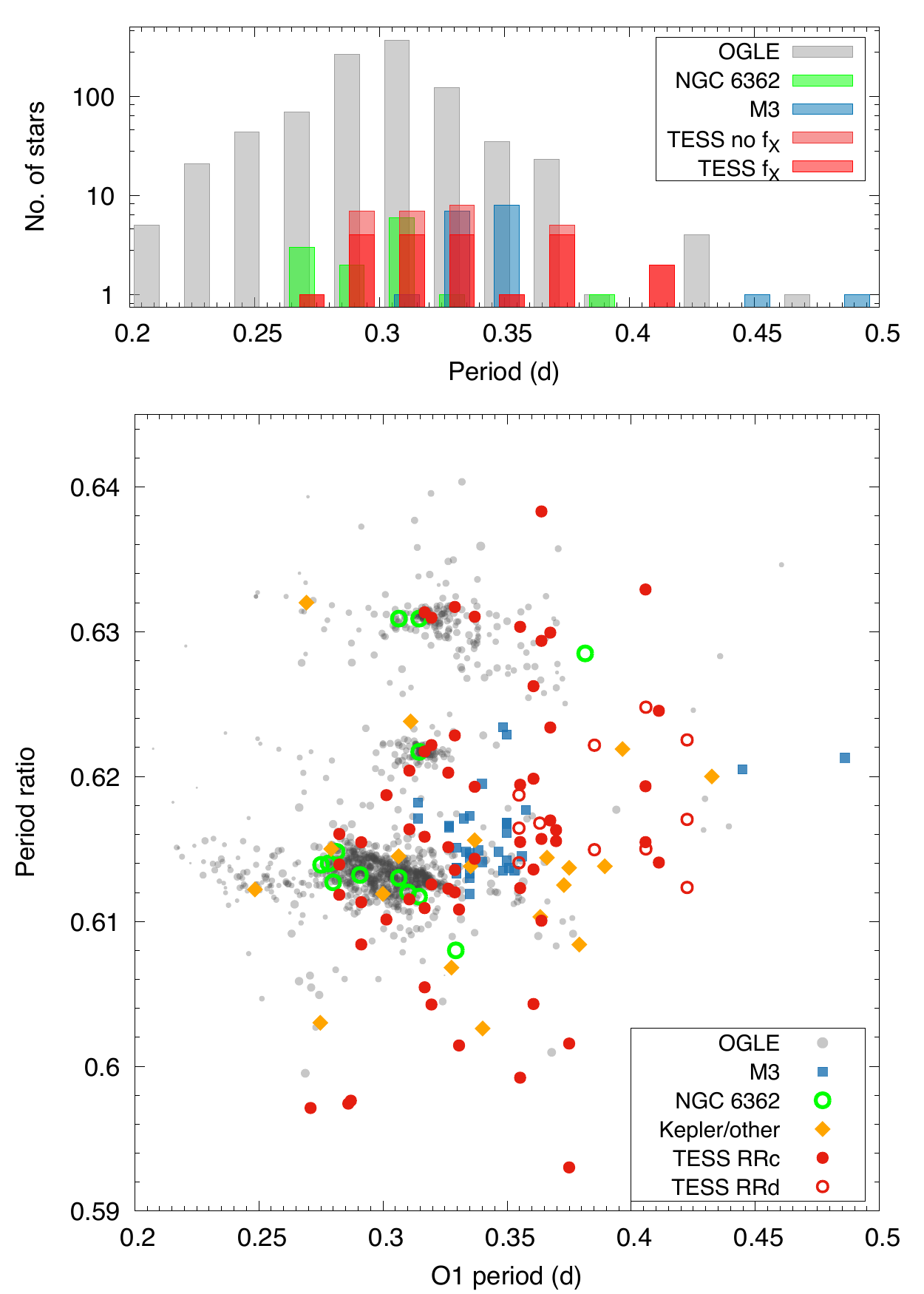}
\caption{Same as Fig.\ref{fig:rrab-petersen} but for the $f_{\rm X}$ modes in RRc and RRd stars Top: period distribution for OGLE (grey), TESS (red) and from two globular clusters, M3 (blue) and NGC 6362 (green). The light red extensions mark the TESS RRc stars where the modes are absent. Bottom: Petersen diagram of the $f_{\rm X}$ or 0.61-type modes. Grey points are the OGLE detections by \citet{netzel2019}; orange diamonds are various stars collected by \citet{moskalik2015} that include other space-based data; blue squares are the M3 stars by \citet{jurcsik2015}; green circles are the NGC 6362 stars \citep{smolec2017}; filled and empty red circles are the TESS RRc and RRd detections. }\label{fig:rrc-rrd-fx}
\end{figure}

The origins of these modes are not yet settled. One possibility presented by \citet{dziembowski} connects the short-period, 0.615- and 0.63-ratio signals to $\ell = 8$ and 9 modes. However, in that scenario, the true mode frequency is the $1/2f_X$ subharmonic, and the $f_X$ and $3/2f_X$ peaks we detect are harmonics: its only the viewing geometry and cancellation effects that make the $f_X$ peak the strongest. The long-period mode is even harder to explain: if we assume that these stars are RR Lyrae variables, the signal would belong to a heavily damped $\ell = 1$ mode. Excitation of such a mode without any other, much less damped neighbors is very hard to explain. Alternatively, it could be the fundamental mode of a binary-evolution pulsator: a low-mass, stripped core of a red giant that could accommodate such a low $P_{\rm O1}/P_{\rm FM}$ period ratio. These are known to exist but are very rare \citep{bep}. Furthermore, some stars show both types of extra modes, and the low-mass models cannot accommodate the high-$\ell$ $f_{\rm X}$ modes since those would also fall to different period ratios than in normal RR Lyrae stars. 

\begin{figure}[]
\centering
\includegraphics[width=1.0\columnwidth]{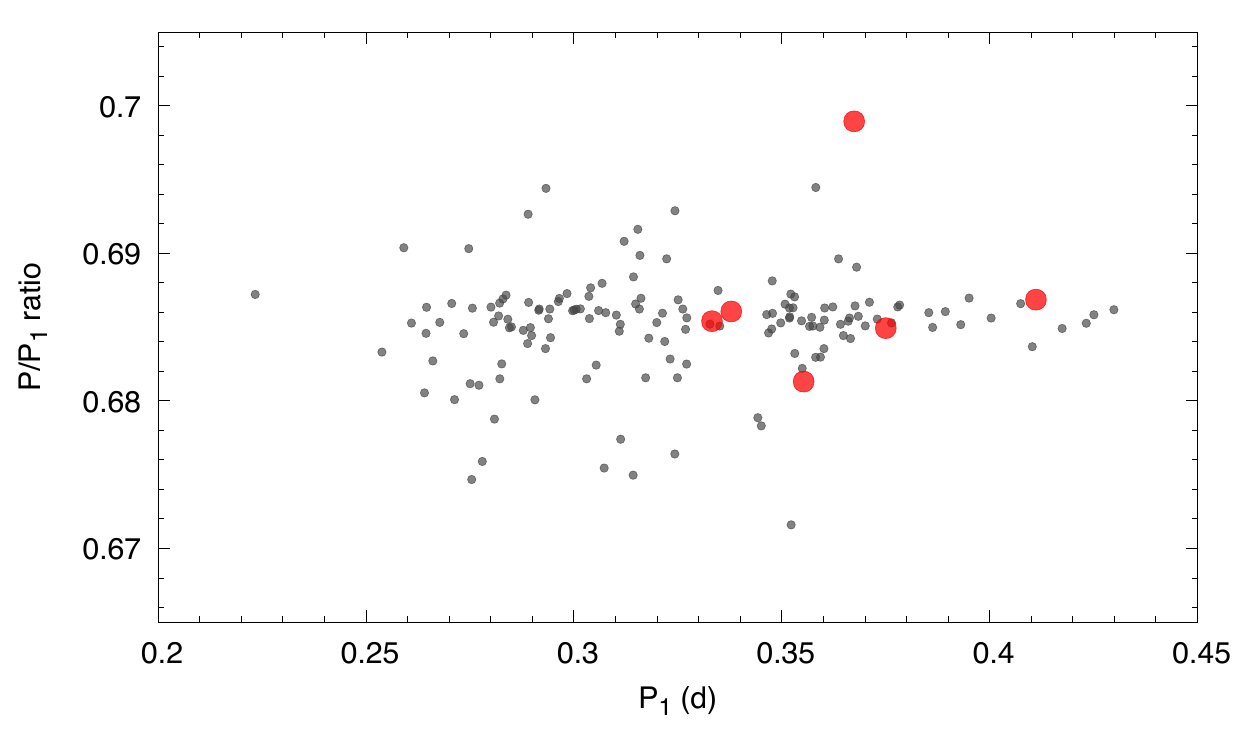}
\caption{The 0.68--type modes: small black circles are the OGLE stars from \citet{netzel2019}, large red circles are the TESS detections. }\label{fig:rrc-068}
\end{figure}

After the analysis of the 31 TESS stars, we concluded that although additional modes are frequent in RRc stars, they are not universal. We identified 6 stars (19\%) among the group to be pure overtone pulsators, down to the mmag level. We found the incidence rate of the $f_{\rm X}$ and 0.68-type modes to be 65\% and 16\% (20/31 and 5/31), respectively, and one more marginal case that might be a 0.68-type mode, but the period ratio is somewhat at higher at $P/P_0 = 0.699$ (Fig.~\ref{fig:rrc-068}). These numbers are clearly much higher than the 8.3\% and 1.3\% incidence rates found by \citet{netzel2019}, based on the entire OGLE bulge data that were limited by data quality. The incidence rate of the $f_X$ was found to be 27\% and 63\% in high-cadence OGLE fields and in the globular cluster NGC~6362, respectively: the latter matches our result \citep{netzel2015a,smolec2017}. Both types of modes were identified simultaneously in two stars (6\%) within the TESS RRc sample, in ASAS J213826-3945.0 and NSVS 14632323. The detection rate of the 0.68-type mode is $\approx5\%$ in the overall RR Lyrae sample, which is considerably higher than the estimated 0.8\% contamination from binary evolution pulsators calculated by \citet{Karczmarek-2017}, suggesting that this mode must have a different origin. 

One peculiar star in the RRc group is BV~Aqr. Beside the $f_{\rm X}$ peaks it also shows two more signals at $f/f_1$ = 0.709 and 0.741 frequency ratios. These are both clearly outside the regimes where the 0.68-type mode and the subharmonics of the $f_{\rm X}$ modes would exist. The latter component, however, could conceivably be the fundamental mode: this value is slightly below the canonical regime at the corresponding period of 0.491~d, but well within the extended region of aRRd stars (grey crosses in Fig.~\ref{fig:rrab-petersen}). However, an important difference between BV~Aqr and aRRd stars is that  in aRRd stars the amplitude of the fundamental mode is almost always higher than, or at least the same order of magnitude as, that of the first overtone, whereas in BV~Aqr the ratio is very low, only $A_1/A_0 = 0.016$. This suggests that BV~Aqr is not an aRRd star, or not yet, but it is rare example of an RRd star with either an emergent or an unusually low-amplitude fundamental mode component instead. Interestingly, this possibility was already raised by \citet{jerzykiewicz-1995}, who identified a single secondary signal at 0.717 ratio, i.e., between the two we identified. A further candidate for this type of mode content is NSV~1432, but there the detection of a signal with 0.740 period ratio to the first overtone remains marginal.

We identified the $f_{\rm X}$ modes in all five RRd stars. The 1/2 and 3/2 $f_{\rm X}$ peaks were also found in all cases but with lower amplitudes. We then searched the RRd stars for further extra modes. Neither the 0.68 mode nor any modes related to the fundamental mode were visible, although in RRd stars the $2f_{\rm O1}-f_{\rm FM}$ combination peak overlaps with the expected position of the $f_2$ group. Notably, we identified $f_{\rm X}/2$ frequencies---proposed to be the true pulsation frequencies by \citet{dziembowski}---in 9 RRc and 2 RRd stars, i.e., in about half of the stars where the $f_X$ modes could be detected.

We then compared the distribution of the $f_{\rm X}$ modes to those of other RRc populations. The upper panel of Fig.~\ref{fig:rrc-rrd-fx} presents the period distribution of the OGLE and TESS stars, plus the RRc stars from the globular cluster M3 that was surveyed by \citet{jurcsik2015}. The TESS sample is currently too small to draw strong conclusions from this histogram alone, except for the lack of short period RRc stars in the TESS and M3 groups. 

The lower panel of Fig.~\ref{fig:rrc-rrd-fx} is a Petersen diagram where we included not only the recent OGLE results but also the additional modes found in the globular cluster M3, and various detections from Kepler, CoRoT, MOST, and other sources \citep{jurcsik2015,moskalik2015,netzel2019}. It is immediately clear from this plot that the field stars seen by TESS are distributed much more evenly and do not form the clear ridges the Bulge population does. Although the detections peak more or less at the same period range (0.28--0.32 d), the Bulge stars extend farther out to shorter periods, whereas the field stars populate the longer side more evenly. The M3 group (blue squares) forms a rather tight clump that does not line up with the ridge of Bulge stars. 

All these differences suggest again that the mode content of RR Lyrae stars is dependent on the physical parameters, probably on the metallicity and age (evolutionary stage) as well. The existence of such dependencies has already been established for the radial modes in RRd stars, from both theoretical and observational approaches
\citep[see, e.g.][]{szabo-rrd-2004,Coppola-2015}. Similarly, the models presented by \citet{dziembowski} show that higher masses and lower Z values may shift the proposed $f_X$ modes to higher period ratios. We also know that the [Fe/H] abundances for RRc stars spans a wide range, from [Fe/H]$\approx-2.5$ up to the recently confirmed, metal-rich end: around --0.5 dex \citep{sneden-2018}. Unfortunately, most RRc stars in the TESS sample do not have  [Fe/H] measurements of sufficient accuracy to test these theoretical predictions. Only 10 of them have photometric [Fe/H] values in \textit{Gaia} DR2, and all have large uncertainties ($\pm0.24$ dex). Further, we identified only 15 RRab and three RRc stars among the recent spectroscopic and photometric [Fe/H] measurements published by \citet{crestani2020,Crestani-2021} and by \citet{Mullen-2021}. Only four and two of those, respectively, feature extra modes, which prevents us from conducting a detailed analysis. 

However, we know that the bulge population clusters around [Fe/H] = $-1.02\pm0.18$~dex, whereas a considerably lower value of [Fe/H] = $-1.43 \pm 0.07$~dex was found for the M3 RR Lyraes \citep{pietrukovicz-2012,cacciari-2005}. Furthermore, a recent, homogeneous study of a large sample of field RR Lyrae stars showed that their [Fe/H] values peak at --1.5~dex \citep{Marengo-2020}. Spectroscopic [Fe/H] measurements for the two RRc stars with extra modes in the sample---AO~Tuc ($-1.69\pm 0.20$) and BV~Aqr ($-1.61\pm0.011$)---match the average halo metallicity \citep{crestani2020}. It is thus evident that differences in chemical composition---and the changes those impart on the stellar structure---govern, in part, mode selection and mode frequencies in RR Lyrae stars,  with our results suggesting that this holds for new modes as well. 

We also compared the positions of the 0.68--type modes to those identified from the OGLE survey by \citet{netzel2019}. These cluster towards the long-period half of the distribution, just like the other modes do. Five stars are within the spread of the OGLE points, but one star, ASAS~J212331-3025.0 lies outside, at a period ratio of nearly 0.70: we treat this sixth signal as a tentative detection. 

\begin{figure*}[]
\includegraphics[width=1.0\textwidth]{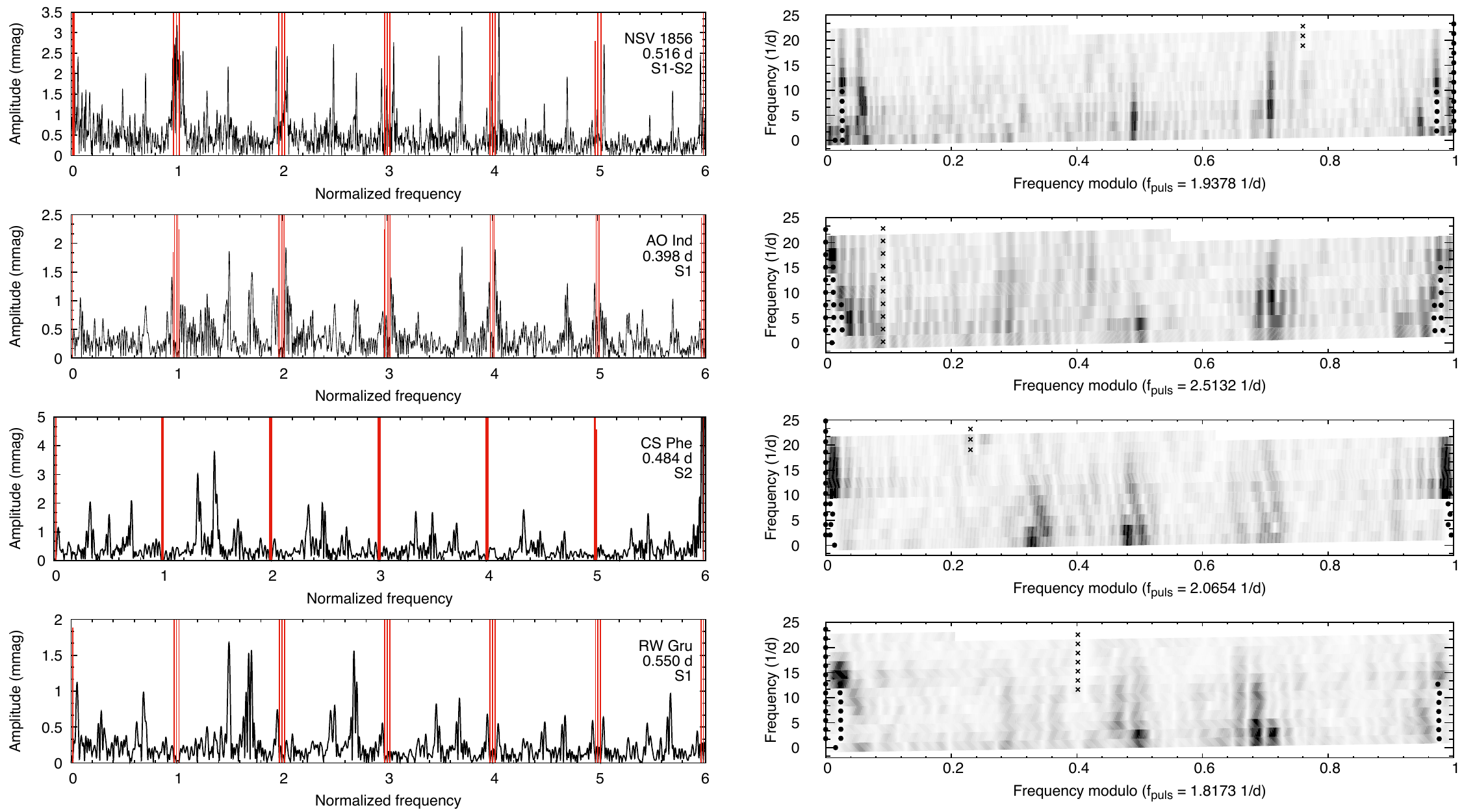}
\caption{Fourier spectra (left) and \'echelle diagrams (right) of selected RRab stars. In the frequency spectra, pulsation and modulation peaks that have been removed are marked with red lines. In the \'echelle diagrams, the locations of these peaks are marked with black dots. Black crosses mark the locations of those pulsation harmonics that are reflected back from the Nyquist frequency. }\label{fig:rrab-echelle}
\end{figure*}

\subsubsection{Degeneracies and RRab \'echelle diagrams}
In most cases additional modes in RRab stars appear clearly between the main pulsation frequency and its first harmonic (between $f_0$ and $2f_0$), along with lower-amplitude combination peaks at $f \pm nf_0$. We list these identifications, along with some peculiar signals, in the Appendix. However, there are exceptions to this pattern, where the highest-amplitude peak falls elsewhere. One example in the K2-E2 sample was already described by \citet{molnar2015}, where the signal was tentatively identified as a low-frequency \textit{g} mode, but the $f_g+f_0$  combination was suspiciously close to the 0.6 period ratio, the position of the second overtone-like peaks. We found similar examples in the present TESS sample as well. We identified multiple stars for which the highest-amplitude peak in the $f_{\rm addtl}+ n\,f_0$ combination series is apparently below $f_0$ or above $2f_0$. Yet, these series have a component that coincides with the positions of the additional mode groups described above. These findings are shown in Fig.~\ref{fig:rrab-petersen}.  

In order to compare the various signal distributions in frequency and amplitude, we generated \'echelle-type diagrams for selected stars. \'Echelle diagrams are a key tool for investigating solar-like oscillations that create (quasi)\-repetitive patterns in the frequency spectrum or power-density distribution. Data are split into segments at the repetition frequency (for solar-like oscillations, the large frequency separation), and the amplitudes are mapped onto a frequency vs. modulo frequency plane. Signals, such as consecutive oscillation modes, form vertical ridges in these plots \citep{bedding2010}. This type of visualization was already used for RR Lyrae stars and Cepheids to plot the distribution of modulation sidepeaks \citep[see, e.g.,][]{sodor2011,guggenberger2012,molnar2014}.

Here, we are using \'echelle diagrams to study the additional modes in a new way. We utilize the repetitions in frequency spectrum caused by the coupling to the large-amplitude, strongly non-linear radial mode. We created the frequency spectra and \'echelle diagrams in Fig.~\ref{fig:rrab-echelle} by first removing the main peak and its harmonics, plus the modulation triplets if necessary: these would otherwise obscure the low-amplitude signals. We then folded the residual spectra into \'echelle diagrams using the pulsation frequency ($f_0$). Unlike in normal \'echelle diagrams, however, the vertical ridges of Fig.~\ref{fig:rrab-echelle} represent the $f+n\,f_0$ combination series of the same mode instead of consecutive modes. 

\begin{figure*}[]
\includegraphics[width=0.98\textwidth]{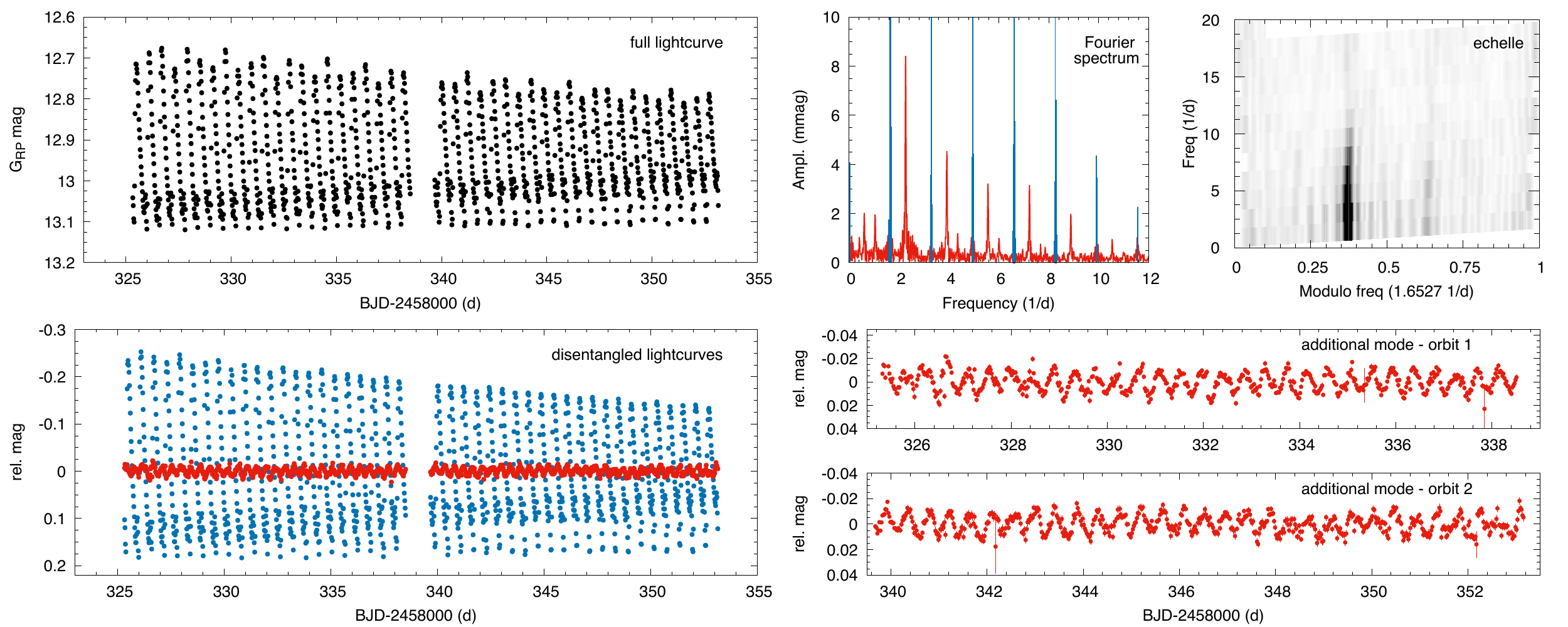}
\caption{CZ~Ind, an RRab star with a strong extra mode. Top row, from left to right: original light curve; frequency spectrum in red, with the prewhitened FM frequency components in blue; \'echelle diagram. Bottom row: disentangled light curves of the fundamental mode (blue) and the extra mode (red), and only the disentangled extra mode, as seen in each of the two orbits during the sector.}\label{fig:cz-ind}
\end{figure*}

Four stars are shown as examples in Fig.~\ref{fig:rrab-echelle}. The residual spectra and \'echelle diagrams show that:
\begin{itemize}
    \item Ridges generally appear at distinct modulus values, even though the position of the strongest peak in the series is not always between $f_0$ and $2f_0$. In NSV 1856, for example, the highest peaks for the two ridges appear between $2f_0$ to $3f_0$ and $3f_0$ to $4f_0$, respectively.
    \item The length of the ridges, indicative of the coupling between the fundamental mode and the additional mode, can be very different: in both NSV 1856 and AO Ind, for example, the one corresponding to period doubling, around $\sim(0.5+n)f_0$ is rather short, whereas the $f_2$ one at $\sim(0.7+n)f_0$ spans almost the full frequency range. (The $\sim0.6$ period ratio translates to $\sim 1.7$ frequency ratio.)
    \item Some ridges have multi-peaked structures that may also change as the frequency increases. In some other cases (CS Phe, DR Dor, SX PsA) we observe skewed ridges, as if the combination peaks were formed with a frequency different from the main peak and its harmonic series. 
    \item Some Blazhko stars, like NSV 14009, are devoid of any additional modes down to 0.3--0.4 mmag amplitudes.
\end{itemize}

The plots clearly show that additional signals in RRab stars are still limited to certain frequency modulus values. The variations in the amplitude distribution of the peaks, however, make it even harder to disentangle the mode identifications. It is unclear whether we seeing the same few modes already described by the Petersen diagram, where other effects (\textit{e.g.}, nonlinear interaction with the fundamental mode or stellar inclination and viewing angles) may change the nature of the strongest frequency component. This scenario would be reminiscent of the proposed nature of the $f_X$ mode in RRc stars, with the first harmonic of the pulsation frequency having higher amplitude.

Remarkably, the \'echelle plots also display a variety of structures, even though we would expect that the series of peaks are simply $f + n\,f_0$ combinations. The skewed ridges, in particular, would suggest that these additional frequency components in the star respond to a slightly detuned fundamental mode that we observe. A simple explanation would be that if we only observe a segment of the Blazhko cycle the pulsation frequency we measure is not modulation-averaged. However, using a fundamental-mode frequency determined from more extended ground-based data has very little effect on the plots. We calculated the frequency shifts necessary to straighten the ridges to be --0.0078 d$^{-1}$ (CS Phe), 0.0085 d$^{-1}$ (DR Dor) and 0.0057 d$^{-1}$ (SX PsA). This amounts to, in relative terms, 0.32--0.38\% of the corresponding pulsation frequencies. 

Clearly, there is much to learn about the additional modes in these stars. This may appear challenging, as the Fourier amplitudes of these modes are in the mmag to sub-mmag level. TESS, however, can help to identify bright targets with stronger extra modes that can then be easily followed up from the ground. One such example is CZ Ind, in which the presence of an O1-type mode is immediately obvious from the beating pattern in the light curve maxima. In Fig.~\ref{fig:cz-ind} we show the frequency spectrum, full light curve and \'echelle diagram of the star (top row), as well as the disentangled pulsation modes: the fundamental mode that also shows the Blazhko effect, and the extra mode (bottom row). The latter reaches 0.01--0.02 mag peak-to-peak amplitude, a signal which is within reach of even moderate-sized ground-based telescopes.

\subsection{RRd stars}
Of the five RRd stars, four were identified as such earlier, based on their All-Sky Automated Survey (ASAS) light curves by independent authors \citep{Wils-2005,Wils-2006,bernhard-2006,szf-2007}. The last one, \textit{Gaia} DR2 6529889228241771264 (shortened to GDR2 71264 from here on), was catalogued as an RRc star both in the Catalina Sky Survey and in \textit{Gaia} DR2 \citep{Drake-2017,clementini2019}. 

\begin{figure}[]
\centering
\includegraphics[width=1.0\columnwidth]{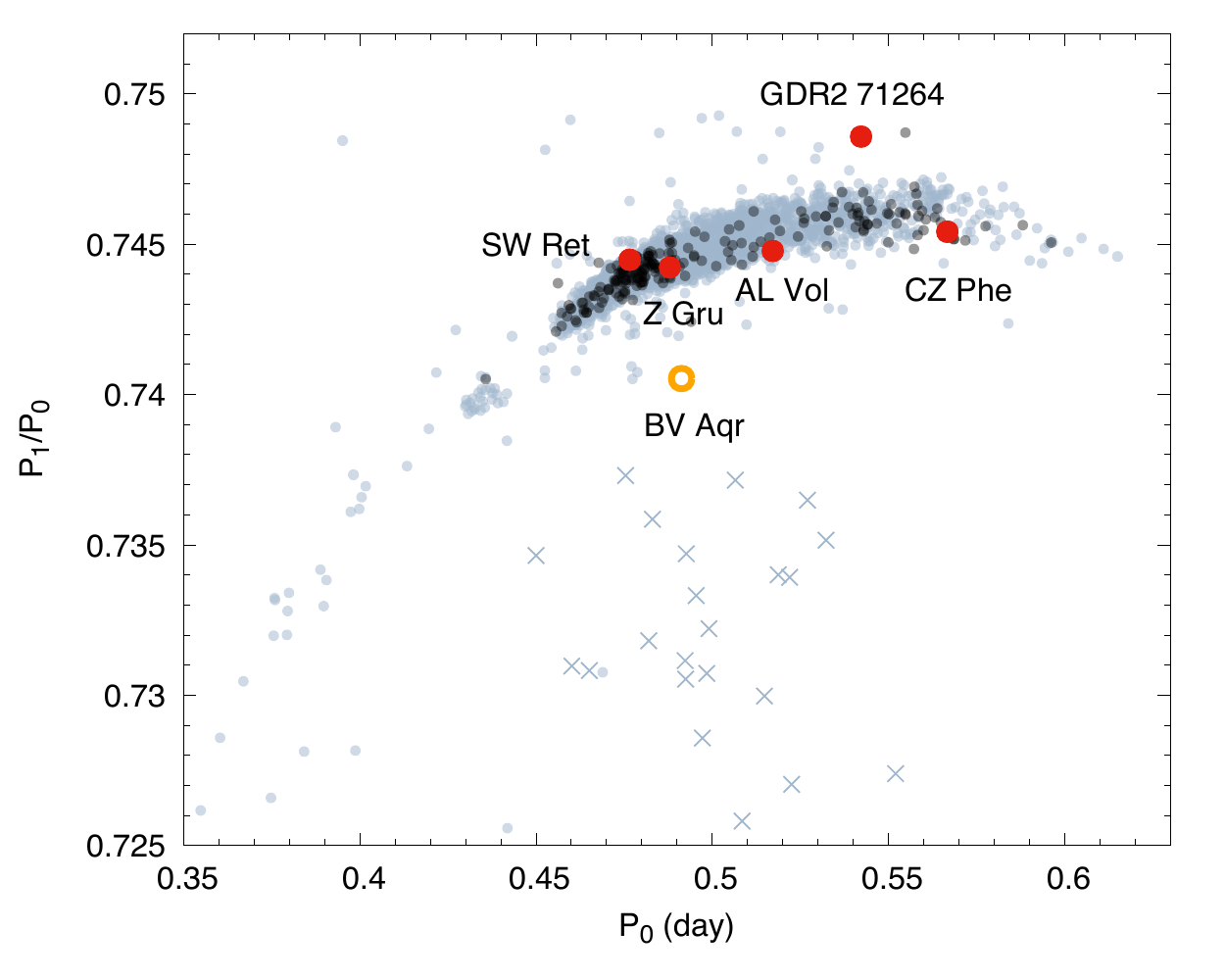}
\caption{Petersen diagram of the five RRd stars we analysed (red). Black dots are various field RRd stars, grey ones are OGLE stars from the bulge and the Magellanic Clouds. Sources are the same as the RRd stars collected in \citet{molnar2015}. Crosses are anomalous RRd stars found in the Clouds \citep{soszynski-2016}. }\label{fig:rrd-petersen}
\end{figure}

Double-mode stars are valuable targets as they provide simultaneous period constraints for multiple modes that can help us determine their physical properties much more accurately than those of single-mode pulsating stars \citep[see, e.g.,][]{molnar2015,Molnar-2019,Joyce-2020}.
If we place these five stars onto the Petersen diagram, four of them fall into the main locus of RRd stars where the other Galactic RRd stars reside (Fig.~\ref{fig:rrd-petersen}). GDR2 71264, however, lies above the main group at a period ratio of $P_{\rm O1}/P_{\rm FM} = 0.7486$, along with a handful of stars that belong to the Magellanic Clouds. A comparison with the non-linear model calculations done by \citet{szabo-rrd-2004} suggests that this star is potentially a low-metallicity, high-mass and high-luminosity RR Lyrae. An order-of-magnitude estimate for its physical parameters suggests $M \approx 0.8 - 0.9 \,M_{\odot}$, L $\sim 60\,L_{\odot}$ and Z $ \sim 10^{-4}$ (the latter being equivalent to [Fe/H] $\approx -2.2$, assuming that [$\alpha$/Fe] = 0). 

Parameters for the rest of the stars in the main group are more ambiguous as the mass and metallicity parameter ranges overlap, and these objects would require more detailed modeling. This can be done, but the selection and calculation of stable double-mode non-linear models is very time-consuming \citep{molnar2015}, therefore we refrained from it in this work. We also note that the physical validity of the double-mode models is not entirely settled yet \citep{smolec-2008}.

\begin{figure*}[]
\includegraphics[width=1.0\textwidth]{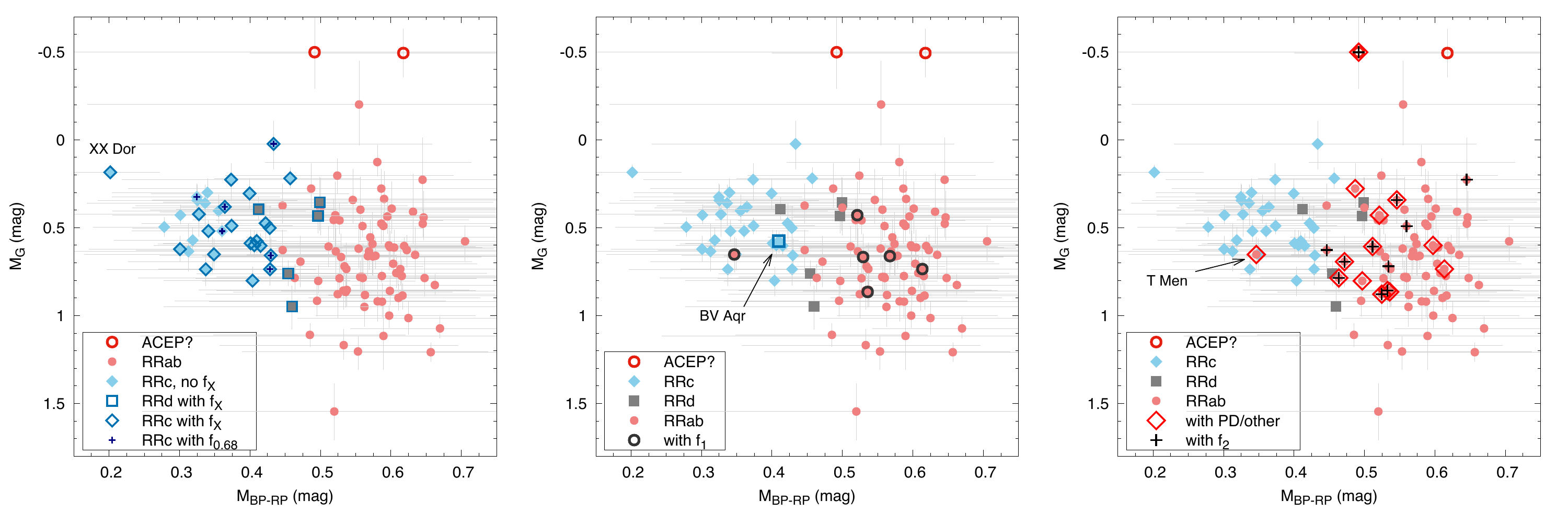}
\caption{Distribution of the stars with additional modes within the RR Lyrae instability strip. Left: the $f_X$ and $f_{0.68}$ modes in RRc and RRd stars. Center: the $f_1$-type modes in RRab stars. Right: the $f_2$ and PD or other type of modes in RRab stars. Two stars marked (XX Dor and T Men) had their interstellar extinction values overestimated and hence appear too blue and too bright. The third star, BV~Aqr, is the extreme RRd candidate.}\label{fig:cmd-modes}
\end{figure*}

\subsection{Distribution of the additional modes in the CMD}
Next, we looked at where the additional modes occur within the RR Lyrae instability strip. We plotted the stars in the \textit{Gaia} CMD and marked those that feature extra modes, focusing on the main mode groups. Upon closer inspection we find two outliers, XX~Dor and T~Men: the first is blueward of the RRc group, whereas the latter, an RRab star, is blueward of the RRab group. Both stars are near the LMC, hence we suspect that the dust map overestimated the interstellar extinction for these stars as well, just to a lesser extent than the three other cases we have shown earlier.

The middle and right panels of Fig.~\ref{fig:cmd-modes} highlight RRab stars where the $f_1$-type (center), $f_2$-type and period doubling or other nearby peaks (right) appear. The sample sizes are rather low, so we cannot draw strong conclusions, but RRab stars with extra modes seem to appear more frequently at the blue part of the RRab cluster: the $f_2$ and PD/other modes clearly favor the RRc/RRab boundary. We marked BV~Aqr, the overtone star which is potentially an aRRd star: this star is also close to the boundary and its position is similar to the normal RRd stars in the sample.

In the left panel of Fig.~\ref{fig:cmd-modes} we plot the occurrence of the $f_X$ and $f_{0.68}$ modes in the RRc and RRd stars. Here it is much more evident that the $f_X$ modes also favor the RRc/RRab boundary. The reddest star where the mode is not detected being at $M_{\rm BP-RP} = 0.36$, from there on such extra signals show up in all RRc/RRd stars we analyzed. The $f_X$ mode has been detected from the ground in a few globular clusters as well. We can therefore compare the distribution of the modes in the CMDs of the clusters with our results. Both in NGC~6362 and M3, the $f_X$ modes were found to cluster towards the redder RRc and the RRd stars \citep{jurcsik2015,smolec2017}. This is in complete agreement with our findings and indicate that the $f_X$ modes are not excited in the hottest and bluest RRc stars.

As we discussed before, all five RRd stars in the sample show various $f_{\rm X}$ modes but apparently none of the additional modes exhibited by RRab stars, associated with the fundamental mode.

\subsection{RRd stars and the Bailey groups}

The first phenomenological classification of RR Lyrae stars dates back to \citet{bailey1902}, who created the groups a, b and c, based on the asymmetry and period of the light curves, as subclasses to the much broader group of short-period variables (Class IV  in the early variable classification scheme proposed by \citealt{Pickering1881}). This was later transformed into the Bailey classes RRab, RRc and RRd, based on physical arguments, referring to fundamental-mode, first-overtone and double-mode pulsators.

However, the discovery of additional modes in a large portion of RR Lyrae stars, some of which may apparently be the first overtone in an otherwise RRab-type light curve, potentially complicates this scheme. Therefore, it is timely to revisit the classical Bailey-type definitions and to reevaluate, how strict the boundaries of the different groups should be. One can argue, for example, that the RRab and RRc classes should refer to pure radial pulsators: this, however would require the creation of subgroups, or the expansion of the double-mode group, the RRd stars, to include all objects that show multiple pulsation modes. We must also consider then that this would make the RRd group very inhomogeneous. Another possibility is that the RRab and RRc groups should refer to stars that are dominated by, but not strictly limited to the fundamental and first overtone modes, respectively. Then the RRd group should refer to stars where these two modes are both present, with amplitudes in the same order of magnitude, dominating over other, much smaller signals. 

\begin{figure}[]
\centering
\includegraphics[width=1.0\columnwidth]{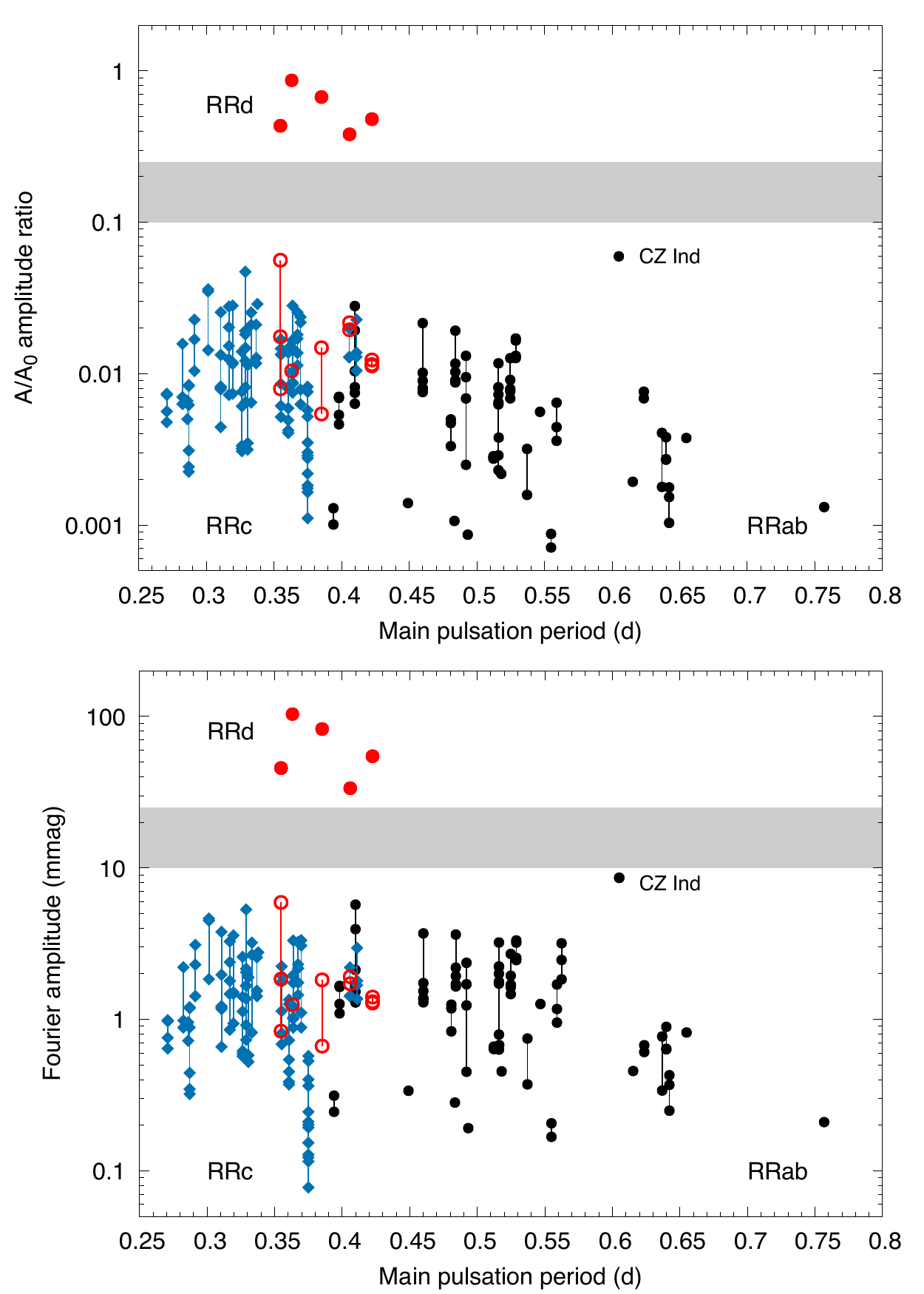}
\caption{Amplitudes of the independent frequency components we found (bottom) and their ratios compared to the amplitude of the strongest frequency peak (top). The grey band is between 10--25 mmag and $A/A_0$ = 0.10--0.25, respectively. Black small circles are additional components in RRab stars; blue diamonds are additional components in RRc stars; red full and empty circles are the ratios between the main peaks of the two radial modes and the additional peaks and the strongest peak in RRd stars, respectively.}\label{fig:amp-ratios}
\end{figure}

In order to test whether a criterion can be devised to separate stars dominated by one or two modes, we calculated the amplitude ratios of the various independent frequency components in our sample and normalized them by the amplitude of the strongest frequency peak. While the RRd sample size in S1 and S2 is rather small, a clear picture emerged: the five RRd stars amplitude ratios between 30--90\%, whereas all the small additional modes are below 10\%, as shown in the lower panel of Fig.~\ref{fig:amp-ratios}. The amplitude ratios steadily decrease towards the longer main pulsation periods, except for a single outlier, but that is also below the 10\% limit. Even if we do not normalize, the separation remains, as the additional modes do not exceed 10 mmag (upper panel of Fig.~\ref{fig:amp-ratios}). Based on this initial sample, we are hopeful that the classical RRab, RRc, and RRd definitions can be upheld, as this approach would also be backwards compatible with a large body of literature. 

\begin{figure*}[]
\includegraphics[width=0.98\textwidth]{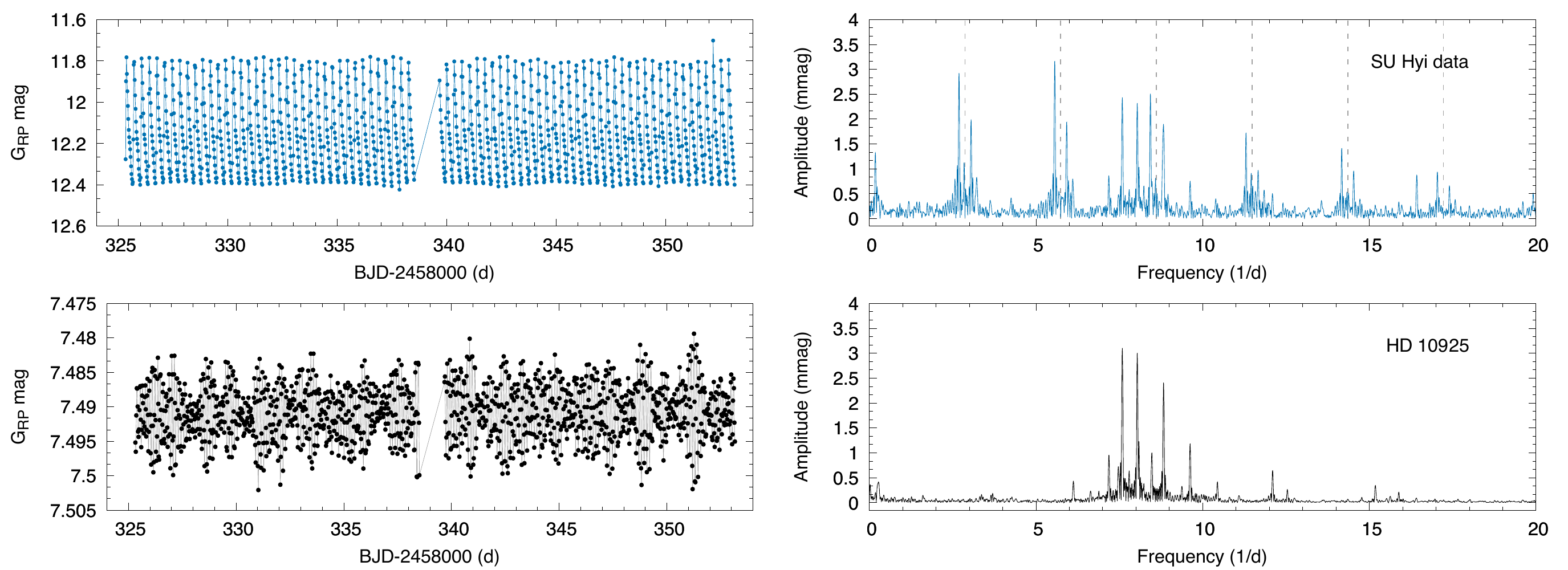}
\caption{A bright $\delta$ Scuti contaminating an RR Lyrae star. Top row: the light curve and Fourier spectrum of SU~Hyi. Dashed lines mark the harmonic series of the pulsation peak that has been removed for clarity. The contamination appears as an extra set of peaks between 6--9 d$^{-1}$. Bottom row: light curve and Fourier spectrum of the nearby bright star HD~10925. The frequency content matches the distribution of excess peaks in the spectrum of SU~Hyi. }\label{fig:tess-contam}
\end{figure*}

\subsection{Contamination}
\label{sect:contam}
One major issue with TESS is the low angular resolution of the images that causes the images of stars to blend or contaminate each other. While differential-image photometry can mitigate this, variations of blended sources will still appear in the light curves. We found one prominent example of this effect. The Fourier spectrum of SU Hyi, as shown in Fig.~\ref{fig:tess-contam}, revealed a complicated structure of peaks around the third harmonic peak. We do not expect strong additional modes to appear in this frequency range, and we could not identify clear combination frequencies with the fundamental mode either. The SIMBAD and \textit{Gaia} DR2 databases did not contain known variable stars close to our target. We therefore extracted the light curves of the nearby, bright stars for closer examination. We discovered that the bright star HD~10925 ($G_{RP} = 7.488$ mag) is in fact a previously unknown $\delta$~Scuti variable star and it is contaminating the light curve of SU Hyi. The two strongest frequency peaks at 7.58919 and 8.03771 d$^{-1}$ (87.838 and 93.029 $\mu$Hz) are separated by 0.449 d$^{-1}$. This is close to, but clearly different from the separation of the modulation sidepeaks ($2f_m = 0.360$  d$^{-1}$). This indicates that the modulation is intrinsic to the RR Lyrae star, and not caused by the presence of the contaminating source.

\section{Comparison with ground-based data and models}
\label{sect:models}
TESS provides accurate but short snapshots for RR Lyrae stars. We were interested in comparing the TESS data to other data sources that provide sparser but potentially much more extended light curves. We also compared the light curve parameters to theoretical calculations. 

\begin{figure*}
\begin{center}
\noindent
\resizebox{82mm}{!}{\includegraphics{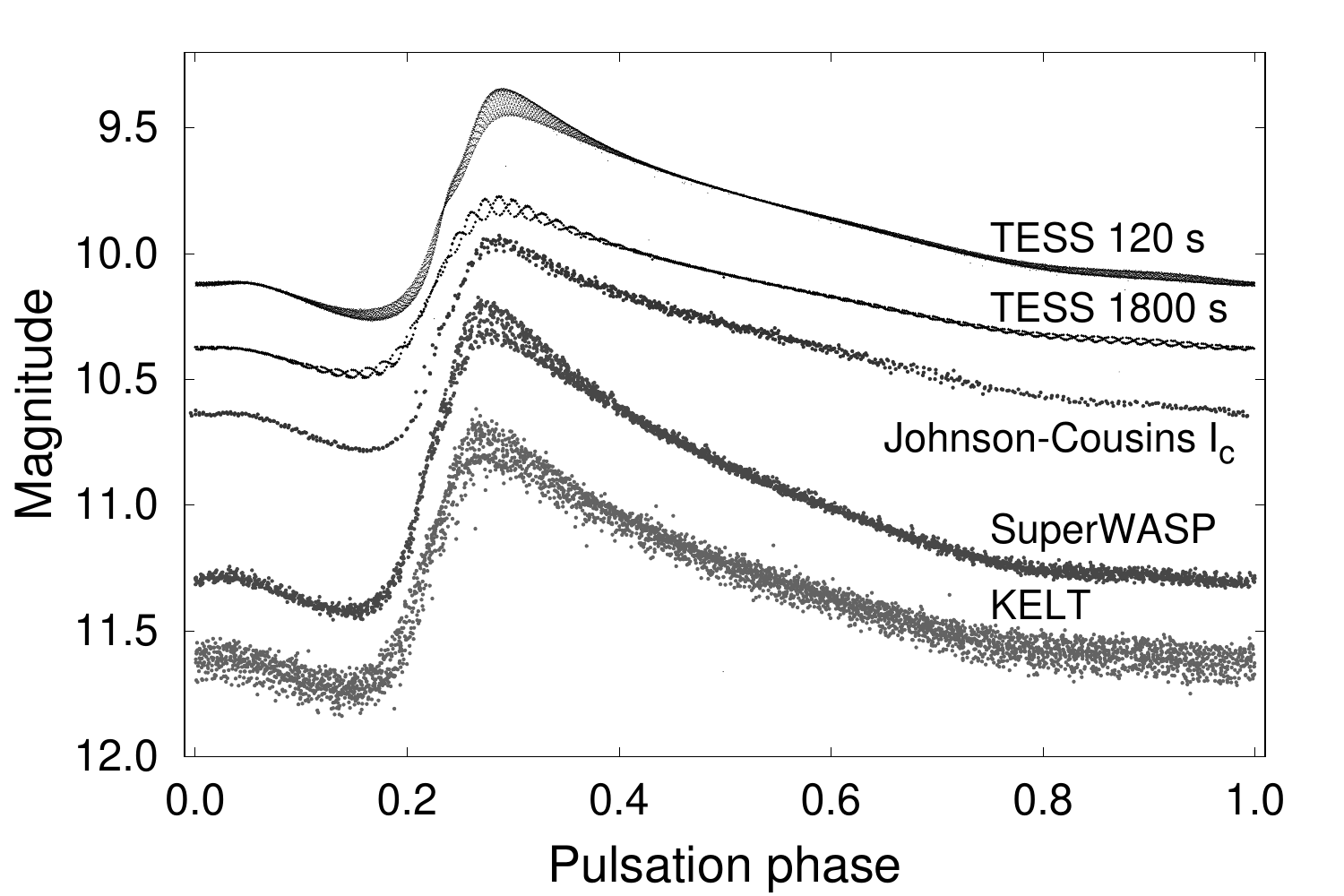}}\hspace*{6mm}%
\resizebox{82mm}{!}{\includegraphics{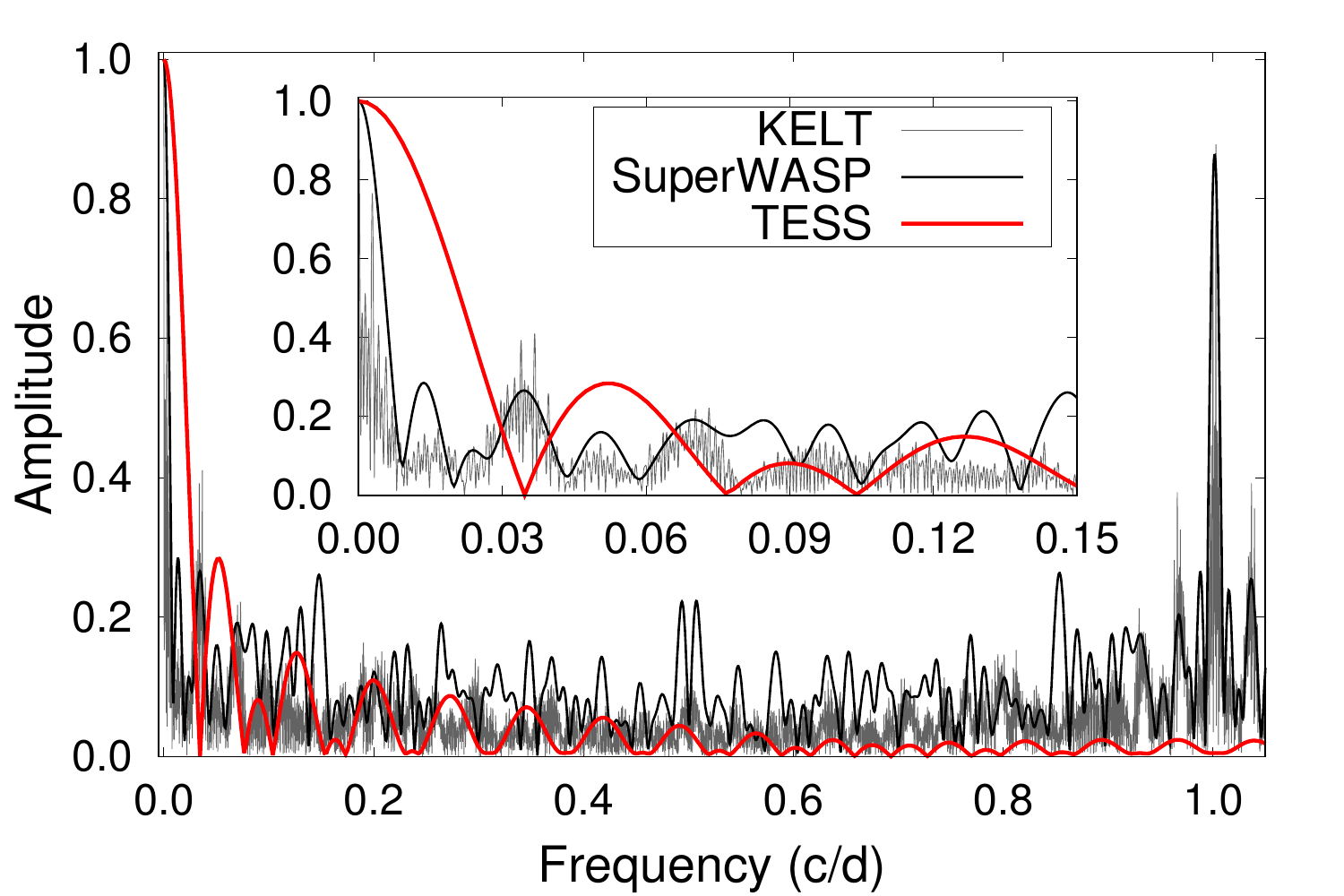}}
\vspace*{3mm}

\noindent
\resizebox{82mm}{!}{\includegraphics{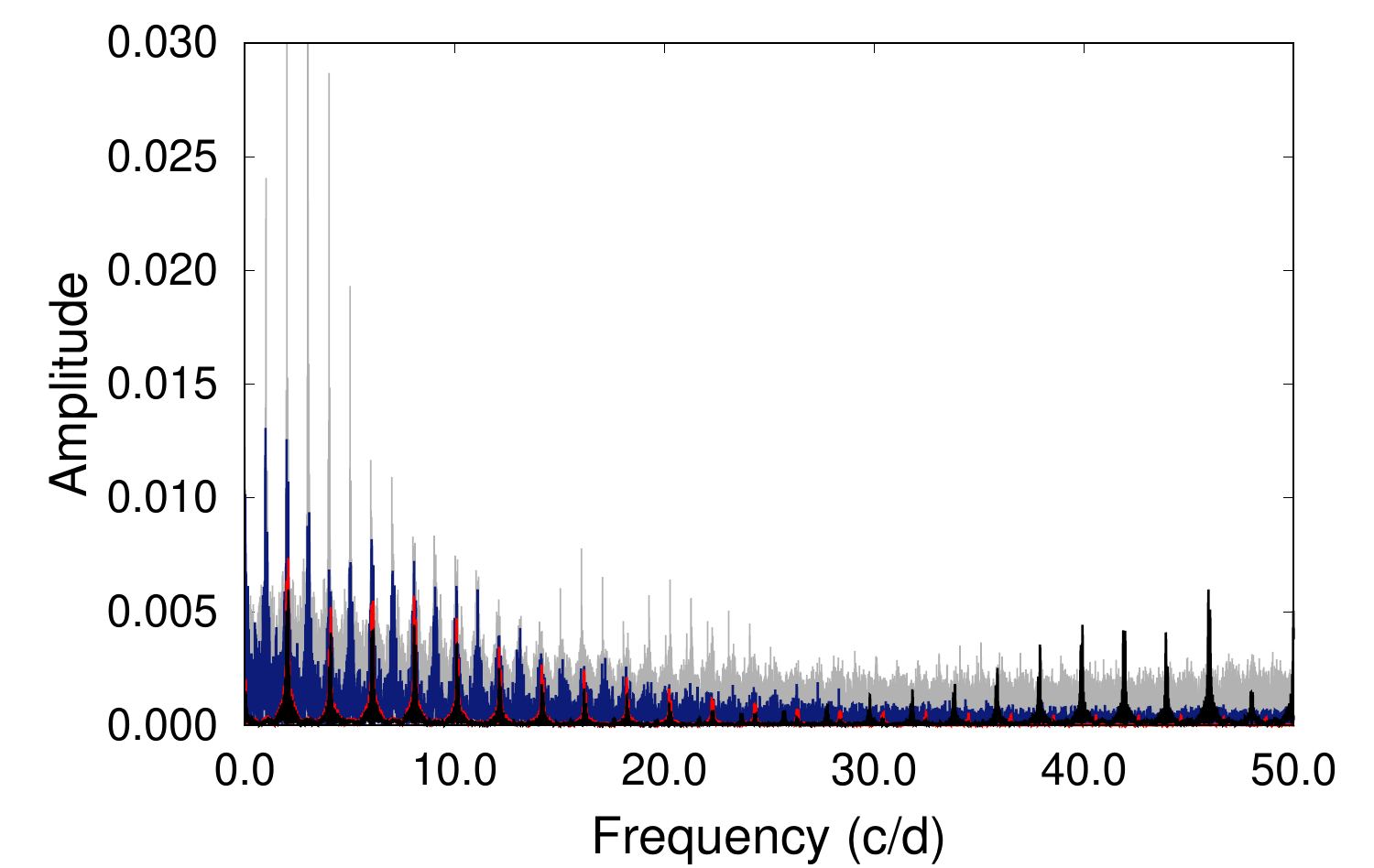}}\hspace*{2mm}
\resizebox{83mm}{!}{\includegraphics{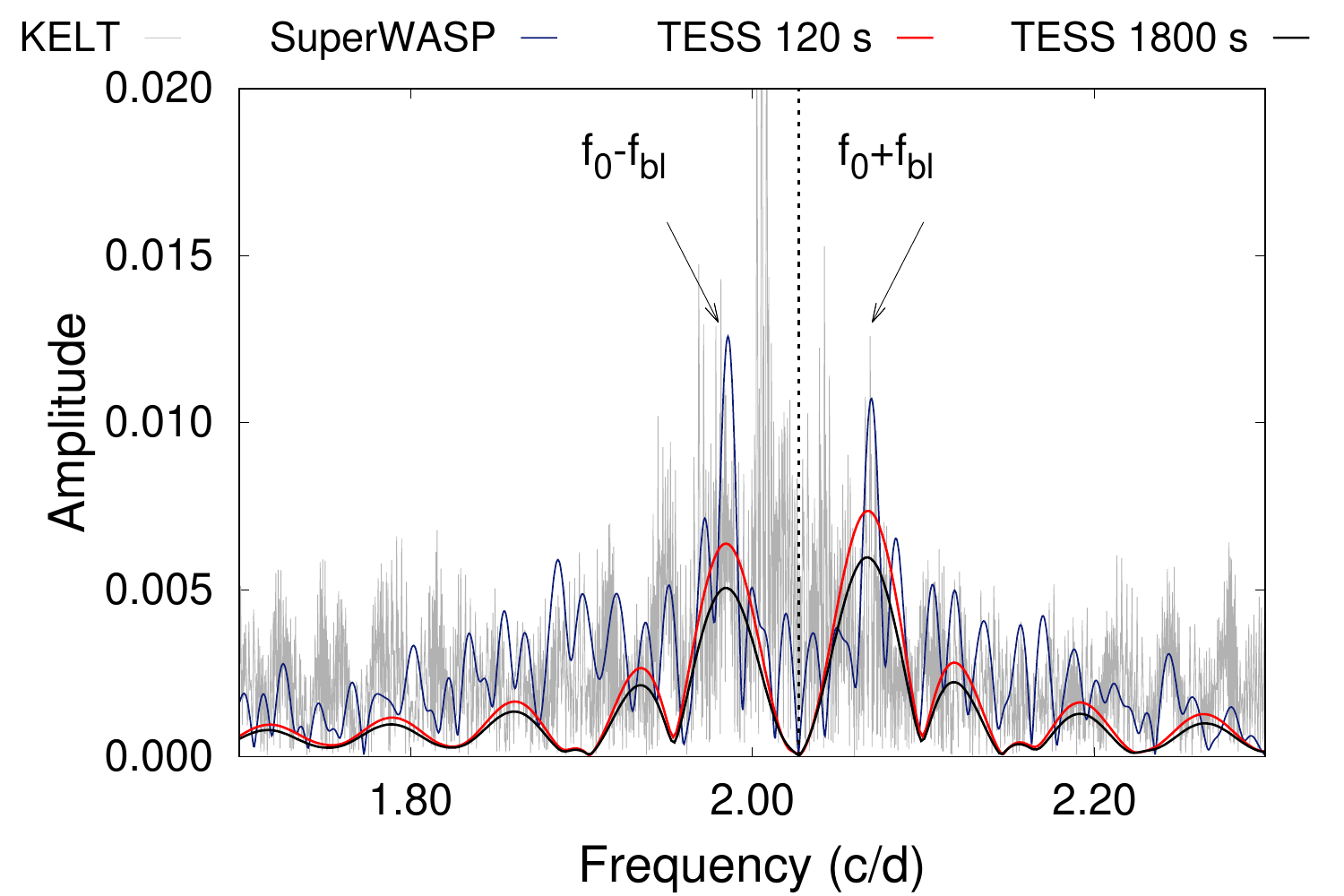}}
\end{center}
\caption{The comparison of the TESS Sector 2 30--min and 2--min cadence data to various observations taken from the ground. Light curves (upper left panel), spectral windows (upper right panel), frequency spectra (lower left panel), and  the modulation sidepeaks around the pulsation frequency (lower right panel) in the different data sets of RU Scl are plotted,  respectively.} 
\label{fig:ruscl-all}
\end{figure*}

\subsection{KELT and other sources}
We extracted photometry from the Kilodegree Extremely Little Telescope (KELT) observations for 33 stars from our sample. KELT uses small robotic cameras with telephoto lenses and a broad, red--pass filter to search for transiting exoplanets \citep{kelt,kelt-south}. For some stars where KELT data were not available, we also used light curves from the public database of another exoplanet survey, SuperWASP (Wide-Angle Search for Planets, \citealt{Butters2010,superwasp}). 

Of the 33 targets, 20 stars had useful amounts of observation in the KELT database. Seven stars were too faint for KELT and we detected only scatter with no signs of pulsation in the data. We were able to detect the modulation in RV Hor and measured it to be 78.93 d long; this is almost a day shorter than the value found by \citet{rvhor2007}.

One of the 2 min targets, RU Scl, was observed in 2017 and 2018 from Chile by one of the authors (FJH). During 13 nights we collected more than 800 points in Johnson-Cousins {\it I$_{\rm c}$} filter with a 40-cm f/6.8 Optimized Dall-Kirkham telescope equipped with an FLI CCD camera with 4k$\times$4k Kodak 16803 chip \citep{Hambsch2012}. 

For RU Scl, we were able to compare five different data sets, bearing in mind that the TESS 2-min cadence data are not independent of the 30-min cadence data, only sampled at a higher rate. RU Scl is a known fundamental-mode Blazhko star with modulation period of 23.9\,days \citep{Skarka2014}. From the variations of the maximum times, \citet{Li2018} marked this star as a potential binary candidate. In Fig.~\ref{fig:ruscl-all}, the comparison of the window functions of TESS, SuperWASP and KELT data of RU Scl are shown. Ground-based data have better resolution, but suffer from many aliases caused by regular (daily and annual) gaps in the observations. 

We detected 55 and 21 pulsation harmonics in the 2-minute and 30-minute data, respectively (see Fig. \ref{fig:ruscl-all}), but only 30 and 9 harmonics in the SuperWASP and KELT data, respectively. After removal of the pulsation frequency, $f_{0}$, and its harmonics, $kf_{0}$, we searched for the peaks corresponding to the Blazhko modulation. We identified the side-lobe peaks up to the 55th harmonic in the 2-min data. The modulation period corresponds to 24.06(4)\,d. As only one cycle was observed during the TESS observations, the period is not determined precisely and the formal error is unrealistically small. No period doubling features (half-integer multiples of $f_{0}$) were identified.

\begin{figure*}[]
\includegraphics[width=1.0\textwidth]{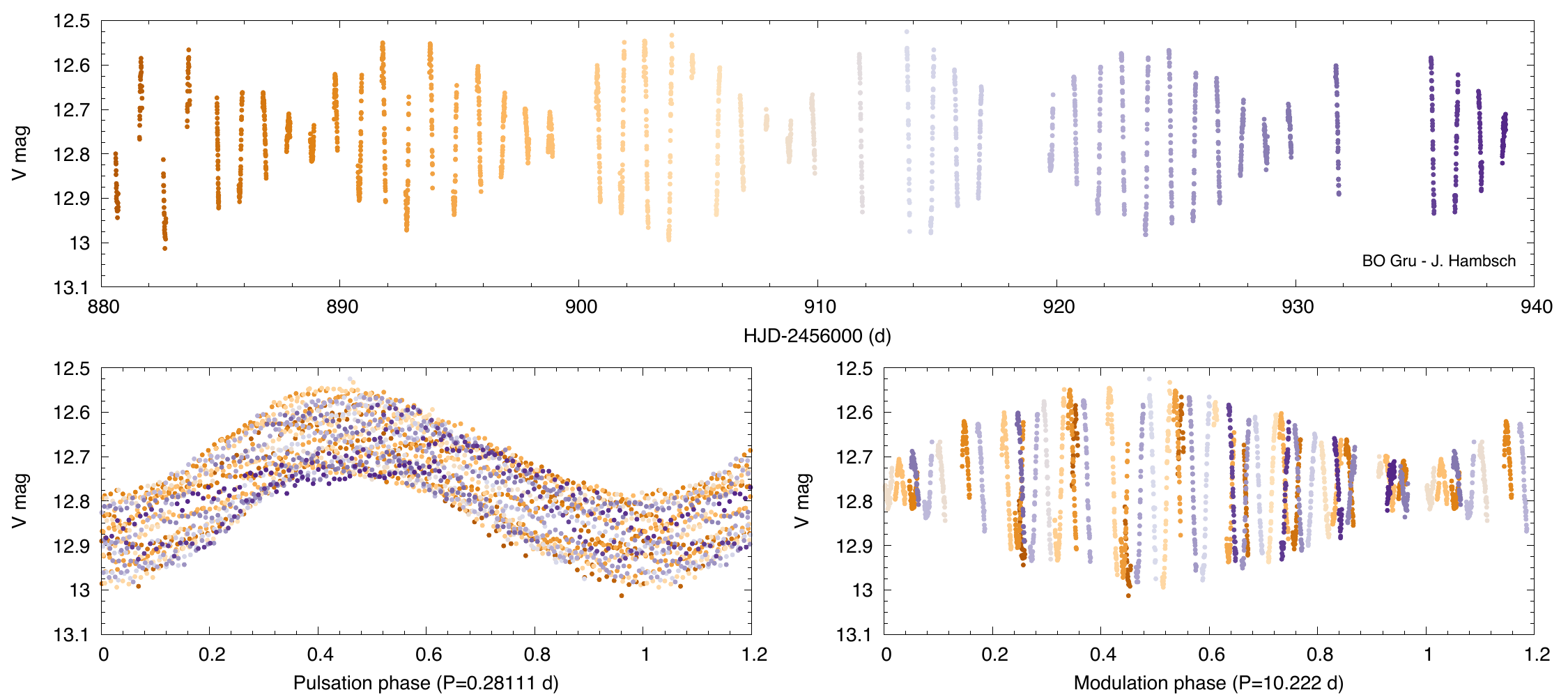}
\caption{Ground-based observations of BO Gru, the strongly modulated RRc star. Top row: the entire light curve. Bottom row: phase curves, folded both with the pulsation and with the modulation periods. Colors follow time of the observations. }\label{fig:bo-gru-h}
\end{figure*}

We also analysed observations of the strongly modulated RRc star BO~Gru. This star was also observed by FJH from Chile, but in Johnson \textit{V} band, in 2014. 2285 data points were collected on 50 nights over a period of 58 days, giving us very dense coverage. The light curve in Fig.~\ref{fig:bo-gru-h} shows the same strong modulation that we observed in the TESS data. We also plotted the light curves folded with the pulsation and modulation periods: the width of the phased light curve in the bottom left indicates that both the amplitude and the phase of the pulsation experienced strong modulation. Meanwhile, the right bottom plot suggests differences between the modulation cycles. These properties can also be detected in the TESS light curve. Overall, the modulation properties of the star appear to be unchanged almost four years apart.

The stability of the modulation in BO~Gru allowed us to compare the temporal evolution of the $A_1$ and $\phi_1$ Fourier components of the two data sets directly. We determined the modulation period to be 10.2221 d, slightly longer than the TESS-only value. We found that the $A_1$ Fourier amplitude of the redder TESS data is 62 \% of the $V$ band $A_1$ amplitude, and this ratio did not change throughout the modulation cycle.

\begin{figure*}[]
\includegraphics[width=1.0\textwidth]{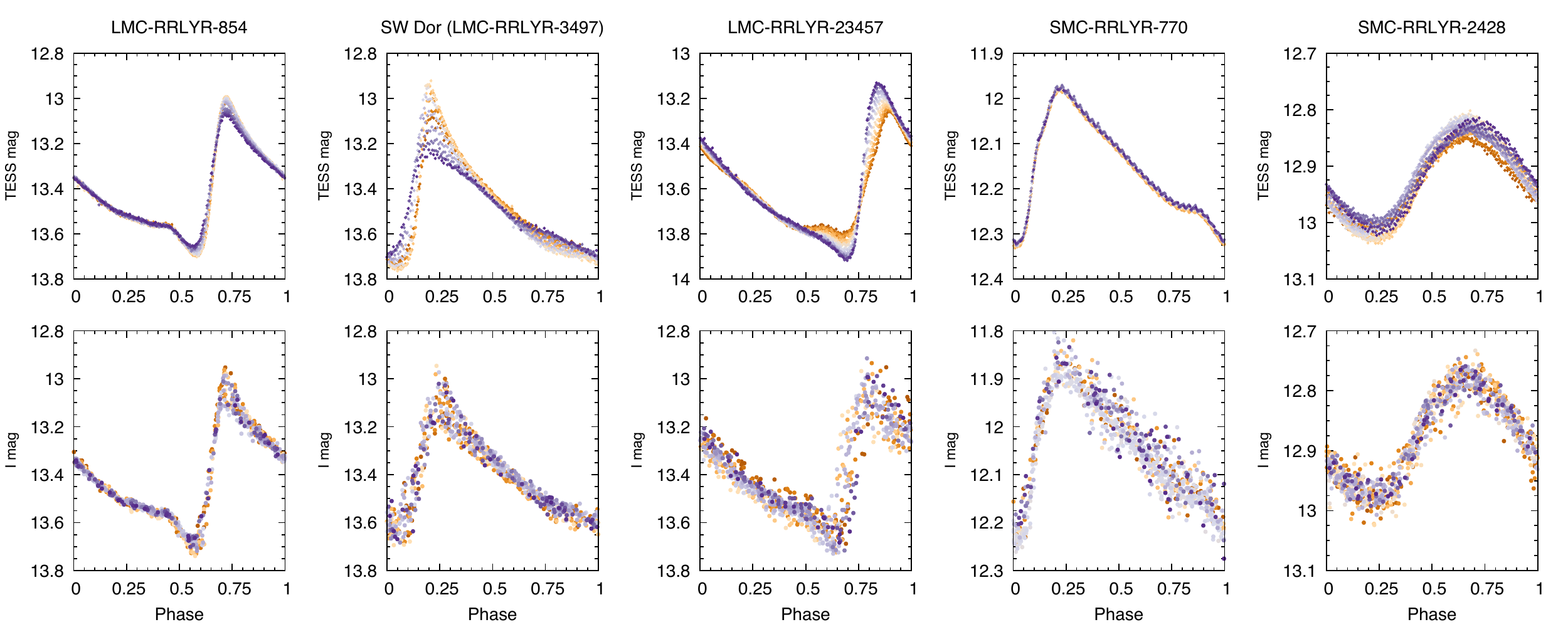}
\caption{Comparison of the TESS (top panels) and OGLE (bottom panels) light curves for the shared targets. Color scale shows progression of time, going from orange to purple.}\label{fig:tessvsogle}
\end{figure*}

\subsection{OGLE stars}
Five stars in our sample have been followed by the OGLE project. We analyzed the OGLE-IV survey data of all stars except SMC-RRLYR-0770 that was only observed in OGLE-III. Folded TESS and OGLE light curves are plotted in Fig.~\ref{fig:tessvsogle}. We did not detect additional-mode signals in the OGLE data, so direct comparison for those is not possible. We were, however, able to determine the Blazhko periods for them from the OGLE light curves, where modulation was present. 

In LMC-RRLYR-00854, we identified a stable Blazhko cycle with a period of $120.55 \pm 0.10$~d. LMC-RRLYR-03497, in contrast, shows multiple modulations, with the following periods: $P_{\rm m1} = 30.418 \pm 0.014$~d, $P_{\rm m2} = 81.80 \pm 0.06$~d, $P_{\rm m3} = 96.52 \pm 0.17$~d. However, we cannot rule out that the third is only a combination in frequency space, $2f_{\rm m3} \simeq f_{\rm m1} - f_{\rm m2} $, and that could indicate interaction, or temporal variability between the two other cycles instead of a third modulation. The data for LMC-RRLYR-23457 suggest a non-stationary Blazhko-cycle with a period of $174.0 \pm 0.3$~d. These modulation cycles are longer than our TESS light curves and thus they are not resolved in the TESS data. The OGLE-based findings, however, reinforce that our assumption to include stars with apparently non-cyclic amplitude and phase changes over the span of the TESS data in the Blazhko category is justified. Moreover, the case of LMC-RRLYR-03497 shows that when short-term data suggest different cycle lengths from the amplitude and phase variations, it is likely the sign of multiple modulation periods in the star. 

We were able to compare TESS and OGLE data for one first-overtone star, SMC-RRLYR-2428. It is a Blazhko RRc star, and one of those stars for which we cannot tell with certainty whether the slow variations are due to beating with a second mode very close to the radial one, or from Blazhko-type modulation but with very asymmetric sidepeak amplitudes, where one side is not detectable. Either way, the OGLE data show that the beat or modulation period, $35.401 \pm 0.015$~d, remained stable over the OGLE-IV observing run. 

\subsection{Pulsation models}
Light curve parameters from TESS can be, in principle, compared to theoretical light curves produced with non-linear pulsation models. One drawback is that models need to be transformed into the TESS passband first, and we cannot rely on existing models that are calculated for, e.g., the Johnson or Sloan photometric systems.

Nevertheless, as a cursory test, we compared the $R_{21}$ and $\phi_{21}$ Fourier parameters to those calculated for the \textit{I} band from the models of \citet{Marconi-2015} for a fixed composition (Z=0.004, Y=0.25). These RR Lyrae models were generated using the one-dimensional non-linear hydrodynamic model that employs time-dependent convection to iterate the pulsating envelope in time \citep{Bono-1994}. 

\begin{figure}[]
\centering
\includegraphics[width=1.0\columnwidth]{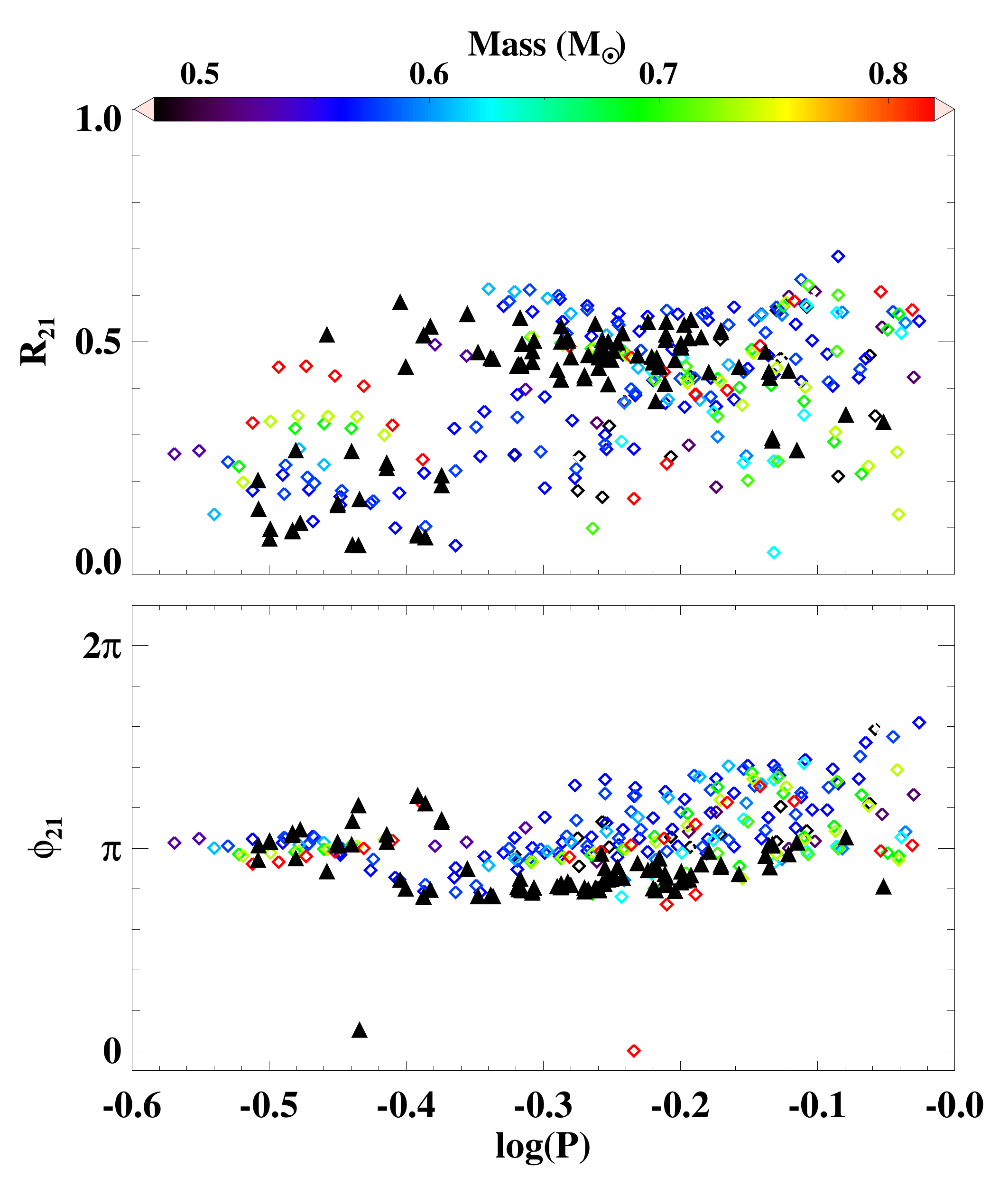}
\caption{Comparison of lower-order Fourier parameters of TESS light curves (filled triangles) with RR Lyrae pulsation models (open squares) from \citet{Marconi-2015} in $I$-band. The colorbar represents different stellar masses for RR Lyrae models.}\label{fig:models}
\end{figure}

Fig.~\ref{fig:models} displays a comparison of Fourier parameters for RR Lyrae from TESS with those for the model light curves in the $I$-band. The comparison clearly shows the same shift in the $\phi_{21}$ phase difference that we observed when comparing the TESS data to the OGLE \textit{I} band values in Fig.~\ref{fig:rrl_fourparam}. However, other differences can be found as well. The calculated $R_{21}$ values spread upwards towards long periods, whereas the upper envelope of the observed values goes downward from about $\log P \gtrsim -0.15$ or $P\gtrsim 0.7$~d. Note that theoretical amplitude parameters for classical pulsators are known to be systematically larger than the observed amplitudes \citep{bhardwaj-2017, Das-2018}. Similarly, phase parameters, like $\phi_{21}$, display a clear dependence on adopted metal-abundances in the $I$ band \citep[see Figure 5,][]{Das-2018}. Moreover, the models do not exactly reproduce all light curves in the overlapping long-period RRc and short-period RRab range. Correcting for these shortcomings in the models would require a finer grid with further fine-tuning of various input physical and convective parameters, an exercise which is beyond the scope of this study. Nevertheless, these discrepancies highlight the broader problem of reproducing observed pulsation amplitudes numerically (see, e.g., \citealt{zhou2021} for solar-like oscillations).

We must also point out that accurate comparison will indeed require transforming the luminosity curves into the TESS passband since the Fourier parameters of both Cepheid and RR Lyrae also vary with wavelength \citep{bhardwaj-2015,Das-2018}. Nevertheless, the benefits are clear: the availability of precise and continuous light curves for thousands of RR Lyrae stars from TESS provides a new opportunity to constrain Blazhko models, for example. TESS light curves, in combination with data from other passbands, can be utilized very effectively to connect photometric properties to physical parameters, such as mass, metallicity, luminosity and radius \citep{Bellinger-2020}. 

\section{Future prospects}
\label{sect:future}
\begin{figure*}
\includegraphics[width=1.0\textwidth]{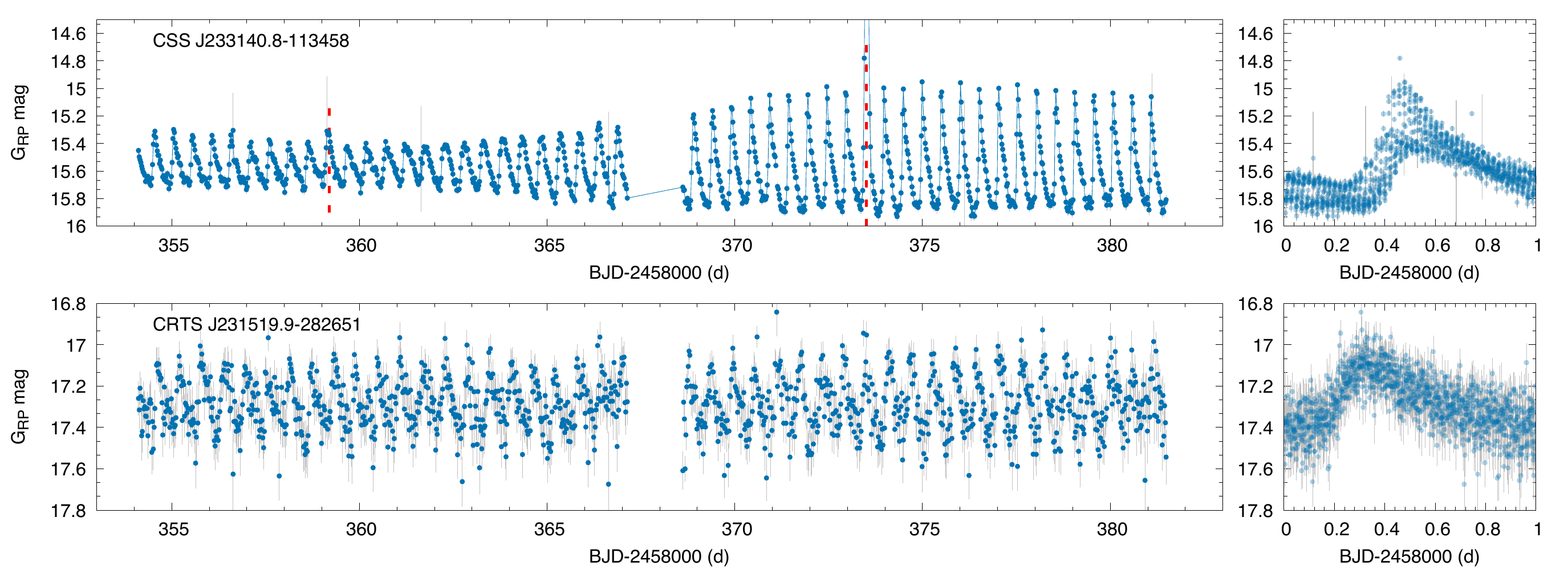}
\caption{Two faint targets from Sector 2. Left: light curves, Right: phase-folded curves. Red dashed lines indicate flux excesses from asteroids crossing the photometric aperture.}\label{fig:tess-faint}
\end{figure*}

By the end of our analysis, TESS has successfully finished its primary mission and has begun the first mission extension. The nearly all-sky data from the first two years constitute an incredibly rich resource for stellar astrophysics. In order to have a better understanding of the capabilities of TESS, we complemented our first-light study with two tests: one aimed at assessing the performance at the faint limit of the telescope, and the other a comparison of the new, 10--minute cadence FFI data of the extended mission to the original 30--min cadence data.

\subsection{Faint targets}
\textit{Kepler} was able to collect useable light curves from RR Lyrae stars at the $Kp\sim 21$~mag level \citep{molnar-leoiv-2015}. Considering the 100-fold reduction in the light correction area between the optical systems of \textit{Kepler} and TESS, we should be able to recover RR Lyrae stars down to 16 magnitudes, if the stellar density is low enough in the images to detect the target. Furthermore, RR Lyrae stars have a color index of [$G-G_{\rm RP}] \approx 0.3-0.4$ mag: since the \textit{Gaia G} and $G_{\rm RP}$ bands closely match the \textit{Kepler} and TESS passbands, respectively, they are 0.3--0.4 mag brighter in the TESS passband than in the \textit{Kepler} one, pushing the faint limit further outward.

To test these assumptions, we selected two stars from Sector 2, CSS J233140.8--113458 at $G_{\rm} = 15.55$~mag, which we expect to produce a lower-precision but clearly recognizable light curve; and CRTS J231519.9--282651 $G_{\rm} = 17.26$~mag, to see if large-amplitude variations remain detectable at this level. The light curves in Fig.~\ref{fig:tess-faint} show that this is indeed the case. The former star produced a clean Blazhko RRab light curve, whereas in the latter case only the existence of a large-amplitude, periodic, asymmetric variation is evident. However, the RRab-type variation is still clearly recognizable in the phase-folded light curve, thus mode classification will still be possible at the 17 mag level. Two short brightenings are also visible in the light curve of CSS J233140.8--113458, the brighter star, marked with red dashed lines. These are caused by a fainter and a brighter asteroid crossing the photometric aperture, respectively. Even with the TESS telescopes avoiding the vicinity of the Ecliptic, large numbers of asteroids cross the fields of view, especially that of Camera \#1, closest to the Ecliptic \citep{Pal2018,Pal2020}.

We then calculated the noise levels in the Fourier spectra of these stars to estimate if additional modes are possible to detect them. In the fainter star, the widely used detection signal-to-noise level of 4 is at 18 mmag, whereas in the brighter one it is around 5-6 mmag. Comparing these values to the Fourier amplitudes in Fig.~\ref{fig:amp-ratios}, we can conclude that at the 15.5 mag level we are able to detect the strongest additional modes just barely. At the 17 mag level, differentiating between RRab, RRc and RRd stars will be possible. However, these predictions are based on well-behaved targets for which neither strong scattered light nor blending with brighter stars affected the photometry. 

\subsection{10--minute cadence improvements}
In the first mission extension of TESS, the cadence of the FFIs was changed from 30\,min to 10\,min. Although the increased sampling lowers the per cadence accuracy, it triples the available frequency range. By the time we finished the various analyses detailed in this paper TESS had already re-observed the areas of S1 and S2 during S27 and S28. We therefore compared the performance of the faster cadence for a few targets. 

For the brighter targets presented, the 10 min sampling comes with clear advantages. In some targets the series of the pulsation harmonics extended beyond the Nyquist frequency ($f_{\rm Ny}$) with the 30-min cadence and thus the $f_{\rm Ny}-nf_0$ reflections of those above this limit showed up below $f_{\rm Ny}$. Furthermore, longer integration introduces phase smearing that lowers the observed amplitudes. This effect is clearly visible in the middle panel of Fig.~\ref{fig:ruscl-all}, where the red frequency spectrum of the 2-min cadence has higher peaks than that of the 30--min cadence data, in black. Moreover, amplitudes are getting attenuated further near the Nyquist frequency. With the 10\,min cadence, we are not facing either of these issues. The benefits will of course start to vanish for fainter targets, where high-frequency components are lost amidst the higher noise level. For sufficiently bright targets, however, amplitudes and relative phases of higher harmonics contain useful information about the precise shape and underlying physics of the light curve, beyond the usual first two $R_{\rm i1}$ and $\phi_{\rm i1}$ Fourier coefficients. Differences in the local minimum of the harmonic amplitudes and the steepness of these curves were highlighted both for RR Lyrae and Cepheid stars by \citet{benko-2016} and \citet{Plachy-Cep-2020}, respectively, but are yet to be utilized further. 

Another benefit is that the wider frequency range makes it possible to construct taller \'echelle diagrams. Finally, fast sampling allows us to track specific features such as the position and depth of the shockwave features (humps and bumps) more accurately in the light curves themselves. 

\section{Conclusions}
\label{sect:conclusions}
In this paper we studied an initial selection of 126 known or candidate bright RR Lyrae stars within Sectors 1 and 2 of the TESS mission, 118 of which turned out to be real pulsators. We were interested in the capabilities of the mission regarding this abundant and well-known class of high-amplitude pulsating stars. A companion paper looking at Cepheid stars was recently published by \citet{Plachy-Cep-2020}. Our results can be summarized as follows:

\begin{itemize}
    \item We created differential-image photometry for our targets from Sectors 1 and 2 with the \textsc{fitsh} software \citep{pal2012}. We were able to extract high-precision light curves for stars in the 9--13.5 magnitude range. We tested the limits of the data and showed that depending on the precision requirements, light curves can be recovered down to 15--17 mag brightness where the number density of Galactic RR Lyrae stars peak \citep{Plachy-TESS-2020}. 
    \item Most TESS light curves are short but can be complemented with ground-based observations, especially with the considerably longer but less accurate (on a per data point basis) OGLE survey data. Comparison with non-linear pulsation models is also promising, but luminosity variations have to be converted into the TESS passband for more detailed studies.
    \item Combining very accurate light-curve data from TESS (periods and Fourier parameters) with homogeneous parallax and hence absolute brightness and color data from \textit{Gaia} EDR3 offers us the best photometric classification scheme for classical pulsators. This is especially true for the RR Lyrae regime, where the $R_{i1}$ and $\phi_{i1}$ Fourier-parameters of binaries, as well as short-period classical and anomalous Cepheids might overlap. We identified 118 \textit{bona fide} RR~Lyrae stars, two anomalous Cepheid candidates and six non-pulsating stars in the sample.
    \item Classification based on the combined light curve shape and parallax information was found to be very effective. This way we are able to confirm stars with unusual light curves (such as BO~Gru), and are able to separate short-period ACEP candidates from genuine RR Lyrae stars. 
    \item We were able to detect modulation and estimate the frequency of the Blazhko effect, which is about 13\% in RRc and between 48--72\% for RRab stars. The light curves are generally too short for more detailed studies of the Blazhko effect: this could be remedied by focusing on the stars within the continuous viewing zones. We also showed that non-modulated RRab stars exist, but estimating the true fraction of those requires very careful data processing at this photometric accuracy. 
    \item While low-amplitude extra modes in RR Lyrae stars are possible to detect from the ground \citep[see, e.g.,][]{jurcsik2015,netzel2019}, space-based photometry is clearly superior in sensitivity. We detected various types of extra signals in a large fraction of stars. We see clear differences in the distribution of extra modes compared to the bulge sample, which suggests that physical properties---likely the metallicity---are affecting mode selection.
    \item In RRab stars, the extra signals can be grouped into three broad categories based on their modulo frequency ratios with the fundamental modes. However, the highest-amplitude peaks of those modes are not confined to the vicinity of the fundamental-mode frequency: they may appear at very different positions in the $f_{\rm addtl}+i\,f_0$ combination series (where $i=-1,0,1,2,3\dots$). Whether all these peaks are simply different modes or combination peak amplitudes are also affected by other effects, shifting the position of the highest peak in the spectrum, is a question for future studies. 
    \item We find that stars with extra modes are more frequent towards the center of the instability strip and the RRab/RRc interface. RRc stars without the $f_X$ mode can be found towards the blue edge of the instability strip, but the mode is present in all of our redder RRc and RRd stars. We describe the distribution in RRab stars for the first time and show that they are more prevalent among bluer, hotter fundamental-mode pulsators. 
    \item So far, RRab and RRc stars with low-amplitude extra modes can be clearly separated from bona-fide double-mode stars based on relative mode amplitudes: we detect an empty region between 10--25\% amplitude ratios relative to the main radial mode. We propose to continue to use the classical Bailey scheme, with RRab, RRc and RRd classes referring to the \textit{dominant} pulsation mode(s) of stars instead of the types of \textit{observed} modes. 
\end{itemize}

While it is not the perfect instrument to study RR Lyrae stars, TESS offers great potential in many respects. We will be able to build up a large and near-homogeneous database of high-precision light curves for thousands of bright RR Lyrae stars, for which spectroscopic and/or multicolor photometric data already exist or are relatively easy to obtain \citep[see, e.g.,][]{crestani2020,Crestani-2021}. Derivation of a well-calibrated photometric metallicity relation, for example, would be very beneficial. This will also undoubtedly help in disentangling the new phenomena we observe, such as the origin and the suspected metallicity dependence of the extra modes. This could also be complemented with data from globular clusters, either via ground-based campaigns, as it was done for M3 \citep{jurcsik2015}, or via processing the globular cluster data obtained in the K2 mission \citep[see, e.g.,][]{plachy-2017,Wallace-rrl-2019}. 

A larger survey of nearby RR Lyrae stars, out to 5--10 kpc, covering most of the sky, and combined with parallaxes would also be a fantastic resource. The synergy between \textit{Gaia} and TESS will make it possible to build up the cleanest sample, devoid of binaries and rotational variables, but preserving unusual or unique RR Lyrae light curves. Furthermore, a thorough search for anomalous Cepheids that may be hiding under different classifications at the moment will provide the opportunity to study these intriguing objects in more detail and to construct the first Galactic period-luminosity relation for them. Finally, an unexpected synergy may be the ability to map the amount of interstellar dust in the Galaxy in directions that are otherwise difficult to probe, such as in front of the Magellanic Clouds, through RR Lyrae stars.

\acknowledgments
\paragraph{Acknowledgements} This paper includes data collected by the TESS mission. Funding for the TESS mission is provided by the NASA Science Mission Directorate. This work has made use of data from the European Space Agency (ESA) mission {\it Gaia}, processed by the {\it Gaia} Data Processing and Analysis Consortium (DPAC). Funding for the DPAC has been provided by national institutions, in particular the institutions participating in the {\it Gaia} Multilateral Agreement.  

L. M. was supported by the Premium Postdoctoral Research Program of the Hungarian Academy of Sciences. The research leading to these results has received funding from the LP2012-31, LP2014-17 and LP2018-7 Lend\"ulet grants of the Hungarian Academy of Sciences and from the `SeismoLab' KKP-137523 \'Elvonal and NN-129075 grants of the Hungarian Research, Development and Innovation Office (NKFIH). The work reported on in this publication has been partially supported by COST Action CA18104: MW--Gaia. 

A.P.\ acknowledges the MIT Kavli Center and the Kavli Foundation for their hospitality during the stays at MIT and the NASA contract number NNG14FC03C. A.B.\ acknowledges a Gruber fellowship 2020 grant sponsored by the Gruber Foundation and the International Astronomical Union and is supported by the EACOA Fellowship Program under the umbrella of the East Asia Core Observatories Association, which consists of the Academia Sinica Institute of Astronomy and Astrophysics, the National Astronomical Observatory of Japan, the Korea Astronomy and Space Science Institute, and the National Astronomical Observatories of the Chinese Academy of Sciences. M.J.\ was supported by the Lasker Data Science Postdoctoral Fellowship of the STScI and by the Research School of Astronomy and Astrophysics at the Australian National University and funding from Australian Research Council grant No.\ DP150100250. She was likewise supported by generous visitation funding from Konkoly Observatory through the LP2014-17 Lend\"ulet grant. C.C.N.\ thanks the funding from Ministry of Science and Technology (Taiwan) under the contract 109--2112--M--008--014--MY3. E.P. was supported by the J\'anos Bolyai Research Scholarship of the Hungarian Academy of Sciences. Z.P.\ acknowledges support by the Deutsche Forschungsgemeinschaft (DFG, German Research Foundation) -- Project-ID 138713538 -- SFB 881 (``The Milky Way System", subproject A11). R.S.\ was supported by the National Science Center, Poland, Sonata BIS project 2018/30/E/ST9/00598. M.S.\ acknowledges the financial support of the Operational Program Research, Development and Education -- Project Postdoc@MUNI (No.\ CZ.02.2.69/0.0/0.0/16\_027/0008360) and the MSMT Inter Transfer program LTT20015. Funding for the Stellar Astrophysics Centre is provided by The Danish National Research Foundation (Grant agreement no.: DNRF106). We gratefully acknowledge the grant from the European Social Fund via the Lithuanian Science Council (LMTLT) grant No.\ 09.3.3-LMT-K-712-01-0103. 

This research has made use of the SIMBAD database, operated at CDS, Strasbourg, France, the International Variable Star Index (VSX) database, operated at AAVSO, Cambridge, Massachusetts, USA, and NASA's Astrophysics Data System (ADS).

%

\vspace{5mm}
\facilities{TESS \citep{ricker2015}, \textit{Gaia} \citep{gaia2016}, SIMBAD \citep{simbad}, VSX}


\software{fitsh \citep{pal2012},
lightkurve \citep{lightkurve}}, Period04 \citep{period04}, gnuplot, mwdust \citep{bovy2016}, galpy \citep{galpy2015}, echelle\_toggler \citep{m_joyce_2021_4427688}, Astropy \citep{astropy:2013, astropy:2018}



\bibliography{tessrrl.bib}



\appendix

\section{List of targets, light curves and classification results}

\begin{figure*}[!hb]
\centering
\includegraphics[width=0.8\textwidth]{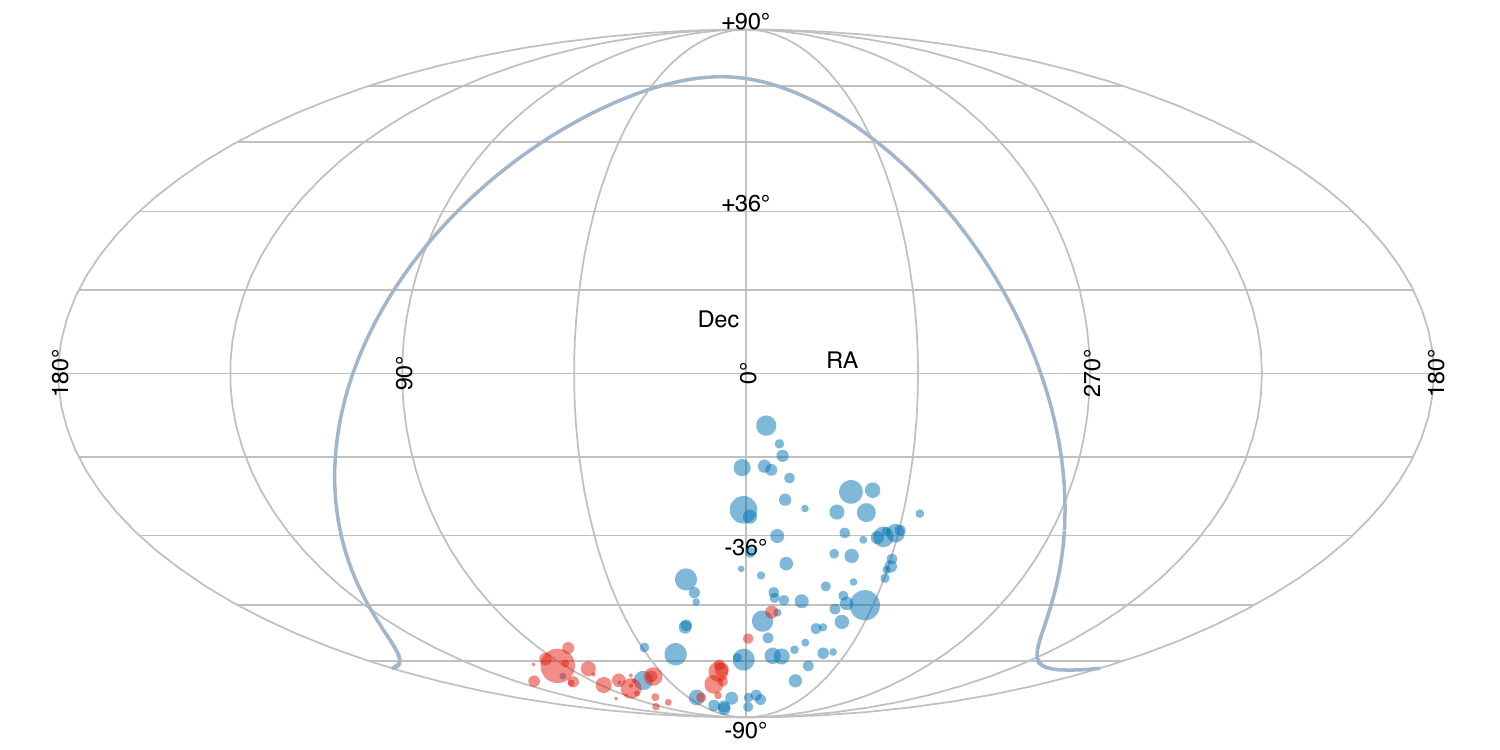}
\caption{Positions of the RR Lyrae targets in the sky. The blue line is the galactic plane. Blue dots are single-sector targets, red dots were observed in both S1 and S2. Sizes correspond to brightness. }\label{fig:skymap}
\end{figure*}

Below we list the stars we identified as RR Lyrae or candidate ACEP stars, as well as updated classifications for some of the stars that were identified as RR Lyrae at some point. In Fig.~\ref{fig:skymap} we provide a map showing the positions of the RR Lyrae targets in the sky. Since target selection was based largely on the \textit{Gaia} DR2 catalog, we did not update the naming scheme with EDR3 identifiers. Nevertheless, we checked the \texttt{gaiaedr3.dr2\_neighbourhood} crossmatch catalog if the identifiers were carried into the new catalog. The only star in the sample whose identification changed is VW~Scl, which is named DR2 4985455994038393088 and EDR3 4985455998336183168 in the two catalogs, respectively. 

The tables in the Appendix contain the following information: 
\begin{itemize}
  \setlength\itemsep{0.1em}
    \item the lists of stars identified as RRab, RRc, RRd stars or ACEP candidates (Tables \ref{tab:rrab_ids}, \ref{tab:rrc_ids}, \ref{tab:rrd_ids} and \ref{tab:acep_ids}),
    \item the list of stars identified as other types of variable stars (Table \ref{tab:non-rrl_ids}),
    \item a sample table of the TESS differential-image photometry data (Table \ref{tab:photometry}, this table is available online in its entirety),
    \item Blazhko identifications and periods (Table \ref{tab:bl})
    \item variation periods, distances, calculated absolute magnitudes and insterstellar extinction values (Table \ref{tab:dist-per-color})
    \item RVs and calculated $U$, $V$ $W$ velocity components (Table \ref{tab:velocities}).
    \item Additional modes and signals identified (Table \ref{tab:extramodes}).
\end{itemize}

\vfill

\startlongtable
\begin{deluxetable*}{cclcccc}
\tablecaption{Stars identified as RRab type in this study. Parameters are based on the \textit{Gaia} DR2 catalog. \textit{Gaia} source identifiers are identical in EDR3 for all stars except for VW~Scl (EDR3 4985455998336183168), marked with asterisk.
\label{tab:rrab_ids}}
\tablehead{\colhead{Gaia DR2 ID}  &  \colhead{Name}  &  \colhead{Sector}  &  \colhead{RA$_{2000}$}  &  \colhead{Dec$_{2000}$}  &  \colhead{G$_{\rm RP}$ mag}  &  \colhead{G mag} }
\startdata
2309225008197193856  &  WW Scl  &  S2  &  1.50821  &  --36.90425  &  13.071  &  13.475  \\
2313783205448950784  &  TW Scl  &  S2  &  358.67366  &  --33.48136  &  12.475  &  12.836  \\
2329707050101123712  &  UZ Scl  &  S2  &  350.69599  &  --30.11941  &  11.993  &  12.369  \\
2334529752915355520  &  TX Scl  &  S2  &  358.84280  &  --26.30201  &  11.986  &  12.413  \\
2336550174250087936  &  RU Scl  &  S2  &  0.70068  &  --24.94538  &  9.991  &  10.392  \\
2385090584663565696  &  V356 Aqr  &  S2  &  348.95810  &  --23.00365  &  12.247  &  12.612  \\
2394898640700233600  &  OW Aqr  &  S2  &  353.07486  &  --17.39756  &  12.276  &  12.645  \\
2408425936552330880  &  V360 Aqr  &  S2  &  350.12811  &  --14.79922  &  12.212  &  12.586  \\
\enddata
\end{deluxetable*}

\startlongtable
\begin{deluxetable*}{cclcccc}
\tablecaption{Stars identified as RRc type in this study.
\label{tab:rrc_ids}}
\tablehead{\colhead{Gaia DR2 ID}  &  \colhead{Name}  &  \colhead{Sector}  &  \colhead{RA$_{2000}$}  &  \colhead{Dec$_{2000}$}  &  \colhead{G$_{\rm RP}$ mag}  &  \colhead{G mag} }
\startdata
2394631356295692032  &  NSVS 14632323  &  S2  &  354.96391  &  -16.74045  &  12.077  &  12.354 \\
2399010298791985280  &  ASAS J231209--1855.4  &  S2  &  348.0382  &  -18.92395  &  12.474  &  12.699 \\
4628067852624828672  &  ---  &  S1--S2  &  69.47026  &  -76.08053  &  13.009  &  13.343 \\
4633341282190404864  &  SX Hyi  &  S1  &  31.91403  &  -77.81319  &  12.304  &  12.614 \\
4643476198121275648  &  BB Hyi  &  S2  &  38.32818  &  -73.61193  &  11.745  &  12.021 \\
4655249601190389632  &  XX Dor  &  S1--S2  &  74.80703  &  -69.59513  &  11.01  &  11.390 \\
4659096066472534272  &  ASAS J055122--6812.7  &  S1--S2  &  87.84083  &  -68.2122  &  11.664  &  11.960 \\
4669338159011904768  &  NSV 1432  &  S2  &  59.55251  &  -66.49459  &  11.217  &  11.523 \\
\enddata
\end{deluxetable*}

\begin{deluxetable*}{cclcccc}
\tablecaption{Stars identified as RRd type in this study.
\label{tab:rrd_ids}}
\tablehead{\colhead{Gaia DR2 ID}  &  \colhead{Name}  &  \colhead{Sector}  &  
\colhead{RA$_{2000}$}  &  \colhead{Dec$_{2000}$}  &  \colhead{G$_{\rm RP}$ mag}  &  \colhead{G mag} }
\startdata
4672117758766686720  &  SW Ret  &  S1--S2  &  52.08425  &  --64.97716  &  12.373  &  12.723 \\
5271444195402177792  &  AL Vol  &  S1--S2  &  124.03922  &  --66.74651  &  12.323  &  12.699 \\
6521271603997336832  &  CZ Phe  &  S1--S2  &  359.09074  &  --53.48948  &  12.514  &  12.851 \\
6529889228241771264  &  ---  &  S2  &  350.10878  &  --43.35176  &  12.578  &  12.904 \\
6562909314771450368  &  Z Gru  &  S1  &  323.65454  &  --49.12469  &  11.966  &  12.289 \\
\enddata
\end{deluxetable*}

\begin{deluxetable*}{cclcccc}
\tablecaption{Stars identified as ACEP candidates.
\label{tab:acep_ids}}
\tablehead{\colhead{Gaia DR2 ID}  &  \colhead{Name}  &  \colhead{Sector}  &  \colhead{RA$_{2000}$}  &  \colhead{Dec$_{2000}$}  &  \colhead{G$_{\rm RP}$ mag}  &  \colhead{G mag} }
\startdata
6508871277499009664  &  ASAS J221052-5508.0    &  S1  &  332.71736  &  --55.13313  &  12.279  &  12.686 \\
6785264788206270336  &  SX PsA  &  S1  &  324.93053  &  -30.89415  &  12.856  &  13.199\\
\enddata
\end{deluxetable*}

\begin{deluxetable*}{lcccccccc}
\tabletypesize{\footnotesize}
\tablecaption{Stars erroneously identified as RR Lyrae stars in previous studies.
\label{tab:non-rrl_ids}}
\tablehead{\colhead{Gaia DR2 ID}  &  \colhead{Name}  &  \colhead{Period (d)}  &  \colhead{Sector}  &  \colhead{Type}  &  \colhead{D (pc)}  &  \colhead{$\sigma$D (pc)} & \colhead{$M_G$} & \colhead{$\sigma M_G$} }
\startdata
4756456588457054848 & UW Dor & 0.33854 & S1 & EW & 559 & +3.4/-3.2 & 4.437 & 0.023 \\
4767858592749306240 & ASAS J055322-5417.9 & 0.39370 & S1-S2 & EW/EB & 839.4 & +11.0/-11.6 & 3.075 & 0.041 \\
4998159961839894656 & UY Scl & 0.36436 & S2 & EW/EB & 187.4 & +1.0/-0.9 & 5.252 & 0.023 \\
6459704308665358848 & -- & 0.36792 & S1 & ROT? & 87.6 & +0.1/-0.1 & 7.205 & 0.005 \\
6559141705035584768 & CRTS J215543.7-500050 & 30.3/2.4523 & S1 & ROT/EB & 1469.8 & +32.6/-25.6 & 2.567 & 0.045 \\
6609938008366400256 & ASAS J225559-2709.9 & 0.31037 & S2 & ROT? & 106.8 & +0.2/-0.2 & 7.097 & 0.007 \\
\enddata
\end{deluxetable*}

\begin{deluxetable*}{lccccccc}
\tablecaption{Sample table of the \textsc{fitsh} differential-image photometry of the FFI targets. The table contains magnitudes, fluxes and average background fluxes from the sky annulus. The entire table is available online.
\label{tab:photometry}}
\tablehead{\colhead{Name}  &  \colhead{BJD (d)}  &  \colhead{$G_{\rm RP}$ (mag)}  &  \colhead{$\delta\,G_{\rm RP}$ (mag)} &  \colhead{Flux (e/s)}  &  \colhead{$\delta\,F$ (e/s)}  &  \colhead{$\overline{F_{\rm bg}}$ (e/s)} & \colhead{$\delta\,\overline{F_{bg}}$ (e/s)} }
\startdata
ad-psa  &  2458325.323520  &  12.5119  &  0.0019  &  1626.50  &  2.85  &  0.18  &  0.62  \\
ad-psa  &  2458325.344350  &  12.5230  &  0.0019  &  1609.88  &  2.82  &  0.06  &  0.61  \\
ad-psa  &  2458325.365180  &  12.5294  &  0.0017  &  1600.49  &  2.56  &  0.01  &  0.55  \\
ad-psa  &  2458325.386020  &  12.5389  &  0.0019  &  1586.48  &  2.77  &  0.01  &  0.60  \\
ad-psa  &  2458325.406850  &  12.5537  &  0.0021  &  1564.98  &  2.99  &  0.16  &  0.66  \\
ad-psa  &  2458325.427680  &  12.5689  &  0.0020  &  1543.29  &  2.87  &  0.29  &  0.63  \\
ad-psa  &  2458325.448520  &  12.5863  &  0.0021  &  1518.67  &  2.98  &  0.31  &  0.65  \\
ad-psa  &  2458325.469350  &  12.6099  &  0.0024  &  1486.03  &  3.24  &  0.29  &  0.72  \\
\multicolumn{8}{l}{\dots}\\
\enddata
\end{deluxetable*}

\begin{deluxetable*}{lcc|lcc}
\tablecaption{Blazhko identifications in the RRab sample, plus the two ACEP (candidate) stars. Notation: BL indicates a known Blazhko star, with the literature modulation period listed in parentheses. BL (new) indicates newly detected modulation in the TESS light curve. BL (new per) indicates a new modulation period detected in a known Blazhko star. Asterisk indicates presence of low-amplitude extra modes.
\label{tab:bl}}
\tablehead{\colhead{ID}  &  \colhead{Blazhko}  &  \colhead{Period (d)}  & \colhead{ID}  &  \colhead{Blazhko}  &  \colhead{Period (d)} }
\startdata
DR  Dor*  &  BL (new)  &  long  &  AR  Oct*  &  BL?  &  ~  \\
RT  Dor  &  BL (new)  &  long  &  TX  Scl  &  BL?  &  ~  \\
SU  Hyi  &  BL (new)  &  5.55 d  &  W  Tuc*  &  BL?  &  ~  \\
AO  Ind*  &  BL (new)  &  $>30$ d  &  ASAS  J065818--7411.7  &  BL?  &  ~  \\
DE  Ind*  &  BL (new)  &  long  &  GDR2 4758746729436076544  &  BL?  &  ~  \\
T  Men*  &  BL (new)  &  $\approx 19$ d  &  NSV  14530  &  BL?  &  ~  \\
SZ  Men*  &  BL (new)  &  long  &  RR  Gru  &  contam  &  ~  \\
AD  PsA*  &  BL (new)  &  long  &  V  Ind  &  contam  &  ~  \\
\enddata
\end{deluxetable*}

\begin{longrotatetable}
\begin{deluxetable*}{lccccccccccc}
\tablecaption{Periods, distances, calculated absolute magnitudes and absorption coefficients of the pulsating star targets. 
\label{tab:dist-per-color}}
\tabletypesize{\footnotesize}
\tablecolumns{12}
\tablehead{\colhead{Gaia DR2 ID}  &  \colhead{Name}  &  \colhead{Period}  & \colhead{Distance}  &  \colhead{$\sigma$D}  &  \colhead{$M_{\rm G}$} &  \colhead{$\sigma M_{\rm G}$} &  \colhead{$M_{\rm BP}$} &  \colhead{$\sigma M_{\rm BP}$ } &  \colhead{$M_{\rm RP}$} &  \colhead{$\sigma M_{\rm RP}$} & \colhead{$A_g$} \\ \colhead{}  &  \colhead{}  &  \colhead{(d)}  & \colhead{(pc)}  &  \colhead{(pc)}  &  \colhead{(mag)} &  \colhead{(mag)} &  \colhead{(mag)} &  \colhead{(mag)} &  \colhead{(mag)} &  \colhead{(mag)} &  \colhead{(mag)} }
\startdata
2309225008197193856 & WW Scl & 0.78492 & 4033 & +279/-237 & 0.392 & 0.149 & 0.602 & 0.179 & 0.001 & 0.165 & 0.042 \\
2313783205448950784 & TW Scl & 0.61678 & 2525 & +129/-120 & 0.780 & 0.119 & 1.003 & 0.154 & 0.444 & 0.135 & 0.034 \\
2329707050101123712 & UZ Scl & 0.44912 & 2486 & +237/-178 & 0.410 & 0.199 & 0.628 & 0.244 & -0.003 & 0.228 & 0.000 \\
2334529752915355520 & TX Scl & 0.72779 & 2114 & +84/-86 & 0.775 & 0.097 & 0.988 & 0.124 & 0.373 & 0.111 & 0.000 \\
2336550174250087936 & RU Scl & 0.49336 & 788 & +18/-16 & 0.826 & 0.060 & 1.105 & 0.081 & 0.443 & 0.081 & 0.000 \\
2385090584663565696 & V356 Aqr & 0.55459 & 2040 & +82/-66 & 0.916 & 0.097 & 1.084 & 0.149 & 0.589 & 0.122 & 0.141 \\
2394631356295692032 & NSVS 14632323 & 0.35554 & 2114 & +84/-79 & 0.737 & 0.093 & 0.885 & 0.117 & 0.456 & 0.104 & 0.000 \\
2394898640700233600 & OW Aqr & 0.65519 & 2676 & +157/-108 & 0.278 & 0.127 & 0.471 & 0.181 & -0.016 & 0.152 & 0.169 \\
\enddata
\end{deluxetable*}
\end{longrotatetable}

\startlongtable
\begin{deluxetable*}{lcccccccccc}
\tablecaption{The $U$, $V$ and $W$ velocity components and radial velocities. The "Source" column refers the following RV data sources: Gaia: \citet{gaia2018}; GALAH: \citet{Buder2020GalahDR3}; Layden: \citet{Layden1994}; RAVE: \citet{Steinmetz2020Rave}.
\label{tab:velocities}}
\tabletypesize{\footnotesize}
\tablecolumns{12}
\tablehead{\colhead{Gaia DR2 ID}  &  \colhead{Name}  &  \colhead{U (km/s)}  & \colhead{$\sigma$U}  &  \colhead{V (km/s)}  & \colhead{$\sigma$V}  &  \colhead{W (km/s)}  & \colhead{$\sigma$W}  &  \colhead{Source} &  \colhead{RV (km/s)}  & \colhead{$\sigma$RV}  }
\startdata
2329707050101123712 & UZ Scl & -208 & 18 & 109 & 10 & 16.9 & 9.6 & Layden & -4 & 10 \\
2336550174250087936 & RU Scl & -144 & 4 & -131 & 4 & -76.9 & 7.9 & Layden & 38 & 8 \\
2438710609949867776 & BR Aqr & -61 & 5 & -28 & 1 & -62.3 & 9.1 & Layden & 29 & 10 \\
4628067852624828672 & -- & -253 & 59 & -157 & 22 & 11.7 & 42 & GALAH & 150 & 75 \\
4631934555845355136 & RW Hyi & -462 & 51 & 252 & 30 & 13.2 & 47.3 & GALAH & 405 & 75 \\
4633341282190404864 & SX Hyi & -131 & 51 & 329 & 29 & -1.6 & 46.7 & GALAH & 212 & 75 \\
4643476198121275648 & BB Hyi & -268 & 51 & 236 & 23 & -97.9 & 49.2 & GALAH & 319 & 75 \\
4643606391466365824 & TW Hyi & -364 & 12 & 41 & 3 & 46.2 & 11.1 & Layden & 227 & 9 \\
\enddata
\end{deluxetable*}

\startlongtable
\begin{deluxetable*}{lcccccccccc}
\tablecaption{Periods and relative Fourier parameters.}
\label{tab:fourparam}
\tabletypesize{\footnotesize}
\tablecolumns{11}
\tablehead{\colhead{Gaia DR2 ID}  &  \colhead{Name}  &  \colhead{Period (d)}  & \colhead{$R_{21}$}   & \colhead{$\sigma R_{21}$} & \colhead{$R_{31}$}   & \colhead{$\sigma R_{31}$} & \colhead{$\phi_{21}$}  & \colhead{$\sigma \phi_{21}$} & \colhead{$\phi_{31}$}   & \colhead{$\sigma \phi_{31}$}  }
\startdata
 \multicolumn{11}{c}{RR Lyrae stars}  \\
2309225008197193856 &  WW Scl &  0.78483 &  0.4683 &  0.0032 &  0.2747 &  0.0052 &  3.0314 &  0.0029 &  6.2151 &  0.0051 \\
2313783205448950784 &  TW Scl &  0.61681 &  0.4856 &  0.0040 &  0.2899 &  0.0062 &  2.9645 &  0.0039 &  6.0443 &  0.0065 \\
2329707050101123712 &  UZ Scl &  0.44914 &  0.5306 &  0.0014 &  0.3750 &  0.0017 &  2.5310 &  0.0014 &  5.3794 &  0.0017 \\
2334529752915355520 &  TX Scl &  0.72777 &  0.4792 &  0.0028 &  0.2905 &  0.0043 &  3.0279 &  0.0028 &  6.2117 &  0.0043 \\
2336550174250087936 &  RU Scl &  0.49336 &  0.5010 &  0.0041 &  0.3695 &  0.0055 &  2.5344 &  0.0043 &  5.3652 &  0.0054 \\
2385090584663565696 &  V356 Aqr &  0.55471 &  0.4957 &  0.0093 &  0.324 &  0.012 &  2.6914 &  0.0086 &  5.718 &  0.012 \\
2394631356295692032 &  NSVS 14632323 &  0.35525 &  0.110 &  0.013 &  0.065 &  0.022 &  3.376 &  0.014 &  0.842 &  0.022 \\
\enddata
\end{deluxetable*}

\startlongtable
\begin{deluxetable*}{lccccccc}
\tablecaption{Simplified identification of extra mode signals, focusing on the region between the dominant pulsation frequency and its first harmonic. Labels refer to the type of frequency groups identified in the text. }
\label{tab:extramodes}
\tabletypesize{\footnotesize}
\tablecolumns{8}
\tablehead{  \colhead{Name}   & \colhead{FM freq.}   & \colhead{FM amp.} & \colhead{O1 freq.}   & \colhead{O1 amp.} & \colhead{Signal freq.}  & \colhead{Signal Amp.} & \colhead{Label}  \\
\colhead{~}   & \colhead{($d^{-1}$)}   & \colhead{(mag)} & \colhead{($d^{-1}$)}   & \colhead{(mag)} & \colhead{($d^{-1}$)}  & \colhead{(mag)} & \colhead{~} }
\startdata
 \multicolumn{8}{c}{RRab stars}  \\
 AD PsA & 1.32132 & 0.15943 & -- & -- & 0.61258 & 0.00021 & PD? \\
 AE PsA & 1.82904 & 0.22569 & -- & -- & 2.49730 & 0.00126 & $f_1$ \\
 AO Ind & 2.51195 & 0.23661 & -- & -- & 3.70060 & 0.00126 & PD? \\
 AO Ind & 2.51195 & 0.23661 & -- & -- & 3.77605 & 0.00164 & PD  \\
 AO Ind & 2.51195 & 0.23661 & -- & -- & 4.24491 & 0.00110 & $f_2$ \\
 AO Ind & 2.51195 & 0.23661 & -- & -- & 4.32216 & 0.00166 & $f_2$ \\
 AQ Gru & 1.57001 & 0.18950 & -- & -- & 1.07763 & 0.00077 & $f_2-f_0$? \\
\enddata
\end{deluxetable*}

\end{document}